\documentclass[12pt]{article}

\usepackage{newtxtext,newtxmath}
\usepackage{graphicx}
\usepackage{lineno}
\usepackage[letterpaper,margin=1in]{geometry}

\renewenvironment{abstract}
	{\quotation}
	{\endquotation}

\date{}

\makeatletter
\renewcommand{\fnum@figure}{\textbf{Figure \thefigure}}
\renewcommand{\fnum@table}{\textbf{Table \thetable}}
\makeatother

\usepackage{scicite}

\usepackage{url}

\providecommand{\farcs}{\mbox{\ensuremath{.\!\!^{\prime\prime}}}} % 0".3
\def\scititle{
	Exoplanets are predominately aligned with exo-Kuiper Belts
}
\title{\bfseries \boldmath \scititle}

\author{
    Jingwen~Zhang$^{1\ast}$,
    Maxwell~A.~Millar-Blanchaer$^{1}$,
    Brendan~P.~Bowler$^{1}$,\and
    Tim~D.~Pearce$^{2,3}$,
    Marvin~Morgan$^{1}$,
    William~Thompson$^{4}$,
    Dori~Blakely$^{4,5}$\and
    \small$^{1}$Department of Physics, University of California, Santa Barbara, Santa Barbara, 93106, CA, USA.\and\
    \small$^{2}$School of Physics, Trinity College Dublin, The University of Dublin, College Green, Dublin 2, Ireland.\and
    \small$^{3}$Department of Physics, University of Warwick, Gibbet Hill Road, Coventry, CV4 7AL, UK.\and
    \small$^{4}$NRC Herzberg Astronomy and Astrophysics, 5071 West Saanich Road, Victoria, V9E 2E7, BC, Canada.\and
    \small$^{5}$Department of Physics and Astronomy, University of Victoria,\\ 
    \small 3800 Finnerty Road, Elliott Building, Victoria,V8P 5C2, BC, Canada. \and
    \small$^\ast$Corresponding author. Email: jwzhang@ucsb.edu
}

\begin{document} 

% Insert the title and author list
\maketitle

% Abstract, in bold
% There are strict length limits, and not all formats have abstracts.
% Consult the journal instructions to authors for details.
% Do not cite any references in the abstract.
\begin{abstract} \bfseries \boldmath

The architectures of planetary systems encode their formation and dynamical histories. In the Solar System, the near alignment of planetary orbits, the Kuiper Belt, and the Sun’s equator supports planet formation in a rotating disk of gas and dust. However, some exoplanet systems show planets on orbits misaligned with their host stars or other planets, raising the question of whether misalignments are primordial or produced by post-formation dynamics. We present the first population-level constraints on the mutual inclinations of 32 planets in 16 systems relative to debris disks. We find that planets across orbital distances from 0.1 to 100 AU are predominantly aligned with the debris disks within 10 deg (30/32 planets). The alignment holds across systems spanning tens of million years to a few gigayears in age. These results indicate that the aligned architecture of the Solar System is representative of the birth conditions of exoplanet systems.

\end{abstract}

% The first paragraph of any Science paper does NOT have a heading
% Nor is it indented

\subsection*{Introduction}
\noindent
The Kuiper Belt is a broad population of icy bodies beyond Neptune, with its main classical belt located around 40-48 AU from the Sun \cite{Edgeworth1949, Kuiper1951, Fernandez1980, Jewitt1993}. It is the largest surviving remnant of the Solar System’s primordial planet-forming disk \cite{Gladman2021}. The Kuiper belt has a dynamically complex structure, consisting of a cold, primordial population and a hot population shaped by resonant and scattering perturbations from Neptune \cite{Hahn2005, Nesvorny2015, Malhotra2019}. The mean plane of Kuiper Belt is mildly warped \cite{Volk2017}. But across the  classical belt, especially within the primordial cold population, it remains broadly aligned to within a few degrees of  the Solar System's invariable plane, the plane perpendicular to the planets’ combined orbital angular momentum vector \cite{Brown2004,VanLaerhoven2019,Matheson2023}. The Solar System's planets likewise show a similar degree of coplanarity, with their orbital planes lying within $7^\circ$ of the invariable plane  \cite{Souami2012}. The near alignment between the Kuiper belt and the orbital planes of the planets is consistent with the nebular hypothesis \cite{Kant1755,deLaplace1796}, in which planets form in an initially flattened accretion disk \cite{Williams2011}. 

In extrasolar systems, the cold widely separated debris disks are the counterparts of the Kuiper belt, with dust continually replenished by collisions among larger  planetesimals \cite{Wyatt2008}. These disks can be resolved in scattered light at visible/near infrared wavelengths or in thermal emission at infrared and longer wavelengths. If exoplanetary systems form like the Solar System, it is expected that they are broadly coplanar, with planets approximately aligned with their exo-Kuiper belts. Recent observations, however, reveal a more diverse architecture, including planets in inclined orbits relative to their host stars' spin axes \cite{Schlaufman2010, Albrecht2012, Huber2013, Zhang2021}, and planets on mutually misaligned orbits \cite{Xuan2020,Zhang2025}. Observations of protoplanetary disks likewise show that modest warps are fairly common \cite{Winter2025}, while large inner-outer disk misalignments are less common but clearly present \cite{Marino2015,Bohn2022}. These findings have motivated two leading hypotheses. First, misalignments among stars, planets and debris disks may result from post-formation dynamical interactions, such as planet-planet scattering \cite{Juric2008}, secular perturbations from planetary and stellar companions \cite{Fabrycky2007}, or disturbances from stellar flybys \cite{Picogna2014}. Alternatively, they could be primordial, inherited from misaligned or warped natal disks \cite{Batygin2012Nature}.

One observable that can help distinguish between these two scenarios is the mutual inclination between planetary orbits and debris disks. To date, however, such measurements are available for only a handful of individual systems, predominantly those hosting directly imaged planets on wide orbits near the debris belt. Recently, the availability of calibrated astrometric accelerations from Hipparcos, Gaia DR2 and DR3  \cite{Thompson2026} now makes it possible to constrain the three-dimensional (3-D) orbits of giant planets detected by radial velocities (RVs) in debris disk systems. This approach can be applied to planets much closer to their host stars and around relatively older stars than those typically accessible to direct imaging observations. Thus, planet-disk alignment tests can be extended to a much wider range of orbital distances and system ages. In this study, we present the first population-level constraints on the mutual inclinations of 32 planets (in 16 systems) relative to their debris disks (exo-Kuiper belts).

\subsection*{Sample description}

Fig.~\ref{fig1}A presents the  radial architectures of the planetary systems in our sample. We construct the sample by cross-matching systems with confirmed planets from the NASA Exoplanet Archive with resolved debris disks listed in the Catalog of Circumstellar Disks and the Resolved Debris Disks Catalog. The initial cross-match yielded 24 systems containing 47 planets with both confirmed planets and resolved debris disks. Of these, 2 systems host transiting planets, 16 host planets detected with radial velocities (RVs), and 6 host directly imaged planets. We then excluded 8 systems that host only close-in, small planets ($ a < 0.5$~AU, $M < 0.1\,M_{\mathrm{J}}$) detected by RVs, for which the orbital inclination cannot be measured currently. We also excluded targets whose host stars are spectroscopic binaries. We include the Solar System in our sample.  Our final sample consists of 32 planets in 16 systems.

The exoplanets in our sample span nearly three orders of magnitude in orbital distance, from 0.1 to 100 AU (see Fig.~\ref{fig1}A). At the smallest separations (0.1-1 AU) are two systems with transiting planets, including the three planets orbiting the 22-Myr-old star AU Mic \cite{Plavchan2020}, whose inclinations are tightly constrained by the transit observations. At intermediate separations (1-10 AU) are giant planets discovered through RVs, whose 3-D orbits can be constrained from orbit fits combining RVs and absolute astrometry from the calibrated  \textit{Gaia} DR3, DR2 and \textit{Hipparcos} catalog (G32H) \cite{Thompson2026}. At the widest separations, typically at 10-100 AU, are directly imaged planets, including well-characterized systems such as $\beta$ Pic and HR 8799, whose orbits are constrained using relative imaging astrometry, absolute astrometry and, when available, RVs. 

All debris disks in our sample are consistently located in outer system regions, from tens to hundreds of AU. These disks are either resolved in scattered light by Hubble Space Telescope (HST)\cite{Woodgate1998}, Gemini Planet Imager (GPI) \cite{Macintosh2014} or Very Large Telescope/SPHERE \cite{Beuzit2019}, in thermal infrared (FIR) by Herschel Space Observatory \cite{Pilbratt2010}, or in submillimeter continuum emission by Atacama Large Millimeter/submillimeter Array (ALMA) \cite{Wootten2009}. We list the stellar and disk parameters in Table~\ref{tab_stellar_parameter} and Table~\ref{oritation_table}.

\subsection*{Planet-debris disk alignment}

First, we quantify the planet-disk misalignment using $\Delta i_{p,d} = |i_p-i_d|$, where $i_p$ and $i_d$ are orbital inclinations of planets and debris disks relative to the sky plane. This quantity reflects the difference in how the planet’s orbital plane and the debris disk are tilted along the line of sight, and thus represents a lower limit on their true mutual inclination. For each planet, we inferred the posterior distribution of the $\Delta i_{p,d}$  by combining the posteriors on the planet's inclination from orbital fitting with the reported disk-inclination distribution from the literature (see materials and methods). Fig.~\ref{fig1}B presents the $\Delta i_{p,d} $ as the function of planet orbital distances. Following the significance framework adopted in recent obliquity studies \cite{Bowler2023,Biddle2025}, we take the maximum a posteriori (MAP) value of $\Delta i_{p,d} $ and define a conservative lower limit as the value $2\sigma$ below the posterior peak (corresponding to $95\%$ confidence). We classify a planet as aligned or misaligned by comparing this lower limit to a chosen threshold. Using a threshold of $10^{\circ}$, we find 30/32 planets are aligned with the debris disks, indicating the most planets are aligned with the debris disks within $10^{\circ}$. At the system-level, we then estimate the inclination of each system's invariable plane, when multiple planets are present. We find 14/16 systems show alignment between the system invariable plane with debris disks assuming a threshold of $10^{\circ}$. For the stricter threshold of $1^\circ$, the aligned fraction decreases to 26/32 for planets and 11/16 for systems. In other words, about 5 out of 16 systems show a modest misalignment with an amplitude of a few degrees. A representative example is $\beta$ Pic, where the two planets are misaligned relative to the debris disk by $3.9^{+0.9}_{-1.2}\,^\circ$ and $3.6^{+1.2}_{-0.9}\,^\circ$, consistent with the observed $3^\circ$--$4^\circ$ warp in the disk. The planet-disk mutual inclination in $\beta$ Pic is therefore modest ($<10^\circ$), yet significantly different from zero. The Solar System provides another example: the mean plane of the Kuiper belt lies within about $0.7^\circ$ of the Solar System’s invariable plane, but is nevertheless distinguishable from it at greater than $99.7\%$ confidence \cite{Matheson2023}.

We visualize the population-level distribution of $\Delta i_{p,d}$ by pooling posterior samples of $\Delta i_{p,d}$ from individual systems and computing a kernel density estimate (KDE, see Fig.~\ref{fig1}C). To avoid overcounting multi-planet systems, we include each system only once, using the orbital inclination difference between the system invariable plane and the debris disk. The KDE peaks at $0^\circ$, with a half-width of about $10^{\circ}$. We then explore two types of underlying population distribution for  $\Delta i_{p,d}$ using hierarchical Bayesian model (HBM): a Rayleigh distribution and a truncated Gaussian distribution. The resulting Rayleigh model yields a characteristic inclination scale of $6.3 \pm 1.1\, ^\circ$. The best-fit truncated Gaussian model shows a peak at $4.5 \pm 2.6\,^\circ$ and a dispersion of $9.3\pm 2.8 \,^\circ$. Together, the KDE and model fits indicate that the $\Delta i_{p,d}$ distribution is concentrated toward low mutual inclinations, with characteristic values below $10^\circ$.

Furthermore, we examine the 3-D mutual inclinations, $\psi_{p,d}$, between the planets and the debris disks. In principle, $\psi_{p,d}$ can be derived with measured orbital inclinations and longitudes of ascending node for both the planets and the debris disks. However, because disk observations reveal only the projected ellipse on the sky, the inferred 3-D disk orientation remains degenerate (see materials and methods). For any individual system, the true solution cannot be uniquely identified without additional information, such as scattering-phase constraints or kinematic observations of gas motion in the disk, which are not always available. For the sample as a whole, however, we account for this ambiguity within the HBM by marginalizing over the degenerate solutions.   Fig.~\ref{fig2} presents the best-fit hierarchical Bayesian models of $\psi_{p,d}$ using a Rayleigh distribution and a truncated Gaussian distribution, respectively. The two models yield similar results: although the $\psi_{p,d}$ distribution  is broader and shifted to larger values than $\Delta i_{p,d}$, they still favor low planet-disk mutual inclination, with a typical scale of $\sim 15^{\circ}$. The Rayleigh model implies a characteristic inclination scale of $15.1\pm 2.5\,^\circ$, while the truncated Gaussian model yields a peak at $15.6\pm 4.5\,^\circ$ and a dispersion of $16.3\pm 3.6\,^\circ$.

\subsection*{Star-planet-debris disk alignment}

We further probe the alignment between star, planet and debris disks in the sample. Stellar inclination measurements are available for 12 systems, inferred from the stellar rotation period $P_{\rm rot}$, projected rotational velocity $v\sin i_*$, and stellar radius $R_*$ (see materials and methods). Similar to $\Delta i_{p,d}$, we define the minimum stellar obliquity as $\Delta i_{p,*}=|i_p-i_*|$, the absolute difference between the planetary and stellar inclinations.  Fig.~\ref{fig3} compares $\Delta i_{p,d}$ and $\Delta i_{p,*}$. To assess the level of agreement, we calculated Lin’s concordance correlation coefficient (CCC), which quantifies how closely two variables match in both correlation and equality about the 1:1 line \cite{Lin1989}. We obtain ${\rm CCC}=0.867$, indicating strong agreement between $\Delta i_{p,d}$ and $\Delta i_{p,*}$. We find 11 out of 12 systems exhibit low inclinations ($<10^{\circ}$) for both the planet-disk mutual inclination and the star-planet obliquity. Overall, these results indicate that the systems in our sample are predominantly consistent with alignment among stars, planets, and debris disks, reminiscent of the architecture of the Solar System.

\subsection*{Misaligned systems}

HD 10647 ($q^{1}$ Eri) and $\kappa$ CrB (HD 142091) are the two exceptions in our sample, showing significant misalignment between their planets and debris disks. HD 10647 is F9V star with an age of $1.4^{+0.9}_{-0.9}$ Gyr \cite{Marmier2013}. $\kappa$ CrB is a K0 sub-giant star with a age of $3.28^{+0.55}_{-0.55}$ Gyr \cite{Sousa2015}. Both systems host a giant planet in the inner system, at 2 AU \cite{Bulter2006} and 2.6 AU \cite{Johnson2008}, respectively. They also host spatially resolved debris disks in the outer planetary system, with characteristic disk radii of $R\sim103$ AU for HD 10647 \cite{Marino2026} and $\sim93$ AU for $\kappa$ CrB \cite{Lovell2022}. The minimum planet-disk mutual inclinations are $\Delta i_{p,d}=43.6^{+8.1}_{-10.4}\,^\circ$ for HD 10647 and $44.3^{+5.8}_{-3.0}\,^\circ$ for $\kappa$ CrB, ruling out alignment within $10^\circ$ at the $>3\sigma$ ($99.7\%$) level. For HD 10647, we can additionally infer the stellar inclination from the stellar rotation period and projected rotational velocity, $v\sin i_*$. This allows us to compare the orientations of all three components in the system: the stellar spin axis, the planetary orbit, and the debris disk. We find that the planet is misaligned with both the star ($\Delta i_{p,*}=37.0^{+11.6}_{-17.4}\,^\circ$) and the debris disk ($\Delta i_{p,d}=43.6^{+8.1}_{-10.4}\,^\circ$), whereas the stellar spin axis is nearly aligned with the debris disk ($\Delta i_{*,d}=8.8^{+3.4}_{-8.6}\,^\circ$).  The two systems represent departures from Solar System-like coplanarity and therefore provide valuable opportunity for probing the origin of orbital misalignments.

\subsection*{Implication for planetary system formation}

Our results reveal that most extrasolar planetary systems resemble the Solar System in exhibiting alignment between the orbital plane of planets and debris disks. This alignment is seen not only for wide-orbit planets near the debris disks, but also for planets in the inner regions, suggesting these systems preserve a common angular-momentum orientation across a wide range of orbital separations. For the subset of systems with stellar inclination constraints, we also find evidence for alignment among the debris disk, planetary orbits, and the stellar spin axis. Together, these results support the basic picture that planets and planetesimal belts form within a common, broadly coplanar protoplanetary disk. We also note that, although most planets are broadly aligned with the debris disks within $10^{\circ}$, many systems, such as $\beta$ Pic, exhibit modest mutual inclinations of a few degree. This is consistent with the observations of protoplanetary disk, where modest warp ($0.5-2^{\circ}$) are common \cite{Winter2025}. Thus, planetary systems may be broadly aligned while preserving small dynamical imprints from disk warps, planet–planet interactions, or other perturbations.

Although rare, we identify two systems, HD 10647 and $\kappa$ CrB, showing significant misalignment between planets and debris disks. Previous observations, including the outer-disk inclination distribution of young dipper stars \cite{Ansdell2020}, and shadows observed in scattered-light images of protoplanetary disks \cite{Bohn2022}, indicate that misalignments between inner and outer disks can arise during the protoplanetary-disk stage. A striking example is HD 142527, which hosts an inner and outer protoplanetary disk with a relative inclination of $\sim 70^{\circ}$ \cite{Marino2015}. If such inner-outer disk misalignments are preserved during planet formation, they could naturally produce misaligned inner and outer planetary architectures, including misalignments between inner planets and outer debris disks. An important question, therefore, is whether the two misaligned systems in our sample reflect this primordial origin, or whether their planets were later driven onto inclined orbits through post-formation dynamical processes. HD 10647 and $\kappa$ CrB are among the oldest stars in our sample, with ages of 1.5 and 3.4 Gyr (see Fig.~\ref{fig_age}), whereas the rest of the sample, spanning a broad range of ages from tens of Myr to a few Gyr, is largely consistent with alignment. Although the sample remains limited, the fact that significant planet-disk misalignments are only observed in older systems may favor a post-formation origin, where the system initially formed in a roughly coplanar configuration and was later misaligned by dynamical events. If large planet-disk misalignments were generally primordial, inherited from natal disks at birth, we would expect to observe them more broadly across systems of different ages, including younger systems. In addition, ref. \cite{Lovell2021} found a south-west ansa asymmetry in the disk of HD 10647, which could be perturbed by a planet with mass from $8\ \rm M_{\oplus}$ to $11\ \rm M_{J}$ at $\sim 60 $ AU. If such a planet exists, it could point to a past interaction with HD 10647 b that excited the planet into a misaligned orbit. Furthermore, the alignment between the debris disk and the stellar spin axis in HD 10647 suggests that the planet was excited onto an inclined orbit at a later stage, rather than inheriting its misalignment from the natal disk. We cannot, however, rule out a primordial origin in individual cases. Even so, the rarity of the observed planet-disk misalignments suggests that primordial misalignment is unlikely to be a dominant pathway shaping planetary-system architectures.

A separate but related question is the origin of stellar obliquity. Previous studies have argued that the observed spin-orbit misalignment could be from primordial misalignment between disk and stars.  Ref.\cite{Biddle2025} has shown that roughly one-third of protoplanetary disks are misaligned with respect to their host stars' spin axes, although the majority remain aligned. In comparison, our results show a tighter alignment between the angular momentum vectors of planetary orbits and debris disks than between those of stars and protoplanetary disks. In our sample, however, stellar obliquity  measurements are available for only 12 systems, and in all of them the debris disks are consistent with alignment with their host stars. As a result, our data do not strongly constrain the frequency of primordial star-disk misalignment at the debris-disk stage for direct comparison. A larger sample with measurements of stellar inclinations, planetary orbits and resolved debris disks will be required to test the connection among all three components. In the future, the release of Gaia DR4 will make the breakthrough by enabling the detection and orbital characterization of thousands more giant planets \cite{Perryman2014} and substantially expanding the sample of systems amenable to such system-wide alignment measurements.

%%%%%%%%%%%%%%%% MAIN TEXT FIGURES %%%%%%%%%%%%%%%
\begin{figure}[b]
\centering
\includegraphics[width=\textwidth]{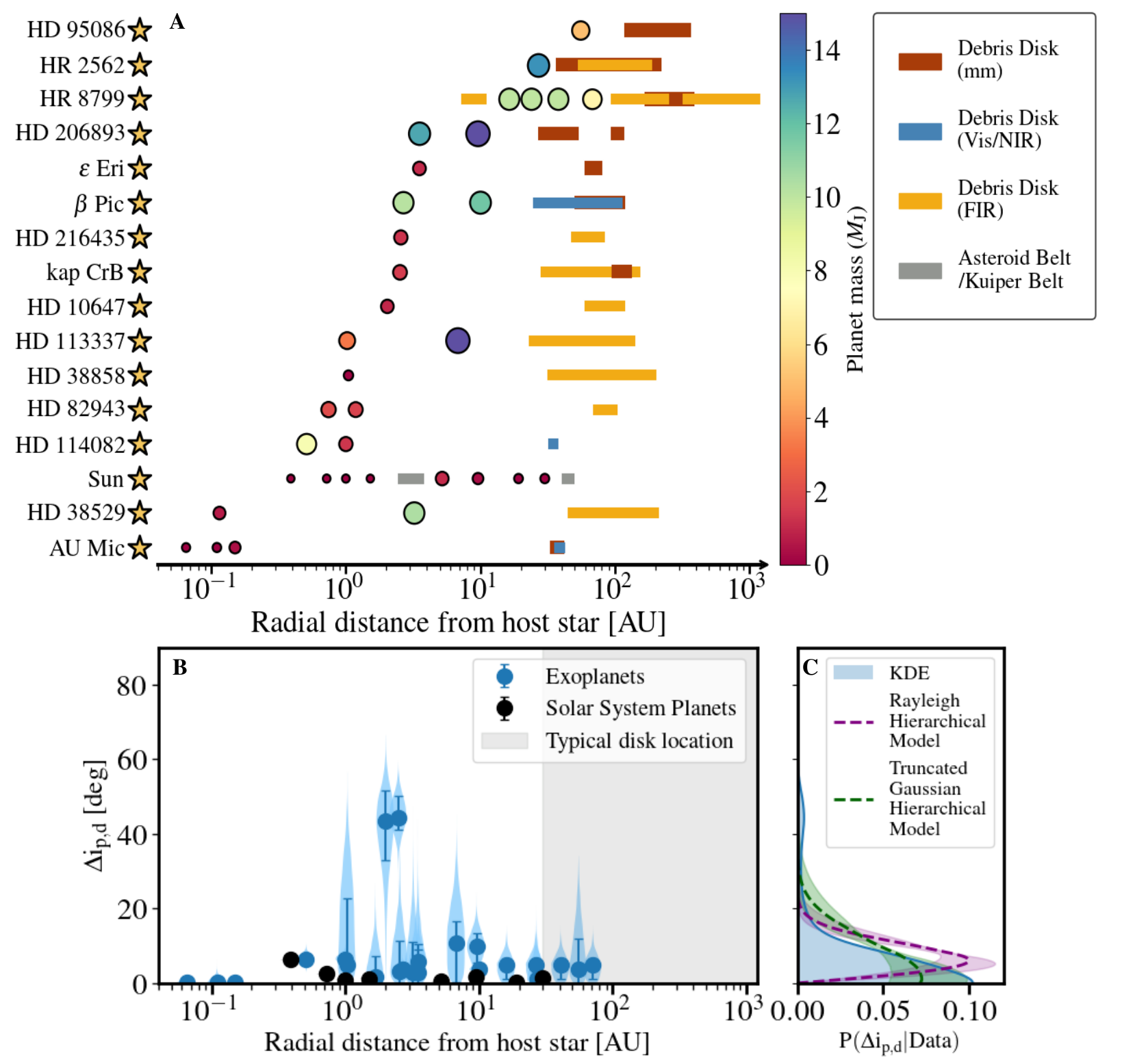}
\caption{\textbf{Overview of the radial architectures and planet-disk alignments across the planetary systems in our sample.} A, Radial architecture of the planetary systems in our sample. Circles represent planets, with colors indicating their masses. Circle sizes scale with planet masses. Bars mark the locations of the debris disks observed at different wavelengths, denoted by colors. Panel A is intended to show the radial locations of the planets and debris disks, not the mutual inclinations between planets and disks. B, Minimum mutual inclination between planets and debris disks, $\Delta i_{p,d}=|i_p-i_d|$, as a function of planetary orbital distance. Violin plots show the full posterior distributions, while circles and error bars indicate the MAP and 1$\sigma$ credible intervals. C, The light blue shaded region shows the KDE of the pooled $\Delta i_{p,d}$ distribution for the full sample. Each system, including the Solar System, is included only once. We present the hierarchical Bayesian models (HBMs) with a Rayleigh distribution and a truncated Gaussian distribution in purple and green, respectively, plotted as the posterior medians (dashed lines) with their $68\%$ credible intervals (shaded regions). }\label{fig1}
\end{figure}
 %The planet sizes are schematic and not to scale.The gray shaded region marks the typical location of the debris disks.

\begin{figure}[!h]
\centering
\includegraphics[width=\textwidth]{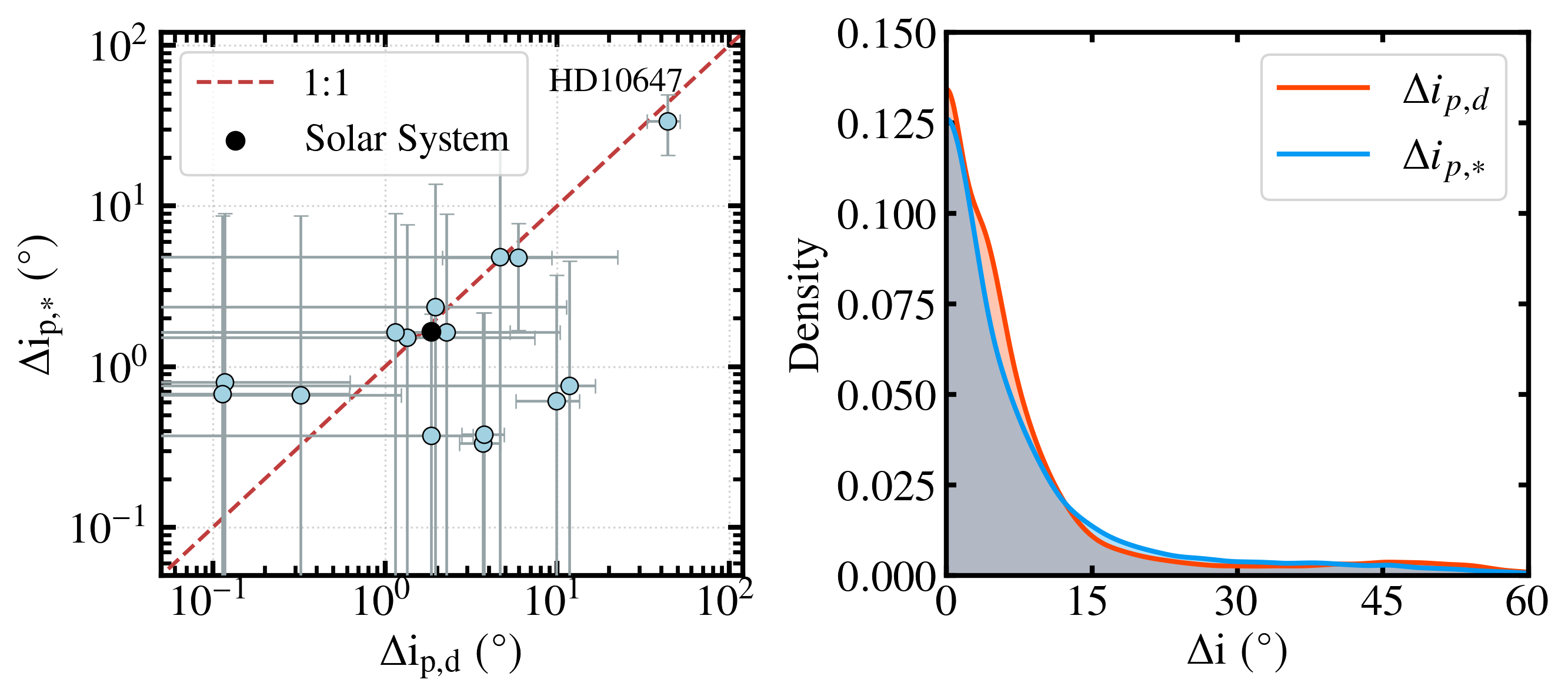}
\caption{\textbf{Comparison of stellar obliquity and planet-disk mutual inclinations.} Left: Minimum stellar obliquity, $\Delta i_{p,*}$, as a function of minimum planet-disk mutual inclination, $\Delta i_{p,d}$, for the 12 systems in our sample with measured stellar inclinations (including Solar System). Right: KDEs of $\Delta i_{p,*}$ and $\Delta i_{p,d}$ across the 12 systems. }\label{fig3}
\end{figure}
\begin{figure}
\centering
\includegraphics[width=0.6\textwidth]{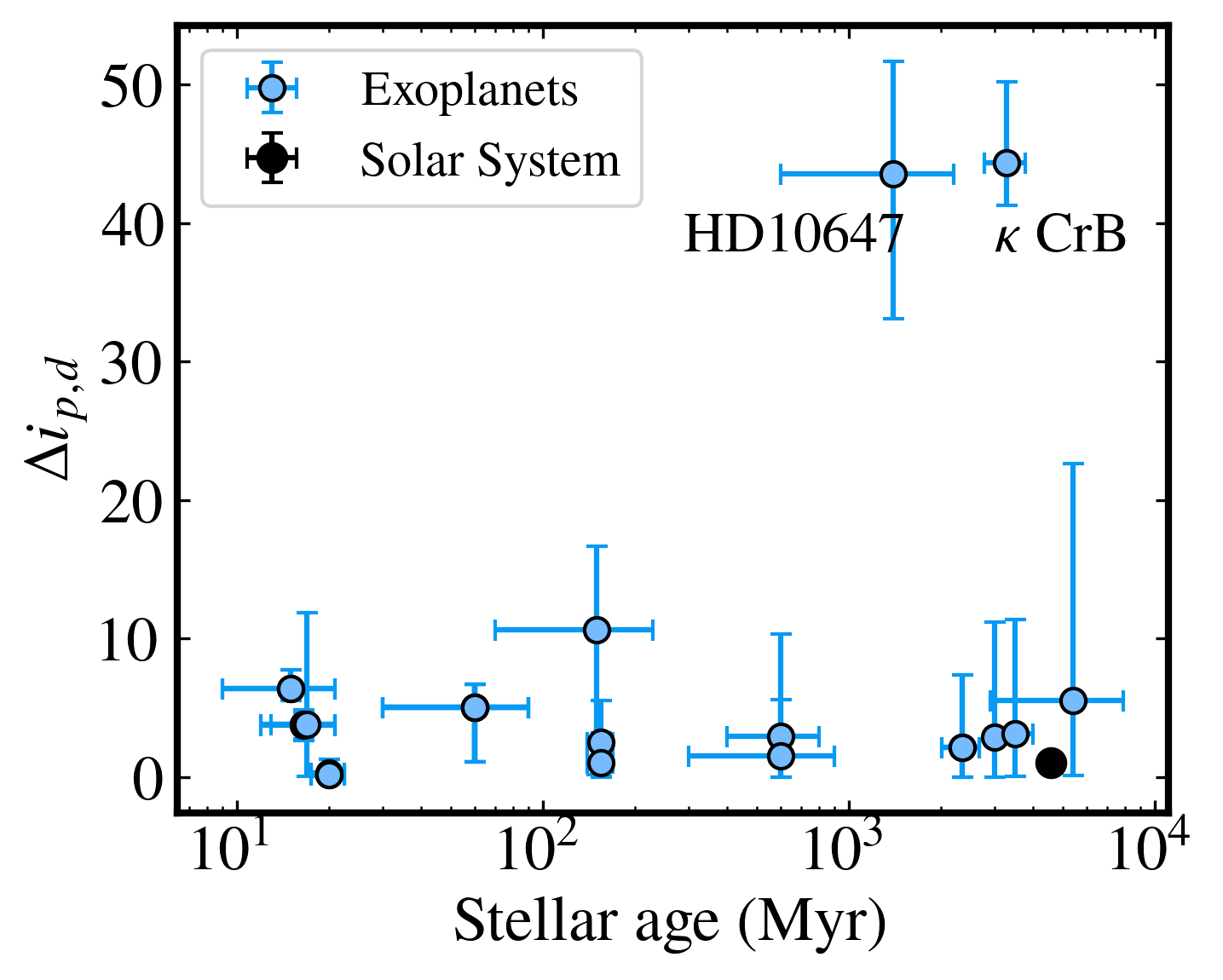}
\caption{\textbf{Planet-disk mutual inclination as a function of stellar age.} Minimum mutual inclination between planet and debris disk, $\Delta i_{p,d}$, is shown versus stellar age for the systems in our sample. The two systems with planets in misaligned orbits relative to the debris disks are labeled.}\label{fig_age}
\end{figure}

%%%%%%%%%%%%%%%% MAIN TEXT TABLES %%%%%%%%%%%%%%%

%%%%%%%%%%%%%%%% REFERENCES %%%%%%%%%%%%%%%

\clearpage % Clear all remaining figures and tables then start a new page

% The list of references goes after the main text and before the acknowledgements
% When preparing an initial submission, we recommend you use BibTeX, like this:
%
\bibliography{science_template} % for a file named science_template.bib
\bibliographystyle{sciencemag}

% After the paper has completed peer review and been revised ready for acceptance,
% you should comment out the lines above and copy-paste the contents of your .bbl
% file here instead. This will help ensure that our conversion software works correctly.
% Remember to re-run BibTeX first - check the timestamp!
%
% Example of the first three entries copy-pasted from science_template.bbl:
%
%\begin{thebibliography}{1}
%
%\bibitem{example}
%A.~N. {Author}, An example reference. \emph{Journal of Improbable Research}
%  \textbf{1}, 67 (2020).
%
%\bibitem{example2}
%F.~M. {Surname}, S.~{Author}, A second example. \emph{Interesting Research
%  Letters} \textbf{32}, 897 (2019).
%
%\bibitem{example_preprint}
%P.~{One}, P.~{Two}, P.~{Three}, {An unpublished preprint}. \emph{preprint}
%  (2021), arXiv:2101.12345.
%
%\end{thebibliography}

%%%%%%%%%%%%%%%% ACKNOWLEDGEMENTS %%%%%%%%%%%%%%%

\section*{Acknowledgments}
J.Z. would like to acknowledge the use of computational facilities purchased with funds from the National Science Foundation (CNS-1725797) and administered by the Center for Scientific Computing (CSC). The CSC is supported by the California NanoSystems Institute and the Materials Research Science and Engineering Center (MRSEC; NSF DMR 2308708) at UC Santa Barbara. This research has made use of the NASA Exoplanet Archive, which is operated by the California Institute of Technology, under contract with the National Aeronautics and Space Administration under the Exoplanet Exploration Program. This research has made use of the Catalog of Circumstellar Disks (circumstellardisks.org), curated by Nicholas Grosso. This work benefited from the 2025 Exoplanet Summer Program in the Other Worlds Laboratory (OWL) at the University of California, Santa Cruz, a program funded by the Heising-Simons Foundation. J.Z. would like to thank Jerry Xuan for his support and helpful scientific discussion. 

\paragraph*{Funding:}

J.Z. and M.A.M. are supported by grant JWST-GO-06012.001-A from the Space Telescope Science Institute under NASA contract NAS 5-03127. B.P.B. acknowledges support from the National Science Foundation grant AST-1909209, NASA Exoplanet Research Program grant 20-XRP20$\_$2-0119, and the Alfred P. Sloan Foundation.
T.D.P is supported by a UKRI Stephen Hawking Fellowship and a Warwick Prize Fellowship, the latter made possible by a generous philanthropic donation.
\paragraph*{Author contributions:}
J.Z. conceived the project, developed the main scientific idea, performed the analysis, and wrote the manuscript – original draft. M.A.M contributed to the conceptualization of the study, interpretation of the results, and funding acquisition. B.P.B and T.D.P. contributed to the conceptualization of the study, interpretation of the results and scientific discussions. W.T. and D.B. provided the astrometric data products, developed the software of orbital fitting and advised on their use. M.M. contributed to methodology and the HBM modeling. All authors reviewed and commented on the manuscript.
\paragraph*{Competing interests:}
There are no competing interests to declare.
\paragraph*{Data, code and materials availability:}
All radial velocity and imaging relative astrometry data used in this analysis (listed in   Table ~\ref{tab:rv_tab1} and \ref{tab_rel_astr}) are from public, archival data with full references to the original data publications. There are no restrictions on the availability of all data used in this work. The absolute astrometry data from Calibrated Gaia DR2, DR3, and Hipparcos Catalog \cite{Thompson2026} are available in \url{https://www.canfar.net/citation/landing?doi=26.0002}. All photometric data from TESS used in this analysis (plotted in Fig.~\ref{fig_tl1}-\ref{fig_tl5}) are available in the Mikulski Archive for Space Telescopes: \url{https://doi.org/10.17909/T9-NMC8-F686} \cite{Huang2020}. 

This work used the open-source package \textit{Octofitter} \cite{Thompson2023} for orbital fitting. Documentation for \textit{Octofitter} is available at \url{https://sefffal.github.io/Octofitter.jl/stable/}.
The code for statistical analysis is available from the corresponding author on reasonable request.
%%%%%%%%%%%%%%%% SUPPLEMENT LIST %%%%%%%%%%%%%%%

% List the contents of your Supplementary Materials, including the numbers of any
% supplementary figures, tables, external data files etc. and any references that are
% cited only in the supplement. In this example, refs. 7-8 are cited only in the supplement.
% Fill out your numbers accordingly and delete any lines that aren't applicable.
\subsection*{Supplementary materials}
Materials and Methods\\
Supplementary Text\\
Figs. S1 to S34\\
Tables S1 to S5\\
References \textit{(50-\arabic{enumiv})}\\ % automatically fills out the last reference number
% (filling out the other numbers automatically is possible but fiddly and liable to break)
%Movie S1\\
%Data S1

%%%%%%%%%%%%%%%% END OF MAIN TEXT %%%%%%%%%%%%%%%

\newpage

%%%%%%%%%%%%%%%% START OF SUPPLEMENT %%%%%%%%%%%%%%%

% Figures, tables, equations and pages in the supplement are numbered S1, S2 etc.
\renewcommand{\thefigure}{S\arabic{figure}}
\renewcommand{\thetable}{S\arabic{table}}
\renewcommand{\theequation}{S\arabic{equation}}
\renewcommand{\thepage}{S\arabic{page}}
\setcounter{figure}{0}
\setcounter{table}{0}
\setcounter{equation}{0}
\setcounter{page}{1} % not 0 as \newpage already started a supplementary page
% References continue the numbering from the main text.

%%%%%%%%%%%%%%%% SUPPLEMENT TITLE PAGE %%%%%%%%%%%%%%%

\begin{center}
\section*{Supplementary Materials for\\ \scititle}

% Author list for the supplement
% Indicate the corresponding authors, but do NOT include institutions here
% It would be nice if the template auto-generated this, but doing so is complicated...

% You can write out first names or use initials - either way is acceptable, but be consistent
Jingwen~Zhang$^{1\ast}$,
Maxwell~A.~Millar-Blanchaer$^{1}$,
Brendan~P.~Bowler$^{1}$,
Tim~D.~Pearce$^{2}$,
Marvin~Morgan$^{1}$,
William~Thompson$^{3}$,
Dori~Blakely$^{3,4}$\\
\small$^\ast$Corresponding author. Email: jwzhang@ucsb.edu
% Joint contributions can be indicated like this

\end{center}

% Fill out the numbers for each type of supplementary material,
% and delete any lines that aren't applicable.
% These are just example numbers that don't match the rest of this template.
\subsubsection*{This PDF file includes:}
Materials and Methods\\
Supplementary Text\\
Figures S1 to S34\\
Tables S1 to S5\\
%Captions for Movies S1 to S2\\
%Captions for Data S1 to S2

%\subsubsection*{Other Supplementary Materials for this manuscript:}
%Movies S1 to S2\\
%Data S1 to S2

\newpage

%%%%%%%%%%%%%%%% MATERIALS AND METHODS %%%%%%%%%%%%%%%

\subsection*{Materials and Methods}

\subsubsection*{Debris Disk Parameters}

We construct this sample by cross-matching systems with confirmed planets from the NASA Exoplanet Archive (\url{https://exoplanetarchive.ipac.caltech.edu/}) with resolved debris disks from the Catalog of Circumstellar Disks (\url{https://www.circumstellardisks.org/index.php})
and the Resolved Debris Disks Catalog (\url{https://www.physik.uni-jena.de/en/21956/catalog-of-resolved-debris-disks}) and other recent compilation of debris disk surveys from ALMA \cite{Marino2026} and SPHERE \cite{Engler2025}. We adopted the measurements  of disk orbital inclination ($i_{\rm{d}}$) and position angle ($\theta_{\rm{d}}$) from literature. For disks where many independent measurements are available, we adopt the highest reported precision. For the disk parameters with asymmetric uncertainties, we adopt the mean of the upper and lower limits. All measurements used in our analysis are with uncertainties below $10^{\circ}$. The adopted stellar parameters and disk parameters are presented in Table~\ref{tab_stellar_parameter} and \ref{oritation_table}. We  provide detailed descriptions of each system in the Supplementary Text.

\subsubsection*{Planet Orbit Fit }

%For transting planets AU mic and xx, we adopted the well chatai orbital incliation of them from litertaures. The uncertie of thse measuremnets both below $1^{\circ}$. 

%\subsubsection*{Data}

We performed joint Bayesian orbit fits using open-source package \textit{Octofitter} \cite{Thompson2023}, simultaneously modeling absolute astrometric measurements from \textit{Gaia} DR3, DR2 and \textit{Hipparcos}  together with long-baseline RVs, or relative astrometry from imaging,  or both when available. We adopted a single-companion model for most systems, except for HD 113337, $\beta$ Pic, HD 206893, and HD 82943, for which we fit two-companion models. Each companion is described by seven parameters: orbital period $P$, planet-to-star mass ratio $q$, eccentricity $e$, argument of periastron $\omega$, cosine of inclination $\cos i$, longitude of the ascending node $\Omega$, and position angle at a reference epoch $\theta$. We also fit additional nuisance parameters, including RV and astrometric noise terms, instrument-specific RV zero points, and astrometric epoch selection (see ref.\cite{Thompson2026} for details). We assume Gaussian priors for host star mass $M_{*}$ based on reported values from literature. 
We adopted a uniform prior between 0 and 0.99  on the eccentricity, and a uniform prior in $\cos{i}$ for orbital inclination. The priors on argument of periastron $\omega$ and longitude of the ascending node $\Omega$ are uniform between 0 to $2\pi$. For companion mass, we adopted a broad log-uniform prior spanning from 0.1 to 100 $M_J$. 

%We also include parallax $(\varpi)$ as a fitted parameters, adopting a Gaussian prior based on the \textit{Gaia} DR3 value and its reported uncertainty.

Following the framework in ref.\cite{Thompson2025, Thompson2026}, the total likelihood is constructed from the product of the RV, relative astrometry from imaging, absolute astrometry from \textit{Hipparcos} and \textit{Gaia}, and \textit{Gaia} astrometric excess noise (unbiased estimator of variance, or UEVA), so that all constraints are fit self-consistently within a single orbit model:

\begin{equation}
ln\mathcal{L}
=
ln\mathcal{L}_{\mathrm{RV}}\,+
ln\mathcal{L}_{\rm{rel.\ ast.}}\,+
ln\mathcal{L}_{\mathrm{abs. \ ast. }}\,+
ln\mathcal{L}_{\mathrm{UEVA}}\,
\end{equation}

We adopted archival RV measurements for ten stars in the sample from a variety of instruments.  For directly imaged planets, we adopted archival relative astrometry from imaging observations. A detailed summary of the RV and relative astrometry data, including the number of measurements and observational baseline for each system, are provided in   Table~\ref{tab:rv_tab1} and \ref{tab_rel_astr}.

For absolute astrometry, we modeled a joint set of calibrated proper-motion observables: the \textit{Hipparcos} proper motion $\mu_{\mathrm H}$, the \textit{Hipparcos}-\textit{Gaia} long-baseline scaled position difference $\mu_{\mathrm{HG}}$, the \textit{Gaia} DR2 and DR3 proper motions ($\mu_{\mathrm{DR2}}$, $\mu_{\mathrm{DR3}}$), and the \textit{Gaia} DR3-DR2 scaled position-difference term $\mu_{\mathrm{DR32}}$.  \textit{Hipparcos} quantities ($\mu_{\mathrm H}$ and $\mu_{\mathrm{HG}}$) were adopted from the DR3-calibrated HGCA solution, whereas \textit{Gaia} DR2 quantities were separately cross-calibrated onto the DR3 frame using global magnitude-dependent rotation, local HEALPix offsets, and uncertainty inflation. We also include the DR2-DR3 cross-covariance to take into account for the shared Gaia transits by DR2 and DR3.

Finally, we included a likelihood from \textit{Gaia} UEVA term, which is inferred from Gaia’s astrometric excess-noise or Renormalized Unit Weight Error (RUWE) information. For each trial orbit, we forward-model the Gaia scan sampling, fit a 5-parameter single-star solution to the perturbed path, and compute the predicted UEVA from the resulting along-scan residuals. We then compare observed and predicted UEVA (typically in a cube-root-transformed Gaussian form). This term adds sensitivity to short-timescale astrometric wobble during the Gaia baseline and helps break degeneracies left by proper-motion anomalies alone.

We sampled the posterior using the stabilized variational nonreversible parallel tempering algorithm implemented in \textit{Pigeons.jl} \cite{pigeons} for $2^{14}$ scans. 
We provide the details of the orbital fitting results for each individual system in Supplementary Text.

%The fitted Keplerian models and joint posterior distributions are presented in Fig.~\ref{fig_ft1}-\ref{fig_cn19}.

\subsubsection*{Stellar Inclination}

Stellar inclination is the angle between the stellar rotation axis and line of sight, for which $i_*=0^{\circ}$ corresponding to pole-on configuration and $i_*=90^{\circ}$ corresponding to edge-on configuration. Typically, when the stellar radius $R_*$ and rotational period $P_{\rm{rot}}$ are known, the equatorial rotational velocities can be calculated as \( v_{eq} = 2\pi R_{*}/P_{\rm{rot}} \). Although it is intuitive to estimate $i_* = \sin^{-1}(\frac{ v\sin{i_*}}{v_{eq}})$, this assumes $v_{eq}$ and $v\sin{i}$ are independent, which ref. \cite{MandW2020M} points out it is not usually valid. We therefore follow the Bayesian probabilistic framework from ref.\cite{MandW2020M,Bowler2023} to derive the stellar inclination. For stars with $v\sin{i_*}>2 \ \rm km s^{-1}$, the stellar inclination can be derived as
 
\begin{equation}{\label{eqn:i}}
P(i_*\mid P_\mathrm{rot}, R_*, v\sin i_*) \propto  \sin i_* \times \frac{e^{- \frac{\big(v \sin i_* - \frac{2\pi R_*}{P_\mathrm{rot}}\sin i_* \big)^2}{2\big(\sigma_{v\sin i_*}^2 + \sigma_{v_\mathrm{eq}}^2 \sin^2 i_* \big)}} }{\sqrt{\sigma_{v\sin i_*}^2 + \sigma_{v_\mathrm{eq}}^2 \sin^2 i_*}},
\end{equation}

where $\sigma_{v\sin i_*}$ and $\sigma_{v_\mathrm{eq}}$ are the uncertainties in the projected velocity $v\sin{i_{*}}$ and the equatorial velocity $v_{eq}$. We estimate the uncertainty of equatorial velocity  $\sigma_{v_\mathrm{eq}}$ from those of stellar rotation period and radii as
\begin{equation}{\label{eqn:veqerr}}
\sigma_{v_\mathrm{eq}} =  \frac{2\pi R_*}{P_\mathrm{rot}} \sqrt{\Big(\frac{\sigma_{R_*}}{R_*}\Big)^2 + \Big(\frac{\sigma_{P_\mathrm{rot}}}{P_\mathrm{rot}}\Big)^2 }
\end{equation}

For slow-rotating stars with $v\sin{i_*}<2\ \rm km s^{-1}$, it is challenging to distinguish macro-turbulence broadening from rotation broadening in the stellar spectra. Therefore, we only adopt a upper limit of $v\sin{i_*}$ and the stellar inclination can be given as 

\begin{equation}
    P(i_\ast \mid P_\mathrm{rot}, R_*, l) \propto \sin i_\ast \times
\left(
\operatorname{erf}\!\left(
\frac{l - \dfrac{2\pi R_\ast}{P_{\rm rot}} \sin i_\ast}
{\sqrt{2}\,\sigma_{v_{\rm eq}} \sin i_\ast}
\right)
+
\operatorname{erf}\!\left(
\frac{\sqrt{2}\,\pi R_\ast}
{\sigma_{v_{\rm eq}} P_{\rm rot}}
\right)
\right)
\end{equation}

where $l$ is the upper limit of $ v\sin{i_*}$ and \(\operatorname{erf}(x)\) is the error function, defined by \(\operatorname{erf}(x)=\frac{2}{\sqrt{\pi}}\int_0^x e^{-t^2}\,dt\).

Across the full sample, 10 systems have published stellar rotation periods, determined from either photometric light curves or stellar activity indices, and the measurements from different sources are broadly consistent. We also derived stellar rotation periods for five stars (see Supplementary Text), obtaining values in agreement with the literature. We adopted the weighted mean rotation period for each star and imposed a minimum uncertainty floor of $10\%$. We compiled $v\sin i_*$ measurements from multiple sources; when reported uncertainties were available, we combined values using a weighted mean and weighted standard deviation, and when uncertainties were unavailable, we estimated a mean and standard deviation from those measurements and included that as an additional constraint. Stellar radii were adopted from the  TESS Input Catalog \cite{TESSinput}, with uncertainties inflated in quadrature by $7\%$.   Table~\ref{supp_table_1} provides a detailed summary of the stellar radii, projected rotational velocities, and rotation periods used in our analysis. We also provide detailed descriptions in Supplementary Text. 

Stellar inclination can also be derived through asteroseismic analysis, because stellar rotation induces splitting of non-radial oscillation modes in a manner that depends on the stellar inclination \cite{Gizon2003,Chaplin2013}. For $\beta$ Pic, we adopted the stellar inclination measurements derived with asteroseismology from ref. \cite{Zwintz2019}. In total, we have 12 systems with stellar inclination measurements in our sample, including the Solar System.

\subsubsection*{Geometry for Mutual Inclination and Stellar Obliquity}\label{sec:geo}

We first assess the  planet–disk mutual inclination using the absolute difference between the orbital inclinations of the planet and the debris disk as 
\begin{equation}
    \Delta i_{p,d}=|i_p-i_d|
\end{equation}
This observable measures the relative tilt between the two plane along the line of sight, and provides a lower limit on the 3-D mutual inclination. For a measured disk inclination $i_d$, the disk angular momentum vector has two possible orientations, $i_d$ or $\pi-i_d$, if the direction of dust's orbital motion of is unknown. The two values correspond to the same plane but opposite normal directions, producing the usual prograde–retrograde degeneracy (see   Fig.~\ref{figg}). In this work, we focus on the angle between the two planes, equivalent to the prograde solution, because distinguishing prograde from retrograde configurations requires kinematic information that is generally difficult for debris disks, which typically contain little or no gas.

We also calculate the 3-D mutual inclination $\psi_{p,d}$ between the debris disk and the planetary orbit as
\begin{equation}
    \cos \psi_{p,d} =\cos i_p \cos i_d+\sin i_p \sin i_d \cos(\Omega_p-\Omega_d)
\end{equation}
where \(\Omega_p\) and \(\Omega_d\) are the longitudes of ascending node. However, debris disk observations only constrain the projected ellipse on the sky, which introduces degeneracies in recovering the true 3D orientation (see   Fig.~\ref{figg}). 
Besides the prograde–retrograde degeneracy, disk imaging usually cannot determine which side of the disk is nearer to the observer, leaving the longitude of ascending node degenerate between $\Omega_d$ and $\Omega_d+\pi$. As a result, each system admits four possible values of the planet-disk mutual inclination:
$\psi^{1}_{p,d}(i_d,\Omega_d)$,
$\psi^{2}_{p,d}(i_d,\Omega_d+\pi)$,
$\psi^{3}_{p,d}(\pi-i_d,\Omega_d+\pi)$, and
$\psi^{4}_{p,d}(\pi-i_d,\Omega_d)$. Two ($\psi^{1}_{p,d}$, $\psi^{2}_{p,d}$) of these correspond to prograde configurations and two ($\psi^{3}_{p,d}$, $\psi^{4}_{p,d}$) to retrograde configurations. These four solutions are not independent: the retrograde solutions are simply the supplementary angles of the prograde ones, such that \(\psi^{3}_{p,d} = \pi - \psi^{1}_{p,d}\) and \(\psi^{4}_{p,d} = \pi - \psi^{2}_{p,d}\), and therefore do not provide additional information. In this work, we are interested in the amplitude of the angle between the planet and disk planes, rather than whether the orbit is prograde or retrograde, which is beyond the scope of this study. We therefore retain only the two prograde solutions and exclude the retrograde solutions from further analysis.

%Kinematic measurements or resolved scattered-light observations, which can reveal  the direction of orbital motion or near and far sides of the disk, can break these degeneracies, but such data are not uniformly available across our sample.corresponding to the four allowed combinations of disk inclination and nodal orientation. 

Similarly, for systems with stellar inclination measurements, we quantity the stellar obliquity with the absolute difference between the stellar inclination and orbital inclination of planets as
\begin{equation}
    \Delta i_{p,*} = \left| i_p - i_* \right|.
\end{equation}

This quantity represent a minimum values for the true stellar obliquity, because the stellar inclination constrains only the angle between the stellar spin axis and the line of sight. The sky-projected stellar obliquity, $\lambda$, remains unknown. The sky-projected stellar obliquity can be measured via the Rossiter-McLaughlin effect, which is typically available for transiting hot Jupiters and not for the majority of planets in our sample. There is also a degeneracy in the stellar inclination: a given angle between the stellar spin axis and the line of sight corresponds to two possible spin-axis orientations, pointing either toward the observer ($<90^{\circ}$) or away from the observer ($>90^{\circ}$). In the same spirit as for the mutual inclination, we therefore focus on the prograde stellar obliquity solution as a measure of the amplitude of the spin-orbit angle, rather than attempting to distinguish between prograde and retrograde configurations.

%This quantity traces only the line-of-sight component of the true stellar obliquity,

\subsubsection*{Kernel Density Estimation of $\Delta i_{p,d}$,  $\Delta i_{p,*}$ and $\psi_{p,d}$ }

For each system, we construct  $\Delta i_{p,d}$ by randomly drawing 10,000 samples of $i_p$ from the posteriors of the orbital fitting and same number of samples of $i_d$ from a Gaussian distribution based on the literature values. For multi-planet systems, we include only a single system-level entry in the population analysis, adopting the inclinations of the invariable plane for each system. To visualize the $\Delta i_{p,d}$ distribution of the full population, we combine individual systems into a pooled sample, and construct a boundary-corrected Gaussian kernel density estimate (KDE) of $\Delta i_{p,d}$. Because $\Delta i_{p,d}$ is restricted to non-negative values, Gaussian kernels centered close to $\Delta i_{p,d}=0$ place part of their weight at negative, nonphysical values and therefore  underestimate the density near the boundary. We therefore apply a reflection correction at $\Delta i_{p,d}=0$ to reduce boundary bias, and re-normalize the resulting KDE over the allowed domain. The KDEs of $\Delta i_{p,d}$ are presented in Fig.~\ref{fig1}C.

Similarly, for systems with stellar inclination measurements, we obtain $\Delta i_{p,*}$ by randomly drawing 10,000 samples of $i_p$ from orbital-fitting posteriors and same number of $i_*$ from the probability density function of the stellar inclination. We then follow the same procedure to construct a boundary-corrected Gaussian KDE of $\Delta i_{p,*}$. The KDEs of $\Delta i_{p,*}$ are presented in Fig.~\ref{fig3}.

We apply the same approach to obtain the pooled KDEs of  $\psi^{1}_{p,d}$ and $\psi^{2}_{p,d}$ by randomly drawing 10,000 samples from the distributions of $i_p$, $\Omega_p$, $i_d$ and $\Omega_d$. We then treat the two degenerate solutions as equally plausible and compute the marginalized population-level distribution as $p(\psi_{p,d}) = \frac{1}{2} p(\psi^{1}_{p,d})+\frac{1}{2} p(\psi^{2}_{p,d})$.  The KDEs of $\psi_{p,d}$ are presented in   Fig.~\ref{fig2}.

\subsubsection*{ Hierarchal Bayesian Modeling of $\Delta i_{p,d}$ }\label{sec:HBA_1}
We infer the underlying population distribution of $\Delta i_{p,d}$ using a hierarchical Bayesian model based on posterior recycling \cite{Thrane2019}. For simplicity, we denote $\theta \equiv \Delta i_{p,d}$ in the following equations. We consider two choices for the latent distribution, which we denote generically as $p(\theta \mid \boldsymbol{\Lambda})$, where $\boldsymbol{\Lambda}$ denotes a set of hyperparameters. 

First, we consider a Rayleigh distribution with $\boldsymbol{\Lambda}=\{\nu\}$, given by

\begin{equation}
    p(\theta \mid \nu) =
    \frac{\theta}{\nu^{2}}
    \exp\!\left(-\frac{\theta^{2}}{2\nu^{2}}\right),
    \qquad \theta \ge 0
\end{equation}
where $\nu$ sets the width of the distribution and characterizes the typical mutual inclination in the population.

The second model we use is a truncated Gaussian distribution with $\boldsymbol{\Lambda}=\{\mu,\sigma\}$, given by

\begin{equation}
p(\theta \mid \mu,\sigma)=
\dfrac{
\exp\!\left[-\dfrac{(\theta-\mu)^2}{2\sigma^2}\right]
}{
\sigma\sqrt{2\pi}\,
\left[
\Phi\!\left(\dfrac{b-\mu}{\sigma}\right)
-
\Phi\!\left(\dfrac{a-\mu}{\sigma}\right)
\right]
},
\qquad a \le \theta \le b .
\end{equation}
where $\mu$ and $\sigma$ characterize the location of peak and dispersion of the truncated Gaussian distribution. $\Phi$ is the cumulative density function of the standard Gaussian
distribution, and we adopt $a=0^{\circ}$ and $b=180^{\circ}$.

For a given model, the goal of the hierarchical inference is to estimate the posterior distribution of the hyperparameters, $p(\boldsymbol{\Lambda} \mid \{d_k\})$, where $d_k$ denotes the observational data for system $k$ and $\{d_k\}$ denotes the full data set for all systems in the sample. For system \(k\), we marginalize over the latent parameter \(\theta\) to obtain the marginal likelihood under the population model,
\begin{equation}
    p(d_k\mid \boldsymbol{\Lambda})=\int p(d_k\mid \theta)\,p(\theta\mid \boldsymbol{\Lambda})\,d\theta.
\end{equation}
Based on Bayes' theorem, this can be rewritten as

\begin{equation}
    p(d_k\mid \boldsymbol{\Lambda})
    \propto
    \int p(\theta\mid d_k)\,
    \frac{p(\theta\mid \boldsymbol{\Lambda})}{\pi(\theta)}\,d\theta,
\end{equation}
where \(p(\theta\mid d_k)\) is the posterior distribution obtained from the orbital fitting of each system, and \(\pi(\theta)\) is the corresponding interim prior, which we assume to be uniform in \(\cos\theta\). We evaluate this integral using posterior recycling by replacing the continuous posterior with a finite set of posterior samples. Specifically, for system \(k\), we denote by \(\{\theta_{k,j}\}_{j=1}^{N_k}\) the \(N_k\) samples drawn from \(p(\theta\mid d_k)\), where \(j\) indexes the samples and \(k\) indexes the systems. The integral can then be approximated as a Monte Carlo average over these samples,
\begin{equation}\label{eqn:pp1}
    p(d_k\mid \boldsymbol{\Lambda})
    \propto
    \frac{1}{N_k}\sum_{j=1}^{N_k}
    \frac{p(\theta_{k,j}\mid \boldsymbol{\Lambda})}{\pi(\theta_{k,j})}.
\end{equation}
%In this way, the posterior distribution from each individual-system fit is represented empirically by its posterior samples, and the system-level marginal likelihood is estimated by reweighting those samples according to the ratio of the population model to the interim prior.

The total likelihood for the population model is given by the product of the marginal likelihoods of all systems,
\begin{equation}
    p_{\rm tot}(\{d_k\}\mid \boldsymbol{\Lambda})
    =
    \prod_{k=1}^{N_{\rm sys}} p(d_k\mid \boldsymbol{\Lambda}),
\end{equation}
where \(N_{\rm sys}\) is the total number of systems. The posterior distribution of the population hyperparameter would be
\begin{equation}
    p(\boldsymbol{\Lambda} \mid \{d_k\})
    \propto
    p_{\rm tot}(\{d_k\}\mid \boldsymbol{\Lambda})\,p(\boldsymbol{\Lambda}),
\end{equation}
where \(p(\boldsymbol{\Lambda})\) is the hyperprior on \(\boldsymbol{\Lambda}\). In our analysis, we adopt a log-uniform hyperprior $p(\boldsymbol{\Lambda})\propto \frac{1}{\boldsymbol{\Lambda}}$. Substituting Eq.~\ref{eqn:pp1} for each system gives
\begin{equation}
    p(\boldsymbol{\Lambda} \mid \{d_k\})
    \propto
    p(\boldsymbol{\Lambda})
    \prod_{k=1}^{N_{\rm sys}}
    \left[
    \frac{1}{N_k}\sum_{j=1}^{N_k}
    \frac{p(\theta_{k,j}\mid \boldsymbol{\Lambda})}{\pi(\theta_{k,j})}
    \right].
\end{equation}

We sample the hyperposterior \(p(\boldsymbol{\Lambda} \mid \{d_k\})\) using the Markov chain Monte Carlo sampler \texttt{emcee} with 32 walkers, running each walker for 50000 steps. We discarded the first 10000 steps as burn-in and thinned the chains by a factor of 10 before constructing the final posterior sample. For each trial value of \(\boldsymbol{\Lambda}\), the posterior samples from the individual-system analysis are reweighted by the importance factor $\frac{p(\theta_{k,j}\mid \boldsymbol{\Lambda})}{\pi(\theta_{k,j})}$, which measures how strongly a given sample is favored by the population model relative to the interim prior.   Fig.~\ref{fig_mcmc1} shows the trace plots and posterior distributions of the hyperparameters for the two models.

%We choose to present $\cos\psi$ rather than $\psi$ because the transformation $\psi=\arccos(\cos\psi)$ is strongly nonlinear near alignment ($\psi\!\approx\!0$, $\cos\psi\!\approx\!1$). In this regime, small symmetric uncertainties in $\cos\psi$ are stretched into asymmetric uncertainties in $\psi$ since
%\[
%\frac{d\psi}{d(\cos\psi)}=-\frac{1}{\sqrt{1-\cos^2\psi}},
%\]
%which diverges as $\cos\psi\to 1$. Equivalently, the density %transforms as
%\[
%p(\psi)=p(\cos\psi)\sin\psi,
%\]
%so probability density is geometrically suppressed at $\psi=0$ and can appear shifted to finite $\psi$ even when the system is nearly aligned. Reporting $\cos\psi$ therefore avoids this coordinate-induced bias and more faithfully represents the inferred alignment.

\subsubsection*{Hierarchal Bayesian Modeling of $\psi_{p,d}$ }

We use a hierarchical Bayesian framework similar to that described in above section to infer the underlying population distribution of \(\psi_{p,d}\). For simplicity, we omit the subscripts below and denote the mutual inclination simply as \(\psi\). We assume two choices for the underlying population distributions $p(\psi \mid \boldsymbol{\Lambda})$: a Rayleigh distribution and a truncated Gaussian distribution, with the same forms as in Eqn.~7 and 8.

In this case, each posterior sample is associated with two degenerate values of the mutual inclination. Rather than selecting a single value for each system, we treat the degeneracy as a nuisance parameter and marginalize over the two possibilities with the equal weight. If \(\psi^{s}_{kj}\) denotes the \(s\)-th degenerate state associated with the \(j\)-th posterior sample for system \(k\), then the marginal likelihood for system \(k\) is
\begin{equation}
p(d_k\mid\boldsymbol{\Lambda})
\propto
\frac{1}{N_k}\sum_{j=1}^{N_k}
\left[
\frac{1}{2}\sum_{s=1,2}
\frac{p(\psi^{s}_{kj}\mid \boldsymbol{\Lambda})}{\pi(\psi^{s}_{kj})}
\right],
\end{equation}
where \(N_k\) is the number of posterior samples for system \(k\). \(\pi(\psi^{s}_{kj})\) is the interim prior and we adot it as uniform in $\cos{{\psi}}$. The full likelihood for the population is the product of all systems 
\begin{equation}
    p_{\rm tot}(\{d_k\}\mid \boldsymbol{\Lambda})
    =
    \prod_{k=1}^{N_{\rm sys}} p(d_k\mid \boldsymbol{\Lambda}),
\end{equation}
where \(\{d_k\}\) denotes the full set of data for all systems. Combining this likelihood with the hyperprior \(p(\boldsymbol{\Lambda})\), the posterior distribution of \(\boldsymbol{\Lambda}\) is
\begin{equation}
    p(\boldsymbol{\Lambda}\mid\{d_k\})
    \propto
    p(\boldsymbol{\Lambda})
    \prod_{k=1}^{N_{\rm sys}}
    \left[
    \frac{1}{N_k}\sum_{j=1}^{N_k}
    \left(
    \frac{1}{2}\sum_{s=1,2}
    \frac{p(\psi^{s}_{kj}\mid \boldsymbol{\Lambda})}{\pi(\psi^{s}_{kj})}
    \right)
    \right].
\end{equation}

\begin{figure}[!h]
\centering
\includegraphics[width=0.85\textwidth]{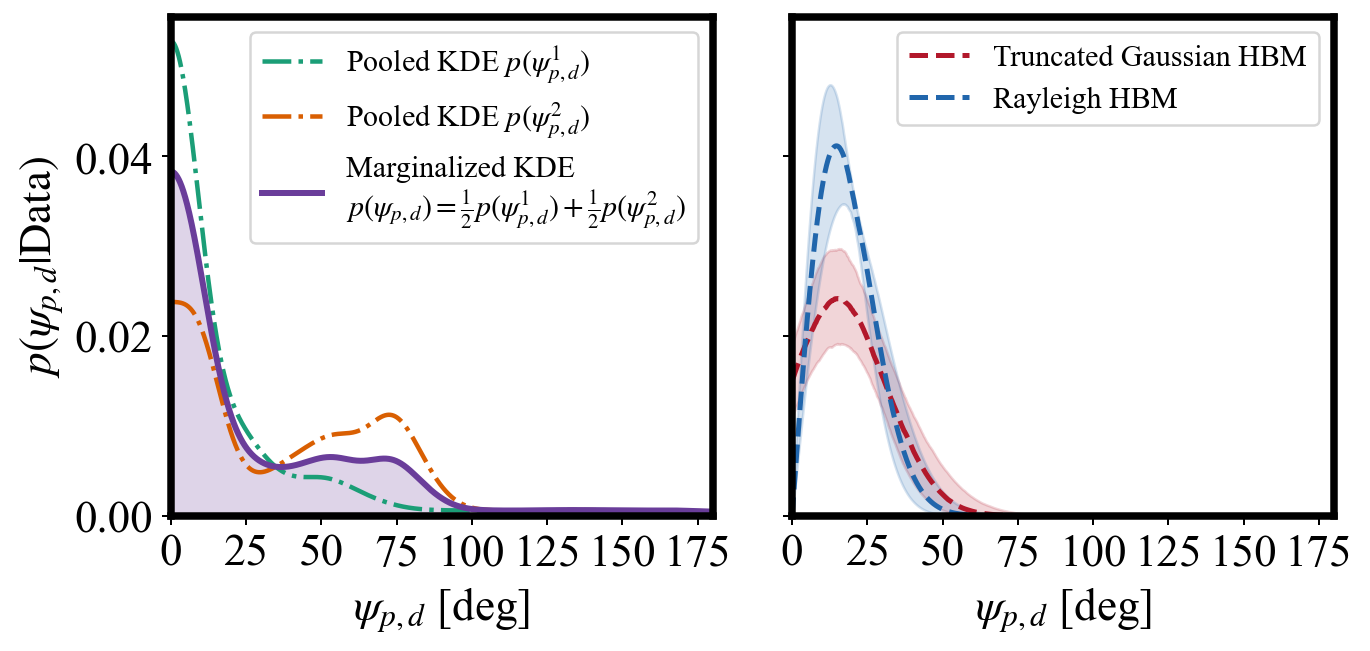}
\caption{\textbf{Distribution of 3-D planet-disk mutual inclinations $\psi_{p,d}$ and hierarchical Bayesian constraints on the underlying population.} Left: KDEs of the pooled posterior distributions for $\psi_{p,d}$. The green and orange dashed lines show the pooled distributions of the two degenerate solutions, $p(\psi^{1}_{p,d})$ and $p(\psi^{2}_{p,d})$.  The purple curve shows the equal-weight marginalized distribution, $p(\psi_{p,d})=\frac{1}{2}p(\psi^{1}_{p,d})+\frac{1}{2}p(\psi^{2}_{p,d})$. Right: HBM fits to $\psi_{p,d}$ assuming underlying Rayleigh (blue) and truncated Gaussian (red) distributions. The dashed lines show the posterior medians, and the shaded regions indicate the corresponding $68\%$ credible intervals. }\label{fig2}
\end{figure}

\begin{figure}
\centering
\includegraphics[width=\textwidth]{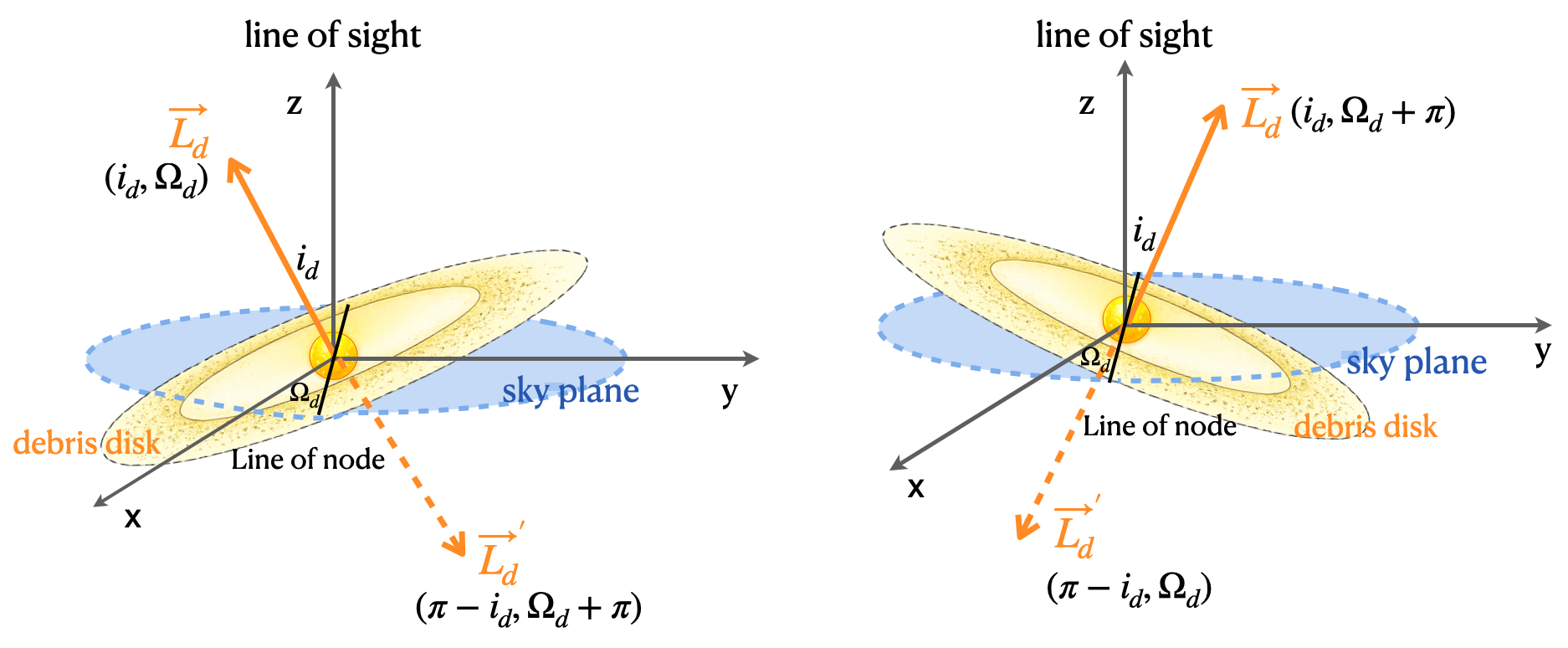}
\caption{\textbf{Geometry of the disk orientation in three dimensions.} The blue circle denotes the sky plane, with the \(z\)-axis pointing along the line of sight. The yellow plane represents the debris disk, and the orange arrows indicate the normal vector of the disk plane. The solid and dashed vectors illustrate the prograde and retrograde degeneracy that arises because the direction of dust motion in the disk is unknown, corresponding to orbital inclinations of \(i_d\) and \(\pi-i_d\), respectively. The left and right panels show the additional degeneracy associated with the unknown near side of the disk, which gives rise to an ambiguity in the longitude of ascending node, \(\Omega_d\) versus \(\Omega_d+\pi\). }\label{figg}
\end{figure}

\begin{figure}[!h]
\centering
\includegraphics[width=0.82\textwidth]{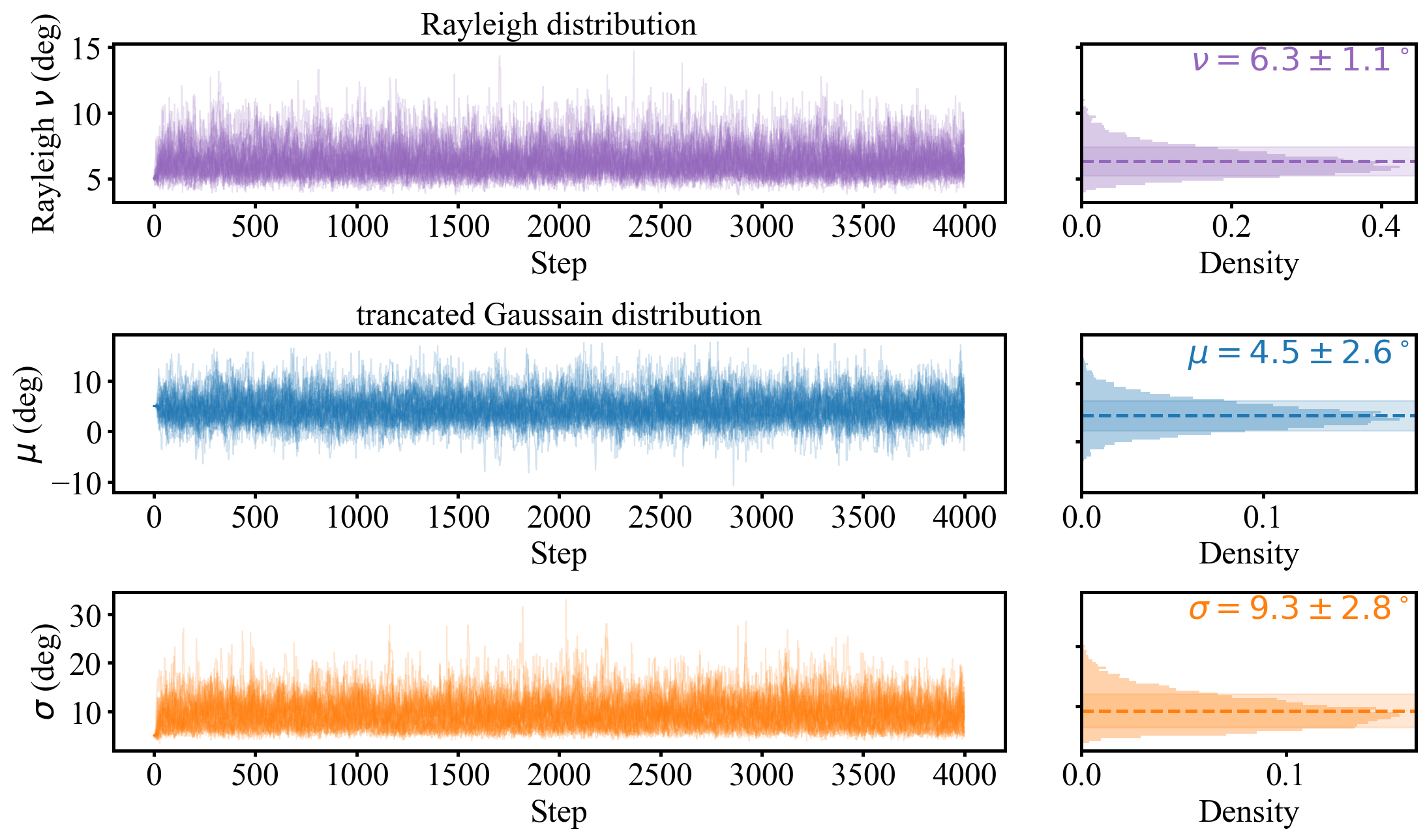}
\caption{\textbf{MCMC diagnostics for the HBM fits to $i_{p,d}$.} The top row shows the results for the Rayleigh population model, and the middle and bottom rows show the results for the truncated Gaussian population models. The left panels show the walker trace plots. The right panels show the corresponding marginalized posterior distributions, where dashed lines mark the posterior means.}
\label{fig_mcmc1}
\end{figure}

\begin{figure}[!h]
\centering
\includegraphics[width=0.82\textwidth]{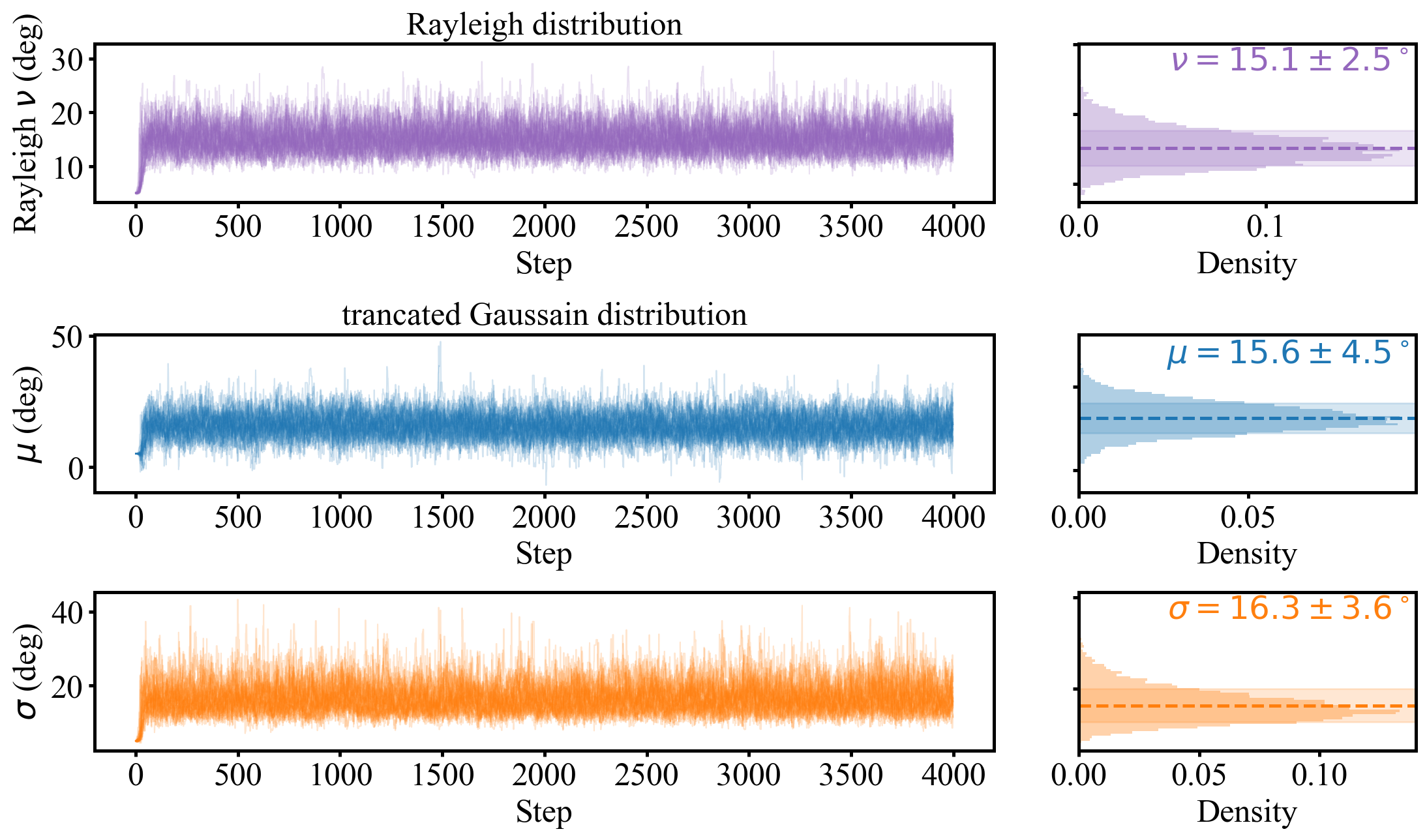}
\caption{\textbf{MCMC diagnostics for the HBM fits to $\psi_{p,d}$.} The top row shows the results for the Rayleigh population model, and the middle and bottom rows show the results for the truncated Gaussian population models. The left panels show the walker trace plots. The right panels show the corresponding marginalized posterior distributions, where dashed lines mark the posterior means.}
\label{fig_mcmc2}
\end{figure}

We sample the posterior distribution of \(\boldsymbol{\Lambda}\) using the \texttt{emcee} Markov chain Monte Carlo sampler with 32 walkers for 50000 steps. We discarded the first 10000 steps as burn-in and thinned the chains by a factor of 10 before constructing the final posterior sample.  Fig.~\ref{fig2} shows the resulting HBM results for $\psi_{p,d}$.   Fig.~\ref{fig_mcmc2} shows the trace plots and posterior distributions of the hyperparameters for the two models. 

\subsubsection*{Observation Bias Test}

In an ideal scenario without observation bias, we would observe the inclination distribution for planets and disk as isotropic, corresponding to be uniform in $\cos{i}$. In practice, however, our sample is not inclination-unbiased. On the planet side, our sample includes two transiting planets, which by construction must have nearly edge-on orbits, and RV detections are also biased toward high inclinations because the RV semi-amplitude scales as $m \sin i$. On the disk side, the ability to detect and characterize disk geometry also depend on inclination. For example, edge-on disks are often easier to spatially resolve in scattered-light imaging, although this dependence is generally weaker in thermal emission. In our sample, four systems(AU mic, HD 114082, $\beta$ Pic, HR 2562) have planet inclinations within $10^\circ$ of an edge-on configuration at $90^\circ$ ($80^\circ < i < 100^\circ$). For an isotropic distribution, the probability of falling in this interval is
\[
P(80^\circ < i < 100^\circ) = \int_{80^\circ}^{100^\circ} \frac{1}{2}\sin i \, di = 0.174,
\]which corresponds to an expected number of $15 \times 0.174 \approx 2.6$ systems in a sample of 15 (excluding the Solar System). The observed count of four systems therefore suggests a mild excess of nearly edge-on orbits from the observational selection effects. 

\begin{figure}
\centering
\includegraphics[width=0.7\textwidth]{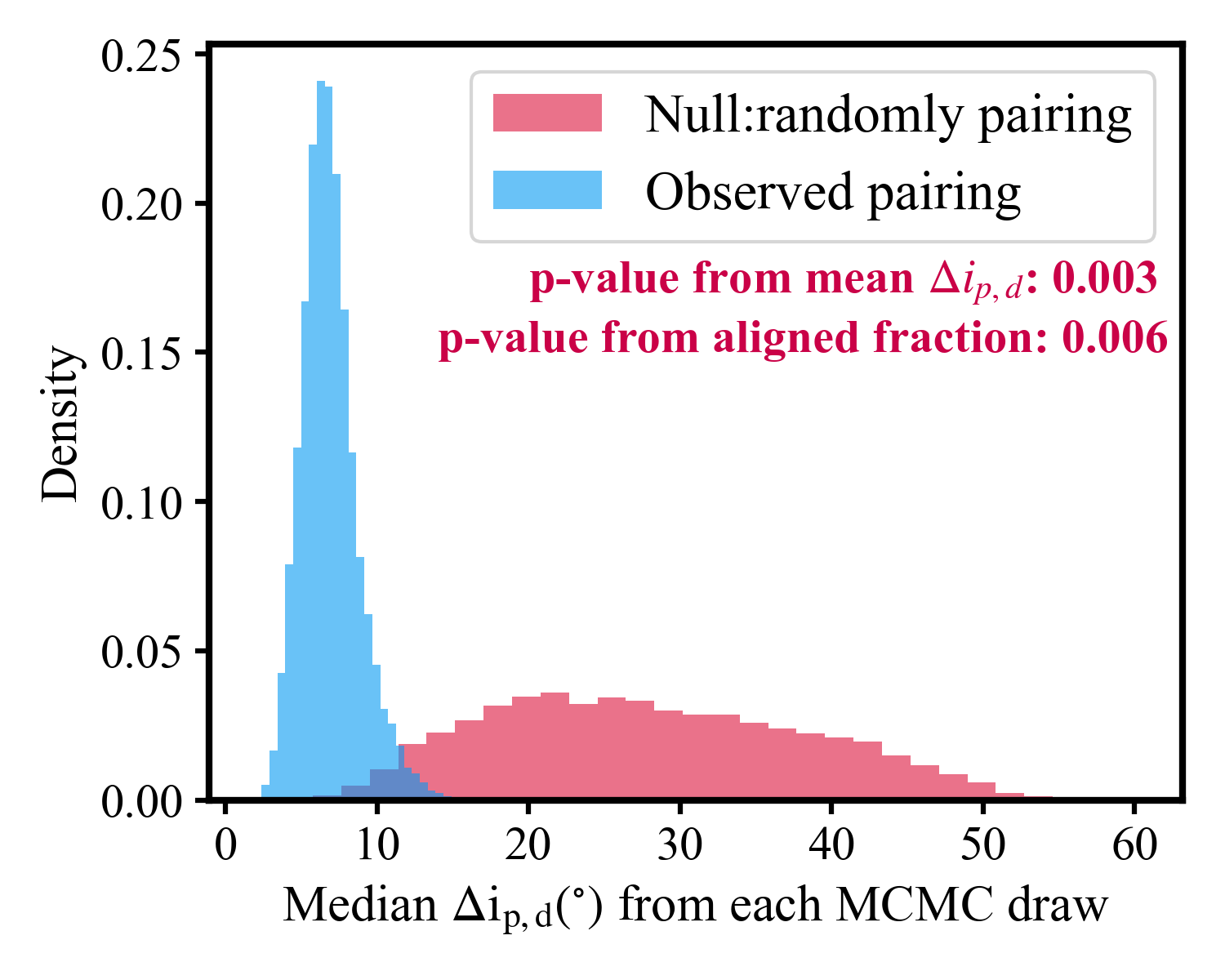}
\caption{\textbf{Results of observation bias test.} The histograms are from 20,000 Monte Carlo trials, with planet and disk inclinations were drawn from their posterior distributions in each trial.  The blue distribution represents the observed sample, in which each planet is paired with the disk of the same system. The red distribution represents the randomly paring sample, in which planet and disk inclinations are randomly reassigned among systems.
}\label{fig_bias}
\end{figure}

We therefore test the null hypothesis that the observed disk-planet alignment solely results from the observed inclination distributions and sample-selection effects using a Monte Carlo permutation test. Specifically, for each Monte Carlo trial, we randomly draw the planet and disk inclinations from their posterior distributions and construct two samples. In the first, the disk and planet inclinations are paired within the same observed system, allowing us to compute the observed $\Delta i_{p,d}$. In the second, the disk and planet inclinations are randomly reassigned between different systems, thereby generating a null sample in which the observation selection effect are preserved but any physical disk-planet connection is removed. We then compare the mean $\Delta i_{p,d}$ of the two samples.   Fig.~\ref{fig_bias} shows the distribution of the mean $\Delta i_{p,d}$ over 20,000 Monte Carlo trials for both the observed and randomly paring samples.

We first define that the randomly paring sample is similar to or more align than observed sample if it has a mean $\Delta i_{p,d}$ smaller than or equal to that of the observed-pair sample. The probability that the randomly paring has a mean $\Delta i_{p,d}$ meets this criteria  is $p = 0.003$, indicating that only $0.3\%$ of randomized trials are at least as well aligned as the observed sample. As a second statistic, we calculate the fraction of systems with $\Delta i_{p,d}<10^{\circ}$ in each sample for each Monte Carlo trial. The probability that randomly paring sample has a same or higher fraction than that of observed sample is $p = 0.006$. Both probability strongly reject the null hypothesis that the observed alignment between disk and planets are from observation bias and sample selection effect.

%%%%%%%%%%%%%%%% SUPPLEMENTARY TEXT %%%%%%%%%%%%%%%

\subsection*{Supplementary Text}
\subsubsection*{Results for individual system}
\subsubsection*{1. HD 38529}

\begin{figure}[!h]
\centering
\includegraphics[width=\textwidth]{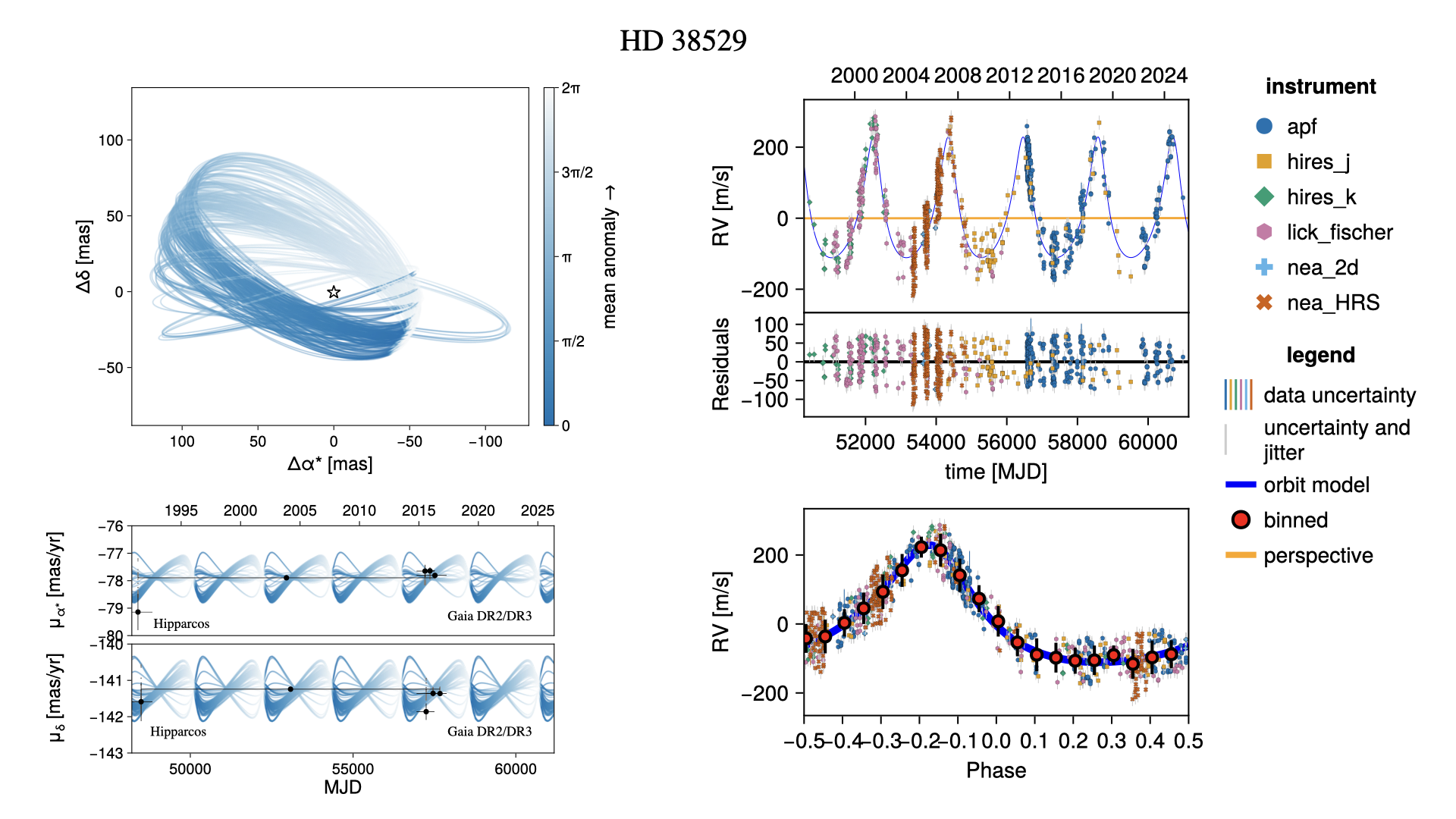}
\caption{\textbf{3-D Keplerian orbital model of HD 38529 c.} Left: astrometric orbits of HD 38529 c projected on the sky plane from our model (top) and observed and fitted absolute astrometry from Hipparcos and Gaia DR2 and DR3 in right ascension and declination (middle and bottom). The blue lines are randomly drawn from the posterior distribution. Right: observed and fitted RVs used in the fitting. }\label{fig_ft1}
\end{figure}

\begin{figure}[!b]
\centering
\includegraphics[width=\textwidth]{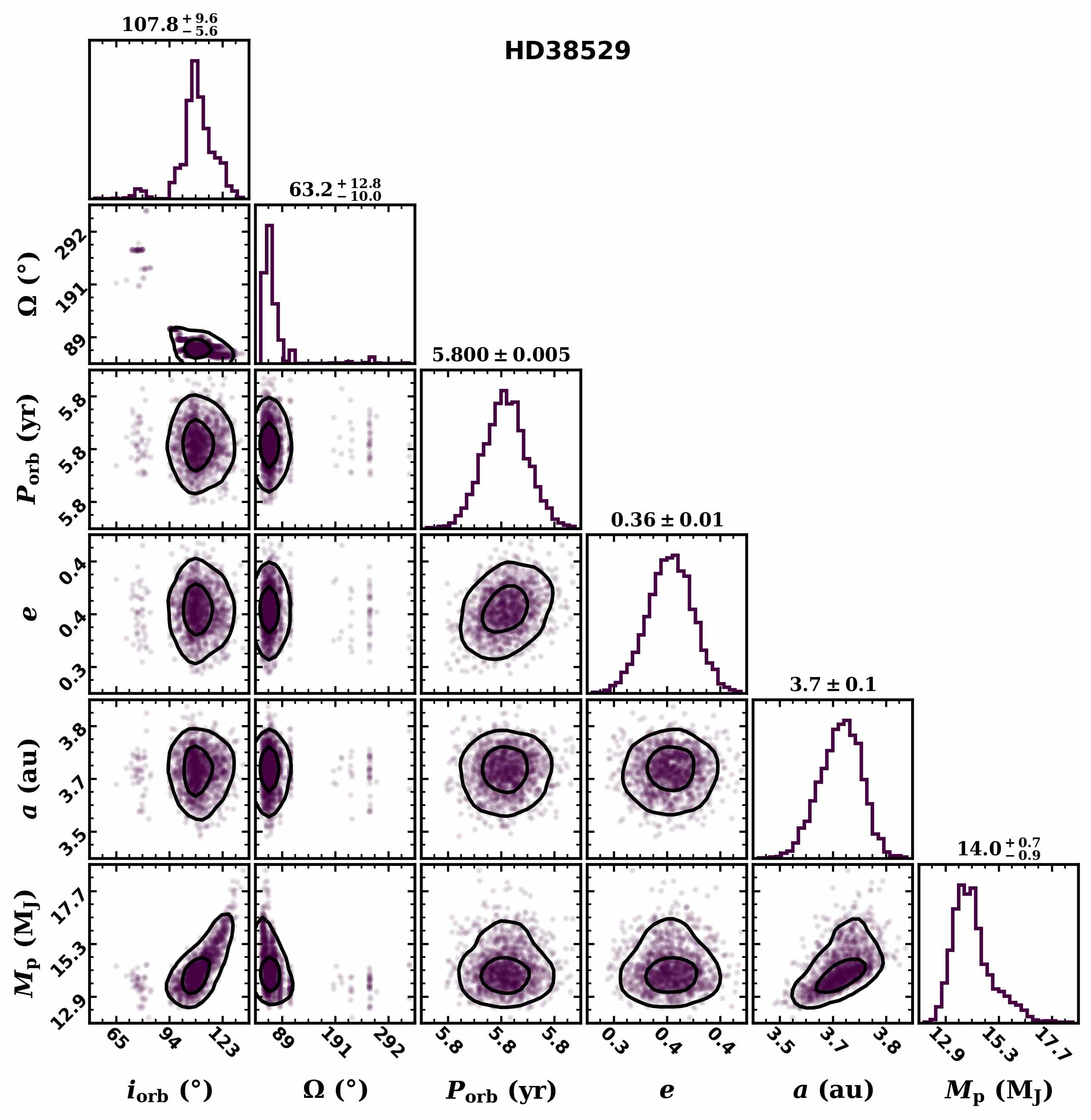}
\caption{\textbf{The joint posterior distributions from the orbital fit of HD 385529 c.} The plotted parameters are orbital inclination $i_{\rm orb}$, longitude of ascending node $\Omega$, orbital period $P_{\rm {orb}}$, eccentricity, and semi-major axis $a$ and planet mass $M_p$.}\label{fig_cn1}
\end{figure}

HD 38529 (HIP 27253) is a bright G4 sub-giant star ($M_*=1.41^{+0.03}_{-0.03}\ \rm{M_{\odot}}$, $R_*=2.84^{+0.14}_{-0.14}\ \rm{R_{\odot}}$, $\rm m_v\sim 5.9$ ) with a age of $3.09^{+0.17}_{-0.17}$ Gyr \cite{Jofre2015,TESSinput}. The star hosts two confirmed planets, both detected with RV observations. HD 38529 b is a close-in, sub-Jupiter mass planet, with an orbital period of $\sim 14$ days and minimum mass of $0.65^{+0.05}_{-0.06}\ \rm M_{J}$ \cite{Fischer2001}. HD 38529 c is a cold Jupiter, orbiting at $\sim3.7$ AU and with a minimum mass of $13.2^{+0.1}_{-0.1}\rm M_{J}$ \cite{Fischer2003}. In the orbital fitting, we ignore the inner planet b because its 14-day orbital period is much shorter than the observing windows of \textit{Hipparcos} ($\Delta t \sim 1278$ days), \textit{Gaia} DR2 ($\Delta t \sim 668$ days), and \textit{Gaia} DR3 ($\Delta t \sim 1038$ days). The astrometric signal of HD 38529 b is effectively averaged out as \textit{Gaia} DR2/DR3 only report mean astrometric quantities across multiple measurements over the observing window. We used the archival RVs from multiple instruments, with a RV baseline over 26 years (see   Table~\ref{tab:rv_tab1}). HD 38529 has a wide stellar companion at a separation of $283^{\prime\prime}$, corresponding to $\sim 12300$ AU \cite{Mason2001}. Because this companion is so widely separated, we do not consider it in the orbital fitting. Fig.~\ref{fig_ft1} presents the data and best-fit model for
the HD 38529 c. We obtained an orbital inclination of $107.8^{+9.6}_{-5.6}\,^{\circ}$ and a longitude of ascending node of $63.2^{+12.8}_{-10.0}\,^{\circ}$ deg. Our inclination constraint is consistent with that from ref. \cite{Feng2022}, which obtained an orbital inclination of $104.5^{+6.4}_{-8.7}\,^{\circ}$. The joint posterior distributions for additional orbital parameters in our fitting are presented in Fig.~\ref{fig_cn1}.

The debris disk around HD 38529 was spatially resolved at $70\ \mu\mathrm{m}$ with \textit{Herschel}/PACS. By modeling the disk geometry, ref. \cite{xuan2020_disk} derived a disk inclination of $71^{+10}_{-7}\,^{\circ}$ and a position angle of $48^{+5}_{-5}\,^{\circ}$. As discussed in Method, the imaging data constrain only the projected disk morphology but not the moving direction of the disk material. The disk inclination is therefore subject to a prograde-retrograde degeneracy, $i_d \rightarrow 180^{\circ}-i_d$. The alternative disk inclination is therefore $i_d = 109.0^{+10.0}_{-7.0}\,^{\circ}$. Considering the prograde configuration between planet and disk, in which the disk angular-momentum vector lies in the same hemisphere as the planetary orbital angular-momentum vector, we find $\Delta i_{p,d} = 2.98_{-2.6}^{+8.6}\,^{\circ}$.

Several independent  measurements of the stellar rotation period of HD 38529 were obtained with different instruments and are in good agreement. From photometric modulation observed with Automated Photoelectric Telescope at Fairborn Observatory, ref. \cite{Fischer2003} identified a rotation period of $P_{\rm rot}=35.7$ d. From HST/FGS photometry, ref. \cite{Benedict2010} measured $P_{\rm rot}=37.0 \pm 0.4$ d.  Ref. \cite{Henry2013} used 12 years of precision photometry and found a consistent rotation period of $P_{\rm rot}=37.1 \pm 0.2$ d. We adopt the average value of the three measurements and derive a stellar inclination of $112.2^{+9.7}_{-14.2}\,^{\circ}$. The stellar obliquity for HD 38529 c projected in line of sight direction is therefore $\Delta i_{p,*} = 3.9^{+9.6}_{-3.9}\,^{\circ}$.

\subsubsection*{2. HD 38588}

\begin{figure}[!t]
\centering
\includegraphics[width=\textwidth]{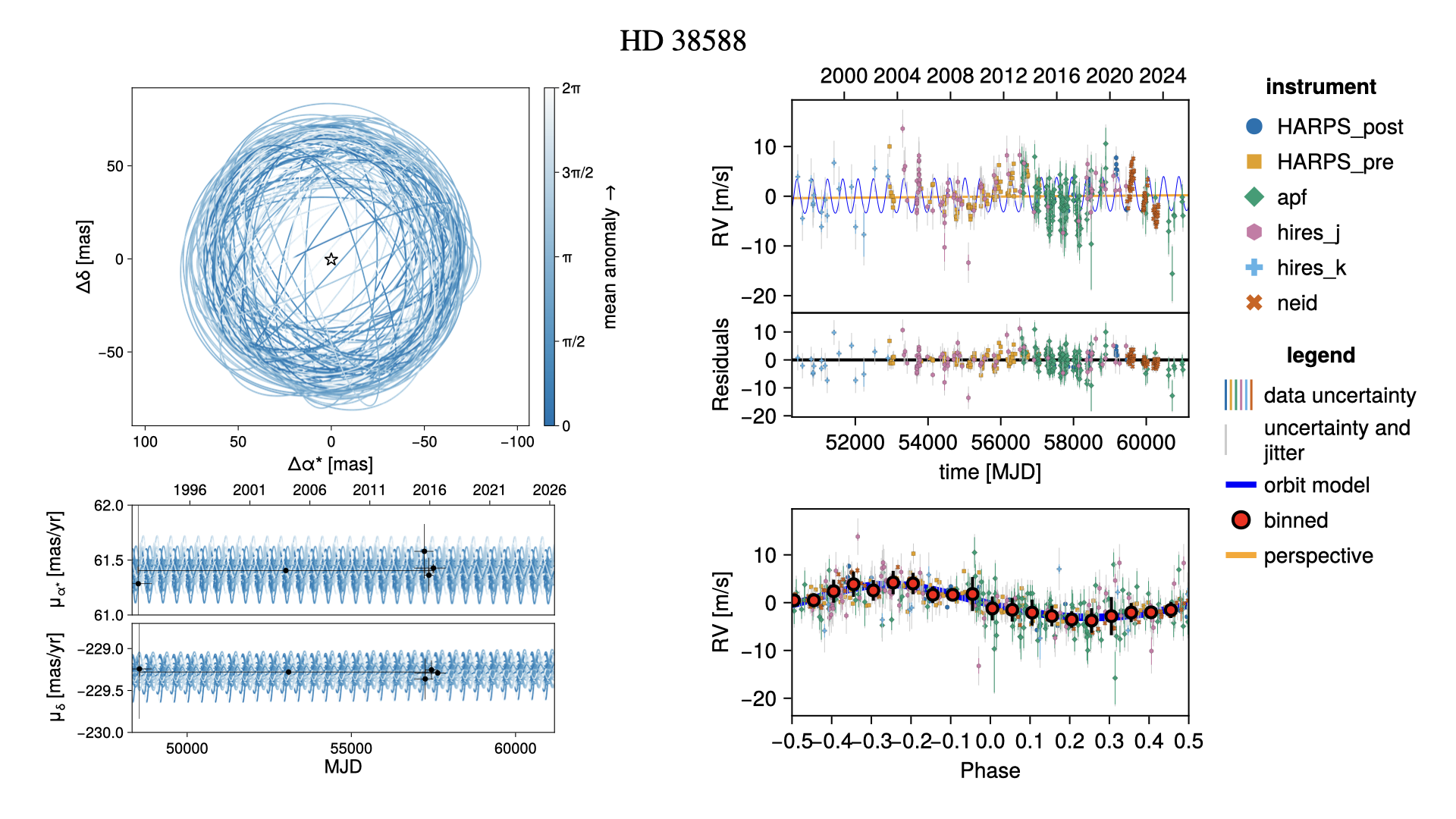}
\caption{\textbf{3-D Keplerian orbital model of HD 38558 b.} Left: astrometric orbits of HD 38558 b projected on the sky plane from our model (top) and observed and fitted absolute astrometry from Hipparcos and Gaia DR2 and DR3 in right ascension and declination (middle and bottom). The blue lines are randomly drawn from the posterior distribution. Right: observed and fitted RVs used in the fitting. }\label{fig_ft2}
\end{figure}

\begin{figure}
\centering
\includegraphics[width=0.85\textwidth]{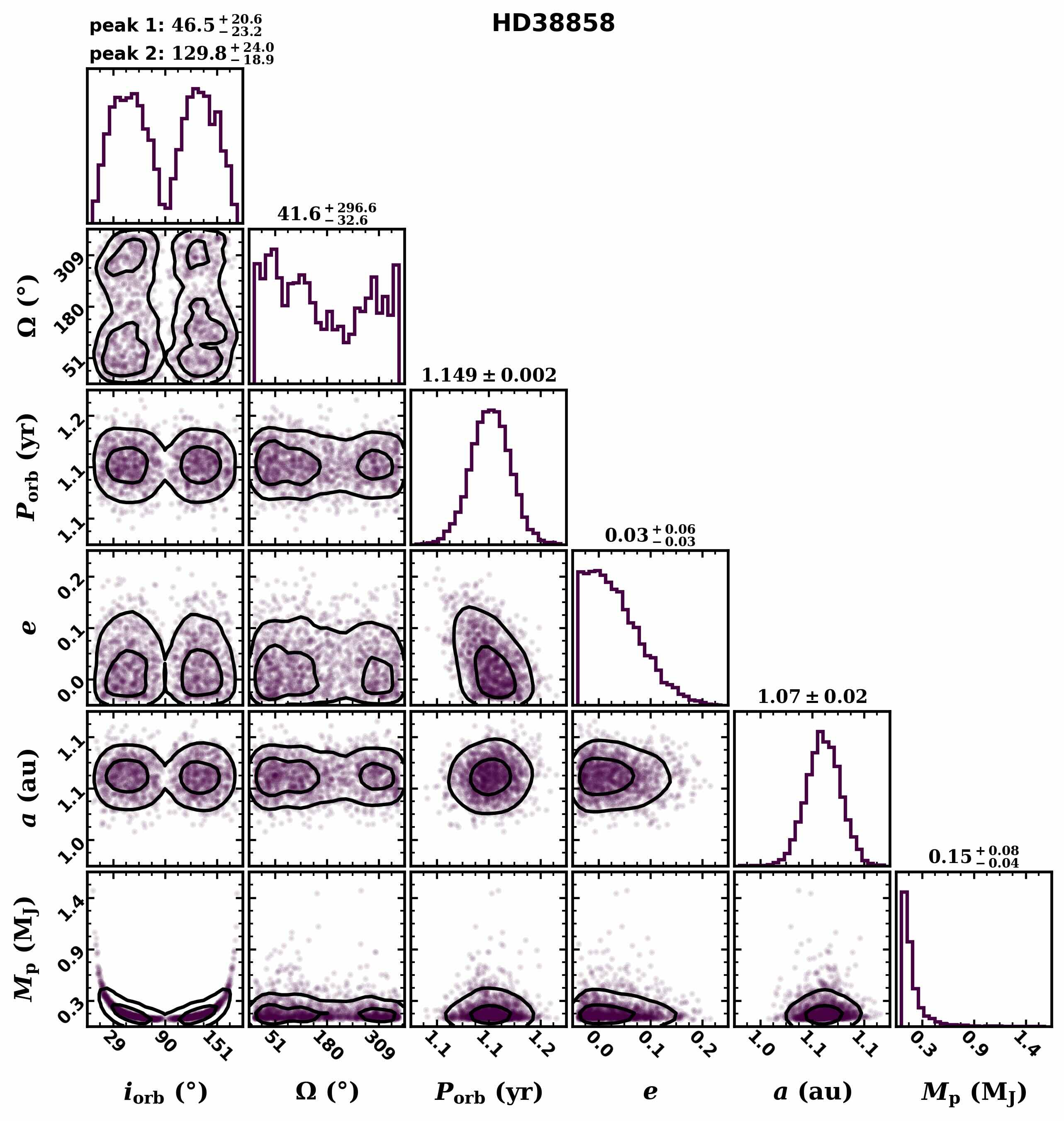}
\caption{\textbf{The joint posterior distributions from the orbital fit of HD 38858 b.} The plotted parameters are orbital inclination $i_{\rm orb}$, longitude of ascending node $\Omega$, orbital period $P_{\rm {orb}}$, eccentricity, and semi-major axis $a$ and planet mass $M_p$.}\label{fig_cn2}
\end{figure}

HD 38858 (HIP 27435) is a sun-like star ($M_*=0.94^{+0.04}_{-0.04}\ \rm{M_{\odot}}$, $\rm m_v\sim 6.0$) with a distance of $\sim15$ pc and an age of $5.4^{+2.7}_{-3.1}$ Gyr \cite{Takeda2007,TESSinput}. The star hosts a confirmed giant planet at $\sim 1 $ AU with a minimum mass of $0.10^{+0.01}_{-0.01}\ \rm M_{J}$ \cite{Mayor2011}. We used archival RVs from several instruments, with a RV baseline over 25 years (see   Table~\ref{tab:rv_tab1}). Fig.~\ref{fig_ft2} presents the data and best-fit model for the HD 38588 b. We obtained a bimodal distribution in the orbital inclination, symmetric along $90^{\circ}$, for which the sum of values at two peaks equals to $180^{\circ}$. These two peaks result from the prograde-retrograde degeneracy, corresponding to that the planet orbits in the same plane but can move in two opposite directions. The two possible orbital inclinations are $46.5^{+20.6}_{-23.2}\,^{\circ}$ and $129.8^{+24.0}_{-18.9}\,^{\circ}$. The joint posterior distributions for additional orbital parameters are presented in Fig.~\ref{fig_cn2}. 

The debris disk around HD 38858 was spatially resolved at $70 \mu\mathrm{m}$ with \textit{Herschel}/PACS. By modeling the disk orientation, ref.\cite{Kennedy2015} derived a disk inclination of $44^{\circ}$ and a position angle of $67^{\circ}$, each with an uncertainty of $5^{\circ}$. Adopting a prograde configuration between planet and debris disk, we find $\Delta i_{p,d} = 4.9^{+17.8}_{-4.9}\,^{\circ}$.

Ref. \cite{Isaacson2010} and ref. \cite{Fuhrmeister2022} independently inferred the stellar rotation period of HD 38858 from chromospheric activity indicators, obtaining consistent values of 24.0 and 24.7 d, respectively. We adopt the average stellar rotation period from their measurements and derive a stellar inclination of $33.0^{+10.9}_{-12.1}\,^{\circ}$. The stellar obliquity for HD 38858 b projected in line of sight direction would be $\Delta i_{p,*} = 8.5^{+18.3}_{-8.4}\,^{\circ}$.

\subsubsection*{3. HD 82943}
\begin{figure}[!t]
\centering
\includegraphics[width=\textwidth]{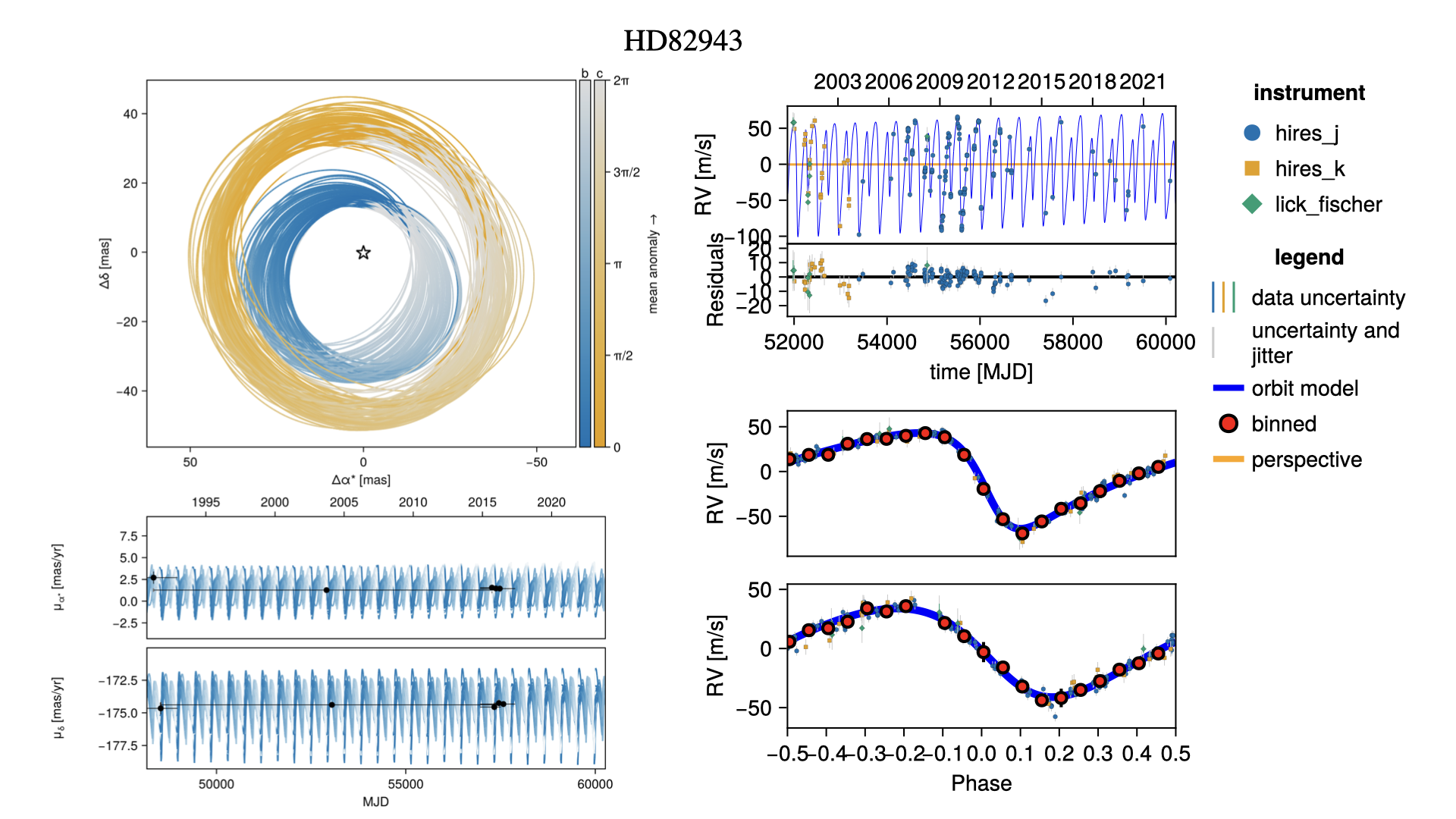}
\caption{\textbf{3-D Keplerian orbital model of HD 82943 b$\&$c.} Left: astrometric orbits of HD 82943 b$\&$c projected on the sky plane from our model (top) and observed and fitted absolute astrometry from Hipparcos and Gaia DR2 and DR3 in right ascension and declination (middle and bottom). The blue and yellow lines are randomly drawn from the posterior distribution. Right: observed and fitted RVs used in the fitting.}\label{fig_ft13}
\end{figure}

\begin{figure}
\centering
\includegraphics[width=\textwidth]{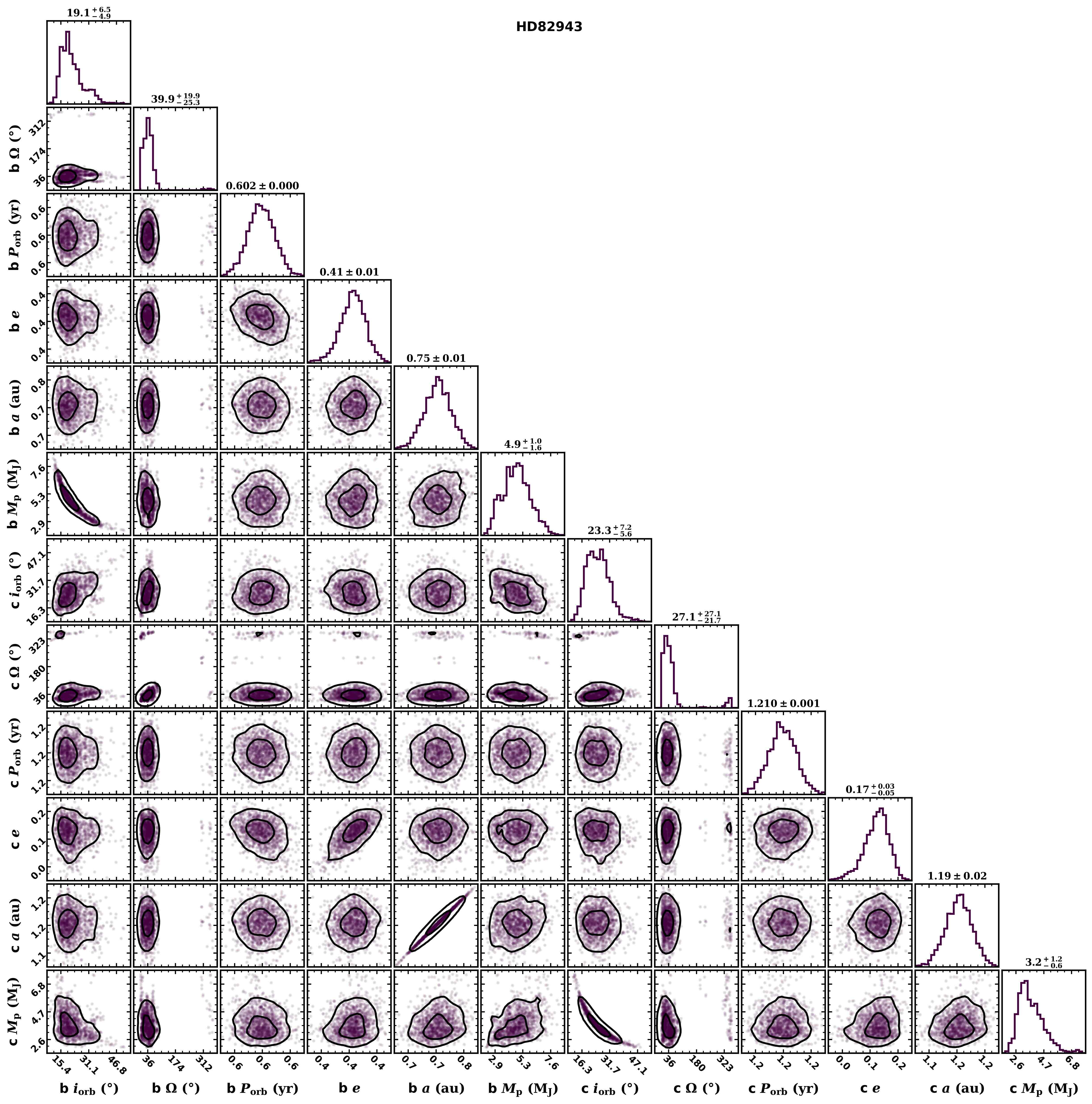}
\caption{\textbf{The joint posterior distributions from the orbital fit of HD 82943 b$\&$c.} The plotted parameters are orbital inclination $i_{\rm orb}$, longitude of ascending node $\Omega$, orbital period $P_{\rm {orb}}$, eccentricity, and semi-major axis $a$ and planet mass $M_p$.}\label{fig_cn13}
\end{figure}

HD 82943 (HIP 47007) is a G2V star ($M_*=1.04^{+0.02}_{-0.02}\ \rm{M_{\odot}}$, $\rm m_v\sim7.4$ ) at a distance of $\sim 36 $ pc and with an age of $2.4^{+0.3}_{-0.3}$ Gyr \cite{Liu2014, TESSinput}. The star hosts two giant planets at $1:2$ orbital resonance, with semi-major axis of $\sim0.74$ AU and $\sim1.2$ AU \cite{Mayor2004}. We used archival RVs spanning from Mar. 2001 to May 2023 (see   Table~\ref{tab:rv_tab1}). We consider a two-planet model in the orbital fitting. Fig.~\ref{fig_ft13} presents the data and best-fit model for the HD 82943 b$\&$c. We obtained the orbital inclination and longitude of ascending node of the HD 82943 b as $19.1^{+6.5}_{-4.9}\,^{\circ}$ and $39.9^{+19.9}_{-25.3}\,^{\circ}$. The orbital inclination and longitude of ascending node of HD 82943 c are $23.3^{+7.2}_{-5.6}\,^{\circ}$ and $27.9^{+27.1}_{-21.7}\,^{\circ}$. Our result is consistent with that of ref. \cite{tan2013}, which inferred an orbital inclination of $\sim 20^{\circ}$ for the HD 82943 b$\&$c by modeling the RV curvature with a dynamical fit that accounts for planet-planet interactions in the mean-motion resonant configuration. Using the measured inclinations and longitudes of ascending node, we find a mutual inclination between HD 82943 b$\&$c  of $6.0^{+9.8}_{-6.0}\,^{\circ}$, indicating that the two giant planets are nearly coplanar. The joint posterior distributions for additional orbital parameters are presented in Fig.~\ref{fig_cn13}.

The debris disk around HD 82943 was resolved at $70 \mu\mathrm{m}$ with \textit{Herschel}/PACS. Ref. \cite{Kennedy2013} modeled the disk geometry and obtained a disk inclination of $27\pm{4}\,^{\circ}$ and a position angle of $152\pm{8}\,^{\circ}$. The minimum planet-disk mutual inclinations for HD 82943 b and c are $6.8^{+4.5}_{-5.4}\,^{\circ}$ and $1.8^{+5.6}_{-1.8}\,^{\circ}$, respectively.

Using the chromospheric activity index, ref. \cite{Mayor2004} and ref. \cite{Isaacson2010} reported stellar rotation periods of HD 82943 as 18 and 17 d, respectively. We also independently identify a stellar rotation period of $18 \pm 12$ d from the TESS light curves. From these constraints, we derive a stellar inclination of $19.9^{+3.7}_{-3.7}\,^{\circ}$.  The stellar obliquity for HD 82943 b and c projected in line of sight direction would be $1.9^{+4.9}_{-1.8}\,^{\circ}$ and $2.3^{+6.4}_{-2.3}\,^{\circ}$.

\subsubsection*{4. HD 113337}

\begin{figure}[!t]
\centering
\includegraphics[width=\textwidth]{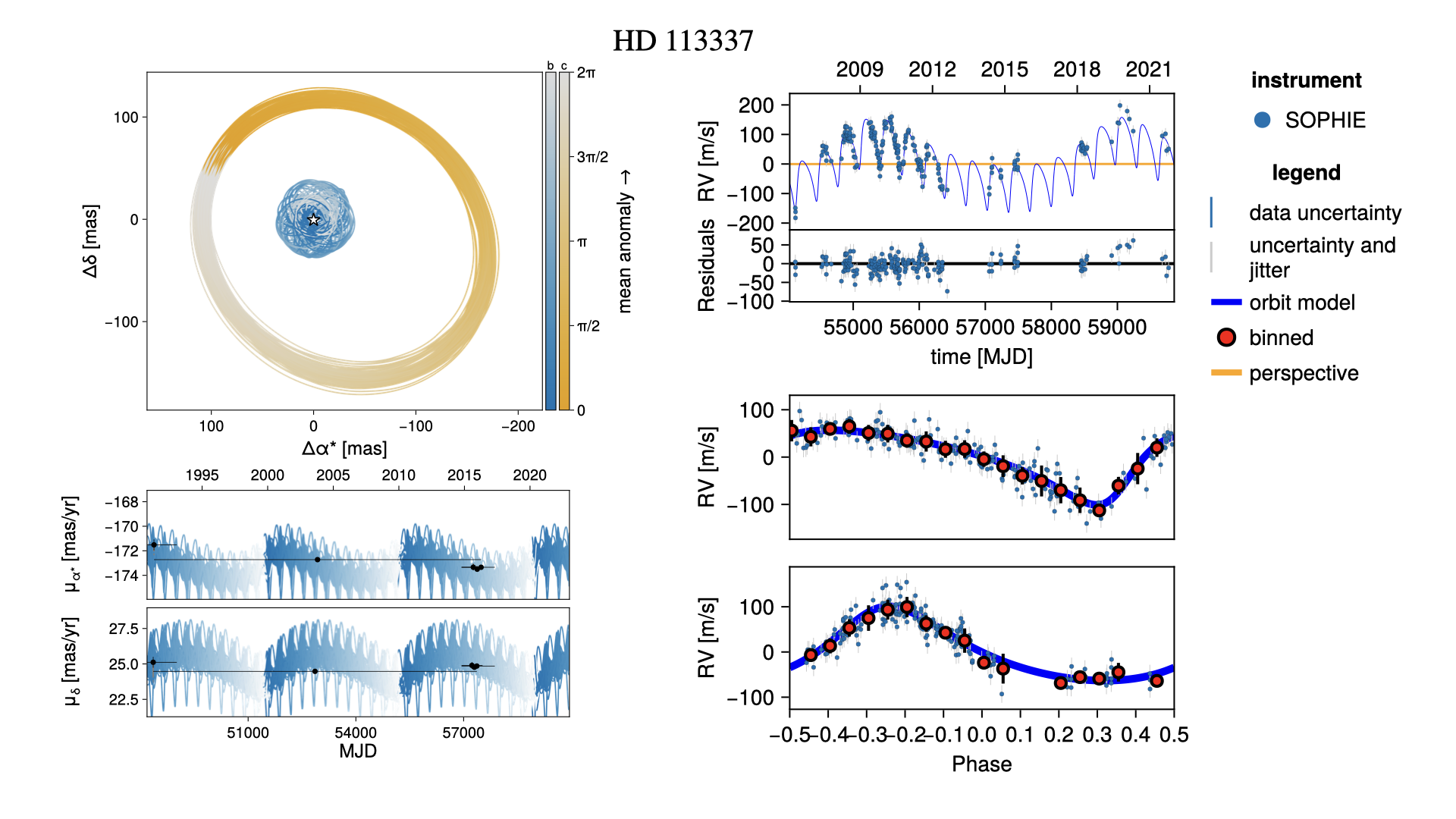}
\caption{\textbf{3-D Keplerian orbital model of HD 113337 b$\&$c.} Left: astrometric orbits of HD 113337 b$\&$c projected on the sky plane from our model (top) and observed and fitted absolute astrometry from Hipparcos and Gaia DR2 and DR3 in right ascension and declination (middle and bottom). The blue lines are randomly drawn from the posterior distribution. Right: observed and fitted RVs used in the fitting. }\label{fig_ft3}
\end{figure}

\begin{figure}
\centering
\includegraphics[width=\textwidth]{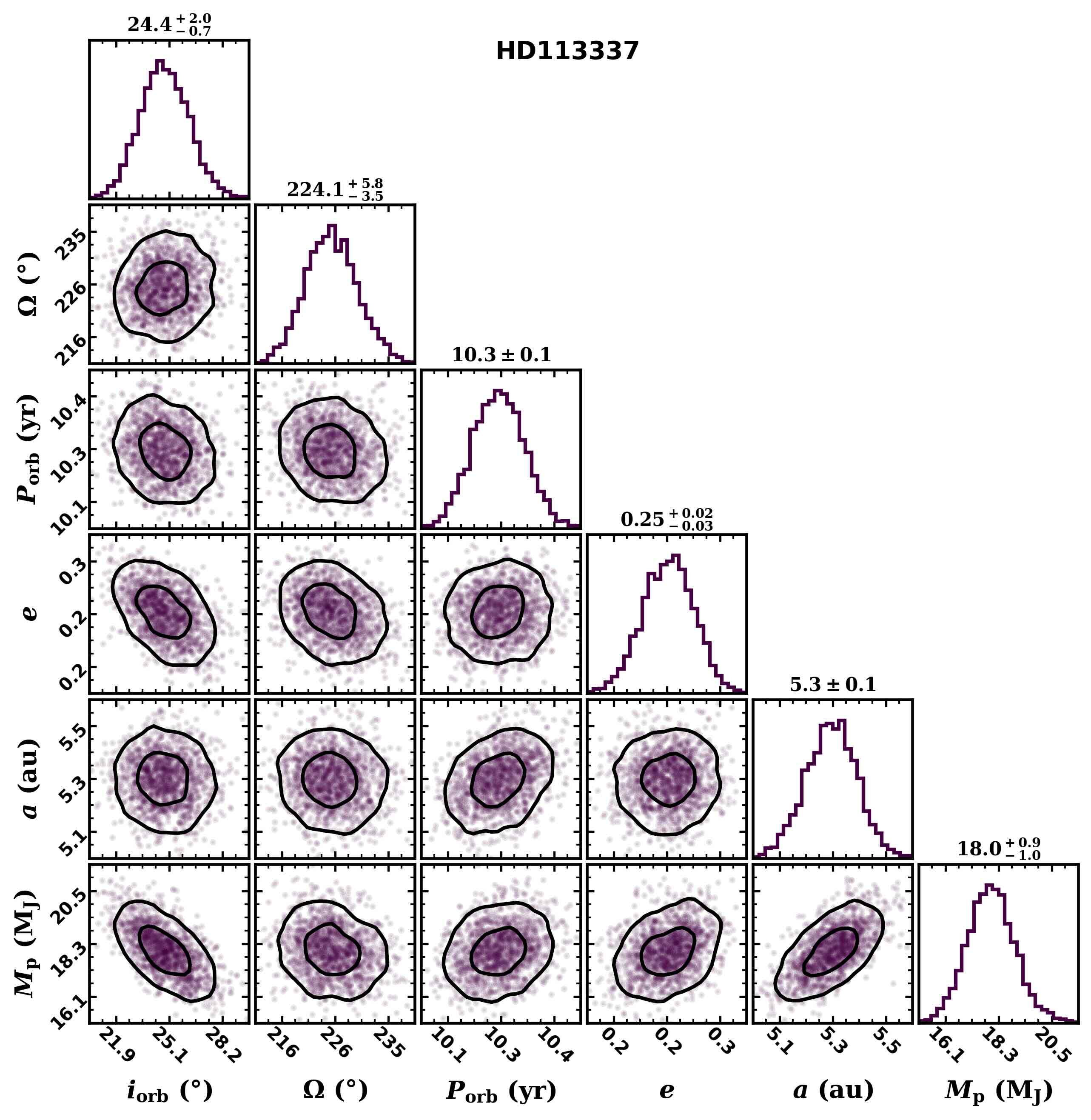}
\caption{\textbf{The joint posterior distributions from the orbital fit of HD 113337 c.} The plotted parameters are orbital inclination $i_{\rm orb}$, longitude of ascending node $\Omega$, orbital period $P_{\rm {orb}}$, eccentricity, and semi-major axis $a$ and planet mass $M_p$.}\label{fig_cn3}
\end{figure}

HD 113337 (HIP 63584) is a bright F6V star ($M_*=1.40^{+0.14}_{-0.14} \rm M_{\odot}$, $\rm m_v\sim 6.0$, $d\sim 37$ pc), with an age of $150^{+100}_{-50}$ Myr \cite{Borgniet2019,TESSinput}. There are two giant planets orbiting around the star, at semi-major axis of $\sim 1$ AU and $\sim 5.3$ AU, respectively \cite{Borgniet2014,xuan2020_disk}. We consider a two-planet model for the orbital fitting. In addition, HD 113337 has a wide M star companion around $120$ arcsec, corresponding to $\sim4400$ AU \cite{Mason2001}. Because this companion is so widely separated, we ignore the wide stellar companion  in the orbital fitting. We used SOPHIE RVs spanning January 2007 to July 2022, corresponding to a baseline of more than 15 years (see   Table~\ref{tab:rv_tab1}). Fig.~\ref{fig_ft3} presents the data and best-fit model for the HD 113337 $b\&c$. We find that the orbital inclination of the inner planet is poorly constrained because its short orbital period ($\sim 317$ d) is shorter than the \textit{Hipparcos} and \textit{Gaia} observing windows, causing its astrometric signal to be diluted by time averaging. But we obtain a well-constrained three-dimensional orbit for planet c, for which the orbital inclination is $24.4^{+2.0}_{-0.7}\,^{\circ}$. Our orbital inclination is consistent with to that ($i_c=31.0^{+5.0}_{-4.0}\,^{\circ}$ deg) from ref. \cite{xuan2020_disk} within $1\sigma$. The joint posterior distributions for additional orbital parameters are presented in Fig.~\ref{fig_cn3}.

The debris disk around HD 113337 was resolved  at $70 \mu\mathrm{m}$ with \textit{Herschel}/PACS.  Ref. \cite{xuan2020_disk} modeled the disk geometry and derived a disk inclination of $13^{+10}_{-9}\,^{\circ}$ and a position angle of $160^{+63}_{-38}\,^{\circ}$. Ref. \cite{Marshall2021} analyzed the same dataset and also reported a consistent disk inclination of $19.0^{+5.3}_{-6.2}\,^{\circ}$ and position angle of $119.8^{+11.5}_{-12.6}\,^{\circ}$. Therefore, we find $\Delta i_{p,d} = 9.8^{+6.4}_{-9.8}\,^{\circ}$. 

Ref. \cite{Borgniet2014} identified a stellar rotation period of $3 \pm 1$ d for HD 113337 from the residual radial velocities and the bisector velocity span. Using the TESS light curves, we independently identify a period of $5 \pm 1$ d (see Fig.~\ref{fig_tl2}), which is consistent with their measurement at the $1\sigma$ level. Adopting a stellar rotation period of $5 \pm 1$ d, we derive a stellar inclination of $25.5^{+4.8}_{-3.6}\,^{\circ}$ and the line-of-sight stellar obliquity for HD 113337 c as  $\Delta i_{p,*} = 1.1^{+3.5}_{-1.1}\,^{\circ}$.

\subsubsection*{5. HD 10647}

\begin{figure}[!h]
\centering
\includegraphics[width=\textwidth]{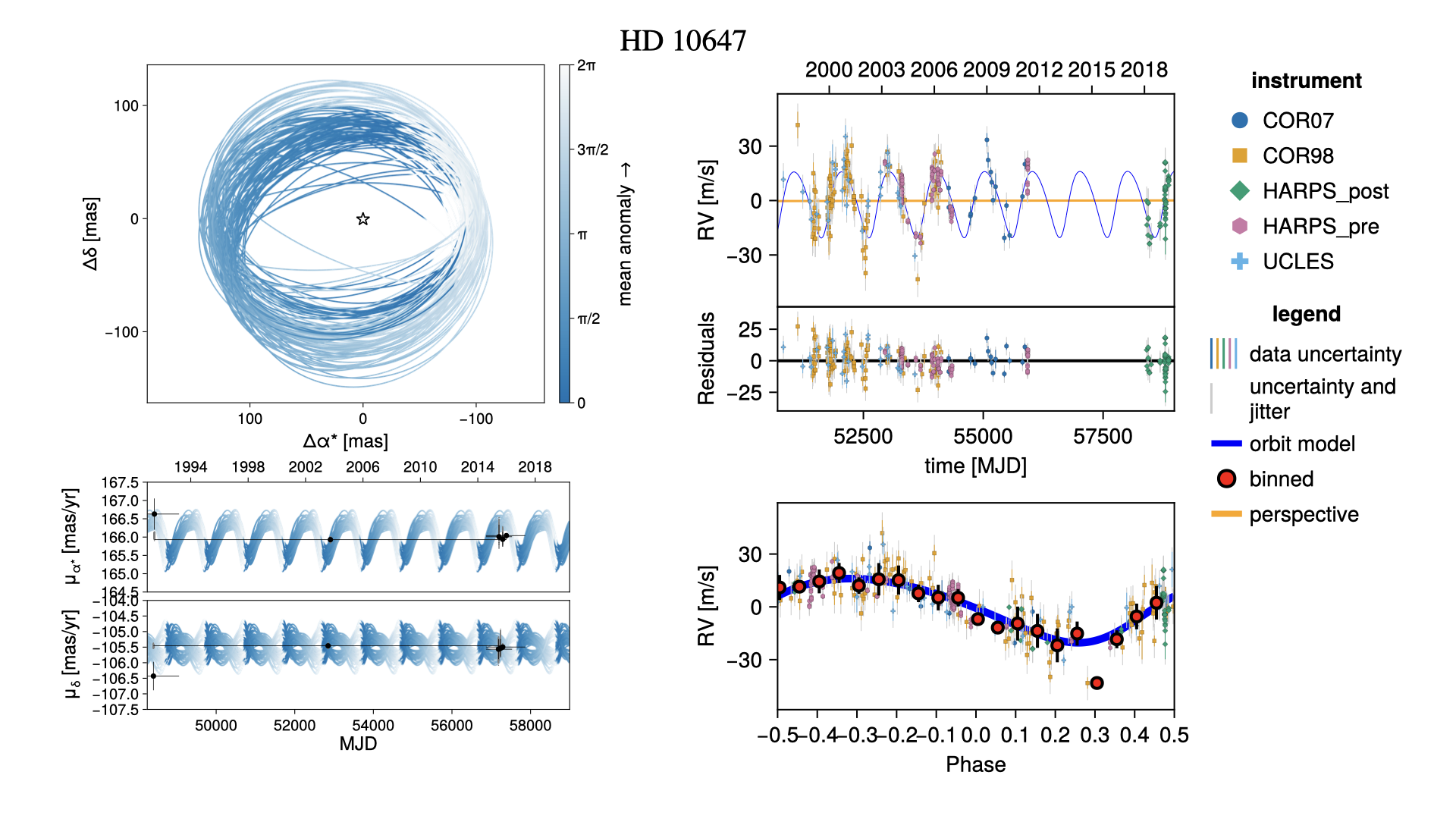}
\caption{\textbf{3-D Keplerian orbital model of HD 10647 b.} Left: astrometric orbits of HD 10647 b projected on the sky plane from our model (top) and observed and fitted absolute astrometry from Hipparcos and Gaia DR2 and DR3 in right ascension and declination (middle and bottom). The blue lines are randomly drawn from the posterior distribution. Right: observed and fitted RVs used in the fitting. }\label{fig_ft4}
\end{figure}

\begin{figure}[!b]
\centering
\includegraphics[width=\textwidth]{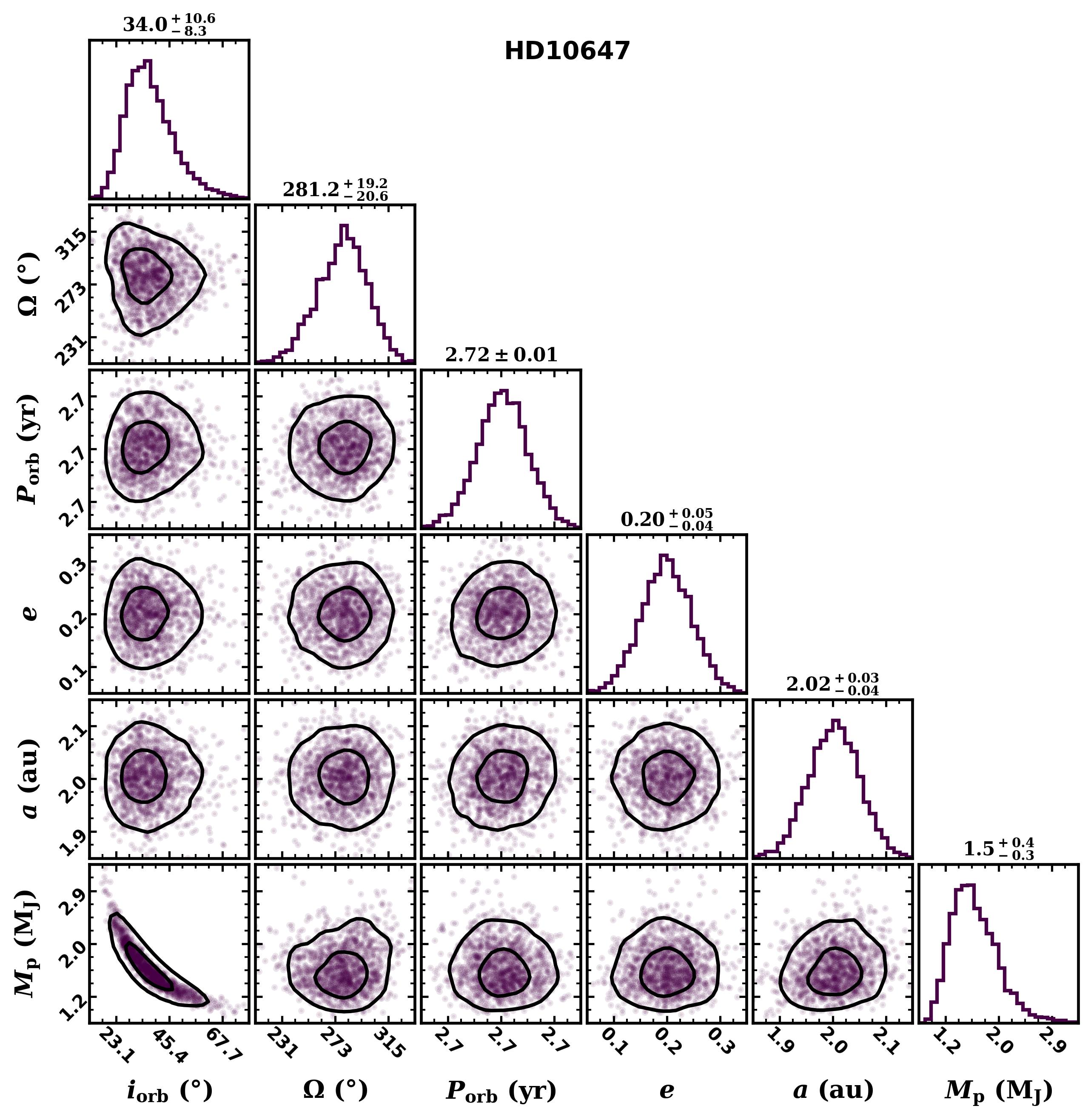}
\caption{\textbf{The joint posterior distributions from the orbital fit of HD 10647 b.} The plotted parameters are orbital inclination $i_{\rm orb}$, longitude of ascending node $\Omega$, orbital period $P_{\rm {orb}}$, eccentricity, and semi-major axis $a$ and planet mass $M_p$.}\label{fig_cn4}
\end{figure}

HD 10647 ($q^{1}$ Eri, HIP 7978) is a nearby F9V star ($M_*=1.11^{+0.02}_{-0.02}\ \rm M_{\odot}$, $\rm m_v\sim 5.5$) with a distance of $\sim 17.3$ pc and an age of $1.4^{+0.9}_{-0.9}$ Gyr \cite{Marmier2013, TESSinput}. It hosts a cold Jupiter at $\sim 2$ AU with a minimum mass of $\sim0.94\ \rm M_{J}$ \cite{Bulter2006}.  We used archival RVs from several instruments, with a RV baseline over 21 years (see   Table~\ref{tab:rv_tab1}). Fig.~\ref{fig_ft4} presents the data and best-fit model for the HD 10647 b. We obtained an orbital inclination of $34.0^{+10.6}_{-8.3}\,^{\circ}$ and longitude of ascending node of $281.2^{+19.2}_{-20.6}\,^{\circ}$.  The joint posterior distributions for additional orbital parameters are presented in Fig.~\ref{fig_cn4}. 

HD 10647 hosts a spatially resolved debris disk that has been imaged in scattered light with HST/ACS \cite{Lovell2022}, in thermal emission at $70$, $100$, and $160\ \mu\mathrm{m}$ with \textit{Herschel}/PACS \cite{Liseau2010}, and at submillimeter wavelengths with ALMA \cite{Lovell2021,Marino2026}. The most recent ALMA analysis from ref.\cite{Marino2026} derived a disk inclination of $77.8^{+0.1}_{-0.1}\,^{\circ}$ and a position angle of $57.3^{+0.1}_{-0.1}\,^{\circ}$. Therefore, we obtained a minimum mutual inclination between planet and debris disk $\Delta i_{p,d}$ as $43.6^{+8.2}_{-10.4}\,^{\circ}$. 

Ref. \cite{Marmier2013} reported a stellar rotation period of 10 d for HD 10647, derived from the chromospheric activity index.  Ref. \cite{Hurt2023} found a consistent stellar rotation period of 9.2 d. Using TESS light curves, we independently identify a stellar rotation period of $9.4 \pm 1$ d, in agreement with these previous measurements (see Fig.~\ref{fig_tl1}). Multiple spectroscopic measurements of the projected rotational velocity are also available in the literature (see   Table~\ref{oritation_table}). Adopting their weighted mean, we obtain $v\sin i = 5.4 \pm 0.2 ,\mathrm{km,s^{-1}}$. From these constraints, we derive a stellar inclination of $68.1^{+13.5}_{-8.8}\,^\circ$. Ref. \cite{Hurt2023} obtained a stellar inclination of $40.6^{+6.4}_{-5.6}\,^{\circ}$, but their estimate was based on a lower value of $v\sin i_* = 3.8 \pm 0.4 \mathrm{km\ s^{-1}}$, which deviates from most other measurements by about $2\sigma$. Our stellar inclination is consistent with the disk inclination, indicating the debris disk is aligned with the stellar spin axis. The minimum stellar obliquity for HD 10647 b is $37.0^{+11.6}_{-17.4}\,^{\circ}$.

  %Our results indicate that the disk is aligned with the  stellar equatorial plane, while planet HD 10647 b is misaligned relative to both the disk and the stellar spin axis.

\subsubsection*{6. $\kappa$ CrB}

\begin{figure}[!t]
\centering
\includegraphics[width=\textwidth]{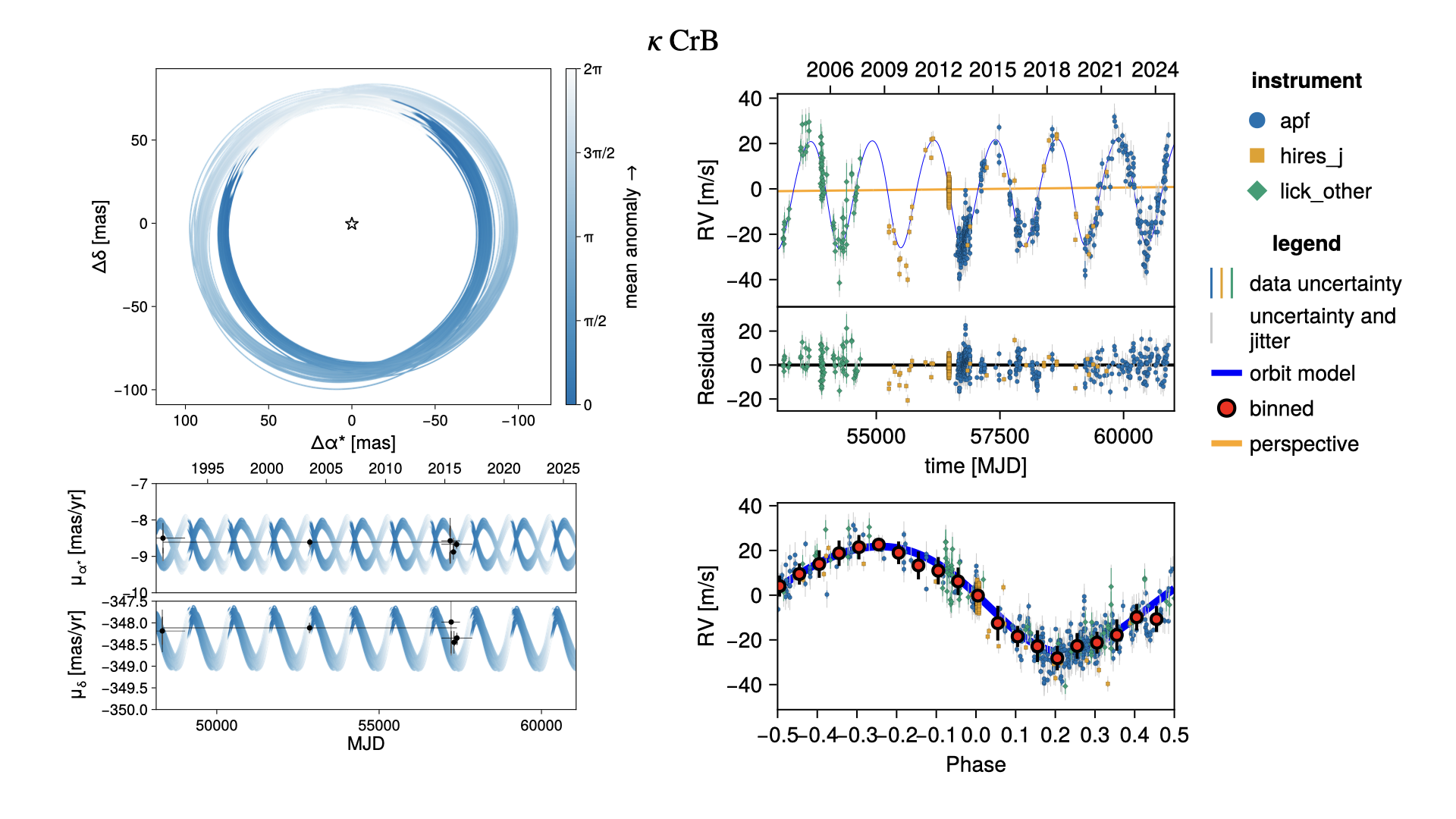}
\caption{\textbf{3-D Keplerian orbital model of $\kappa$ CrB (HD 142091) b.} Left: astrometric orbits of $\kappa$ CrB (HD 142091) b projected on the sky plane from our model (top) and observed and fitted absolute astrometry from Hipparcos and Gaia DR2 and DR3 in right ascension and declination (middle and bottom). The blue lines are randomly drawn from the posterior distribution. Right: observed and fitted RVs used in the fitting.}\label{fig_ft5}
\end{figure}

\begin{figure}
\centering
\includegraphics[width=\textwidth]{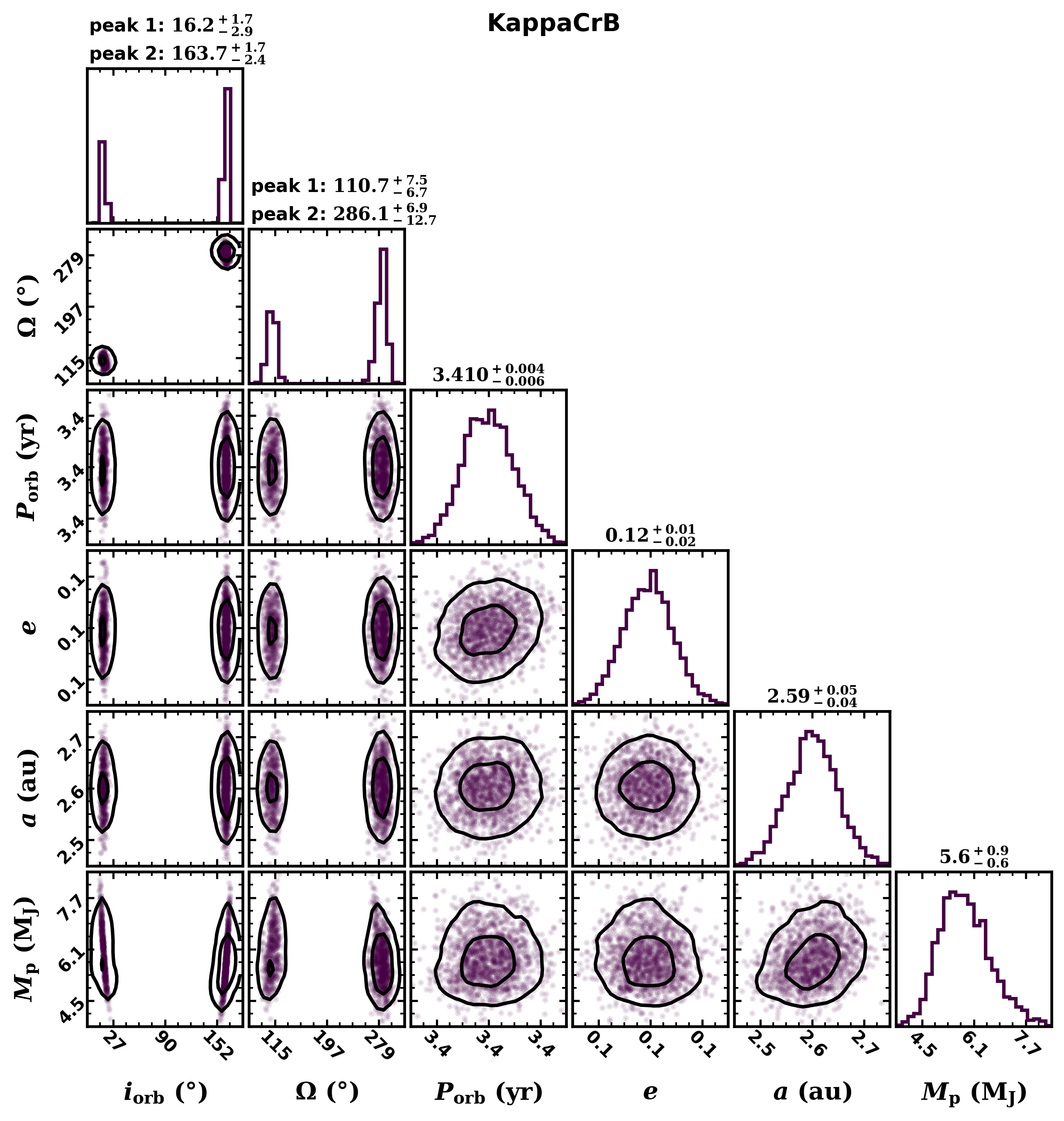}
\caption{\textbf{The joint posterior distributions from the orbital fit of $\kappa$ CrB b.} The plotted parameters are orbital inclination $i_{\rm orb}$, longitude of ascending node $\Omega$, orbital period $P_{\rm {orb}}$, eccentricity, and semi-major axis $a$ and planet mass $M_p$.}\label{fig_cn5}
\end{figure}

$\kappa$ CrB (HD 142091, HIP 77655) is a K0 sub-giant star ($M_*=1.45^{+0.03}_{-0.03}\ \rm{M_{\odot}}$, $R_*=4.68^{+0.84}_{-0.71}\ \rm{R_{\odot}}$, $\rm m_v\sim 4.8$ ) with a age of $3.28^{+0.55}_{-0.55}$ Gyr \cite{Sousa2015, TESSinput}. It hosts a giant planet at $\sim 2.6$ AU with a minimum mass of $\sim 1.8\ \rm M_{J}$ \cite{Johnson2008}. We used archival RVs with a baseline over 18 years (see   Table~\ref{tab:rv_tab1}). Fig.~\ref{fig_ft5} presents the data and best-fit model for the $\kappa$ CrB b.  We obtain a nearly phase-on orbit for the planet. The orbital inclination posterior shows bimodal distribution, which are symmetric around $90^{\circ}$, and sum of values at two peaks equals to $180^{\circ}$. These two peaks result from the prograde-retrograde degeneracy, indicating that the planet orbits in the same plane but can move in two opposite directions. The two possible orbital inclinations are $16.2^{+1.7}_{-2.9}\,^{\circ}$ and $163.7^{+1.7}_{-2.4}\,^{\circ}$. The joint posterior distributions for additional orbital parameters are presented in Fig.~\ref{fig_cn5}. 

$\kappa$ CrB hosts a debris disk that has been spatially resolved at $100$ and $160,\mu\mathrm{m}$ with \textit{Herschel}/PACS \cite{Bonsor2013} and at submillimeter wavelengths with ALMA \cite{Lovell2022}. Ref. \cite{Bonsor2013} reported a disk inclination of $59 \pm 10^{\circ}$ and a position angle of $142 \pm 10^{\circ}$ using \textit{Herschel} observations. Based on ALMA dataset, ref. \cite{Lovell2022} found consistent values of disk inclination $62.3 \pm 3^{\circ}.4$ and position angle $148 \pm 5^{\circ}$. Adopting the disk parameters with higher precision from ALMA observations \cite{Lovell2022}, we find a minimum disk-planet mutual inclination of $45.6^{+3.5}_{-5.5}\,^{\circ}$. For $\kappa$ CrB, we could not identify a reliable stellar rotation period measurement in the literature, nor did we detect one from the TESS light curve; therefore, the stellar inclination is unavailable. 

\subsubsection*{7. HD 216435}

\begin{figure}[!t]
\centering
\includegraphics[width=\textwidth]{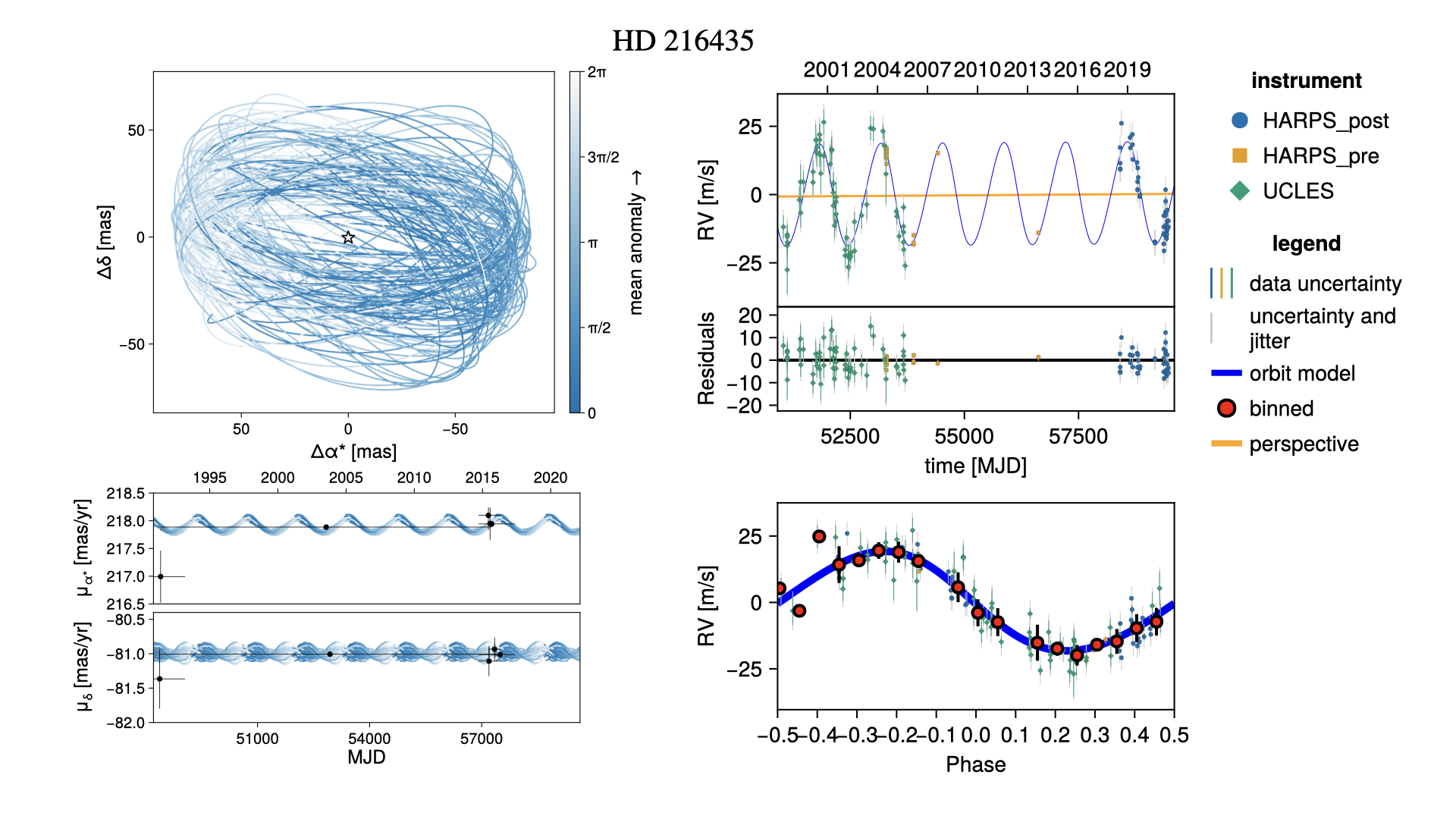}
\caption{\textbf{3-D Keplerian orbital model of HD 216435 b.} Left: astrometric orbits of HD 216435 b projected on the sky plane from our model (top) and observed and fitted absolute astrometry from Hipparcos and Gaia DR2 and DR3 in right ascension and declination (middle and bottom). The blue lines are randomly drawn from the posterior distribution. Right: observed and fitted RVs used in the fitting.}\label{fig_ft6}
\end{figure}

\begin{figure}
\centering
\includegraphics[width=\textwidth]{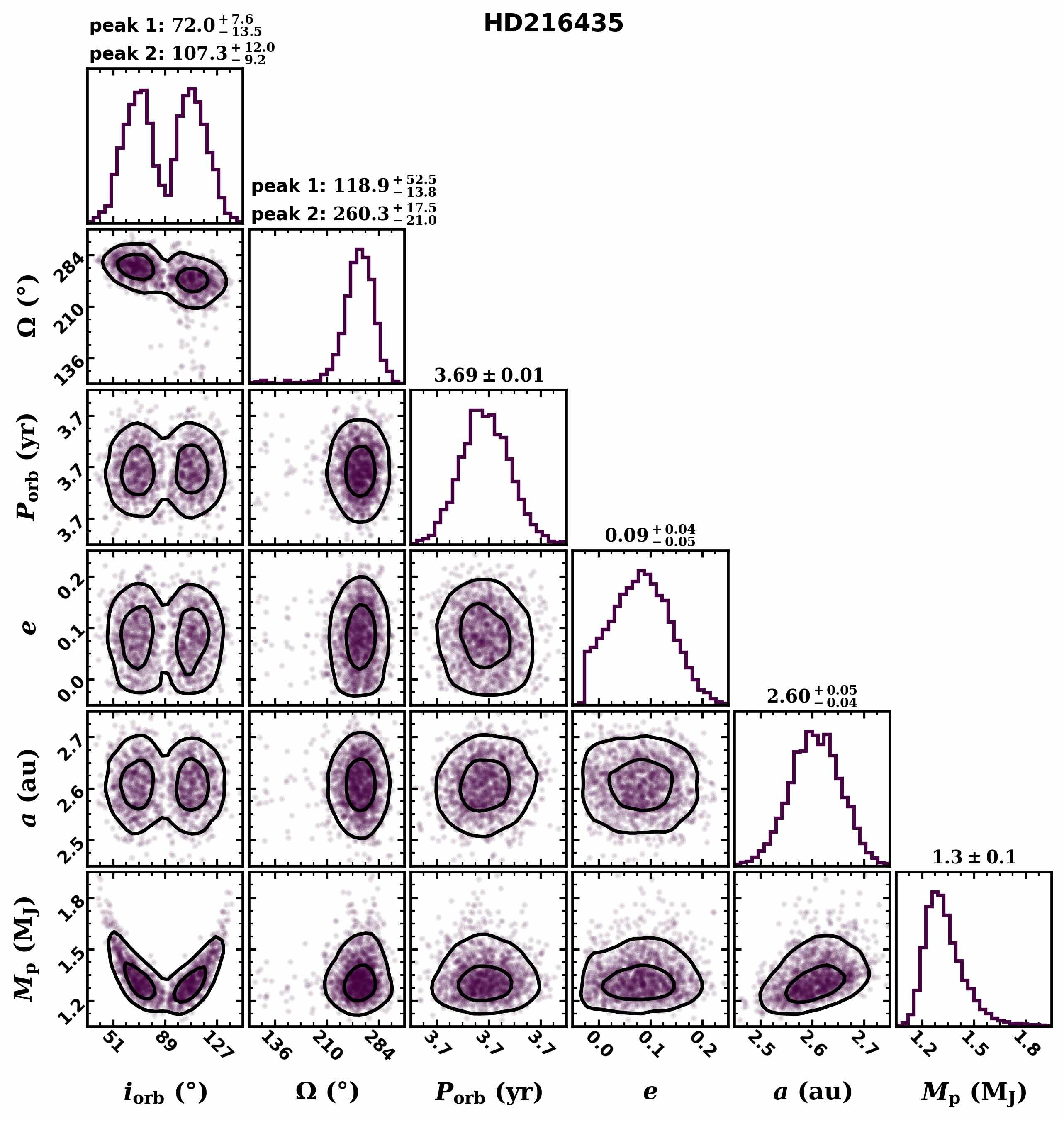}
\caption{\textbf{The joint posterior distributions from the orbital fit of HD 216435 b.} The plotted parameters are orbital inclination $i_{\rm orb}$, longitude of ascending node $\Omega$, orbital period $P_{\rm {orb}}$, eccentricity, and semi-major axis $a$ and planet mass $M_p$.}\label{fig_cn6}
\end{figure}

HD 216435 ($\tau^{1}$ Gruis, HIP 113044) is a sunlike star (G0V, $M_*=1.30^{+0.10}_{-0.10}\ \rm{M_{\odot}}$, $\rm m_v\sim 6.0$) at the distance of $\sim 33$ pc and an age of $3.5^{+1.5}_{-0.5}$ Gyr \cite{Ghezzi2010,TESSinput}. The star hosts a giant planet at $\sim2.56$ AU with a minimum mass of $\sim1.26\ \rm M_{J}$ \cite{Jones2003, Bulter2006}.  We used archival RVs with a baseline over 23 years in the orbital fitting (see   Table~\ref{tab:rv_tab1}). Fig.~\ref{fig_ft6} presents the data and best-fit model for the HD 216435 b. We obtained a bimodal distribution for the orbital inclination, which corresponds to  $72.0^{+7.6}_{-13.5}\,^{\circ}$ and $107.3^{+12.0}_{-9.2}\,^{\circ}$. These two peaks are symmetric around $90^{\circ}$, which results from the prograde-retrograde degeneracy. The joint posterior distributions for additional orbital parameters are presented in Fig.~\ref{fig_cn6}. 

The debris disk around HD 216435 was spatially resolved by at $70 \mu\mathrm{m}$ with \textit{Herschel}/PACS \cite{Marshall2021}, yielding a disk inclination of $68.5^{+3.6}_{-5.5}\,^{\circ}$ and position angle of $66.1^{+6.1}_{-5.1}\,^{\circ}$. Therefore, we obtained a minimum planet-disk mutual inclination of $\Delta i_{p,d}=3.1^{+8.4}_{-3.1}\,^{\circ}$. Because we could not identify a reliable stellar rotation period measurement in the literature, nor did we detect one from the TESS light curve for HD 216435, the stellar inclination is not available. 

\subsubsection*{8. $\beta$ Pic}

\begin{figure}[!t]
\centering
\includegraphics[width=\textwidth]{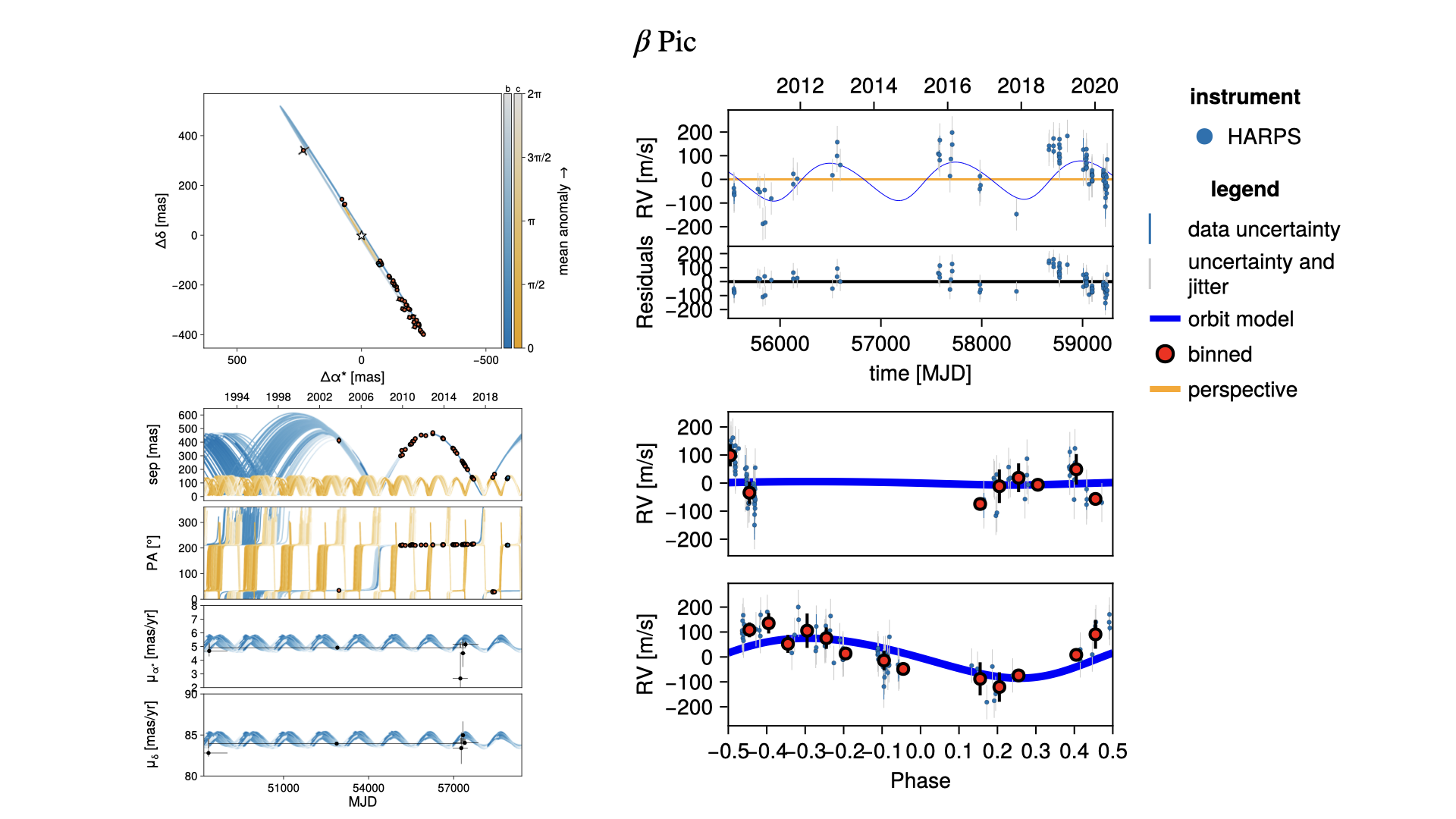}
\caption{\textbf{3-D Keplerian orbital model of HD 216435 b.} Left: astrometric orbits of $\beta$ Pic b$\&$c projected on the sky plane from our model with observed relative astrometries of the two planets (top) and observed and fitted absolute astrometry from Hipparcos and Gaia DR2 and DR3 in right ascension and declination (middle and bottom). The blue and yellow lines are randomly drawn from the posterior distribution. Right: observed and fitted RVs used in the fitting.}\label{fig_ft7}
\end{figure}

\begin{figure}
\centering
\includegraphics[width=\textwidth]{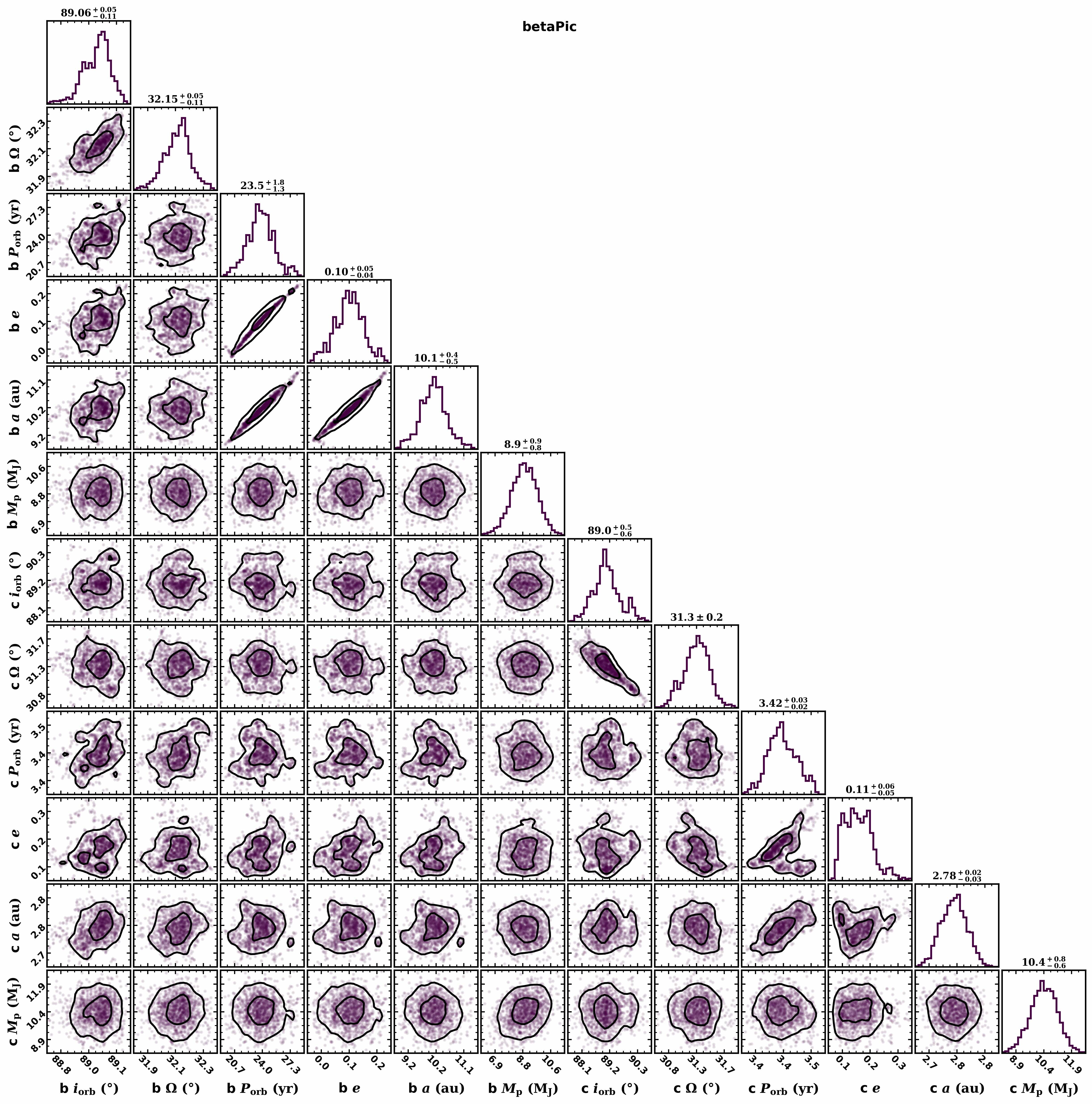}
\caption{\textbf{The joint posterior distributions from the orbital fit of $\beta$ Pic b$\&$c.} The plotted parameters are orbital inclination $i_{\rm orb}$, longitude of ascending node $\Omega$, orbital period $P_{\rm {orb}}$, eccentricity, and semi-major axis $a$ and planet mass $M_p$.}\label{fig_cn7}
\end{figure}

$\beta$ Pic (HD 39060, HIP 27321) is a bright A5V star ($M_*=1.76^{+0.18}_{-0.17}\ \rm{M_{\odot}}$, $m_v\sim3.9$) with a distance of $\sim 50$ pc \cite{Nielsen2014, TESSinput}. It is an young star with an age of $\sim16.5^{+4.5}_{-4.5}$ Myr. The star hosts two giant planets from direct imaging observations, at a semi-major axis of $\sim9$ AU and $\sim2.7$ AU \cite{Lagrange2009,Lagrange2019}. In the orbital fitting, we  used the archival RVs for the host star from HARPS spanning from Dec. 2010 and Jan. 2021 and the relative astrometry for the two planets from the direct imaging observations (see   Table~\ref{tab:rv_tab1} $\&$ \ref{tab_rel_astr}).   Fig.~\ref{fig_ft7} presents the data and best-fit model for the $\beta$ Pic b$\&$c. We obtained very similar orbital inclination and longitude of ascending nodes for the two planets, indicating they are nearly coplanar. The orbital inclination for $\beta$ Pic b and c are $89.1^{+0.1}_{-0.1}\,^{\circ}$ and $89.2^{+0.2}_{-0.4}\,^{\circ}$, respectively. The  longitude of ascending nodes for $\beta$ Pic b and c are $32.1^{+0.1}_{-0.1}\,^{\circ}$ and $31.3^{+0.1}_{-0.1}\,^{\circ}$. Our results are consistent with previous results from ref. \cite{MMB2015, Nowak2020}. The joint posterior distributions for additional orbital parameters are presented in Fig.~\ref{fig_cn7}. 

$\beta$ Pic hosts a nearly edge-on debris disk that was first resolved in the infrared by the Infrared Astronomy Satellite \cite{Smith1984}. From scattered-light images obtained with \textit{HST}/ACS, ref.\cite{Heap2000} identified a $\sim4^{\circ}$ warp in the disk, which was later attributed to dynamical perturbations from the planets. Using the polarimetric mode of the Gemini Planet Imager, ref. \cite{MMB2015} resolved the disk from $0\farcs3$ ($\sim6$ au) to $1\farcs7$ ($\sim33$ au), and measured a disk inclination of $85.3^{+0.2}_{-0.2}\,^{\circ}$ and a position angle of $30.4^{+0.1}_{-0.1}\,^{\circ}$. Our results show that the disk-planet mutual inclination  for $\beta$ Pic b$\&$c to be $3.9^{+0.9}_{-1.2}\,^{\circ}$ and $3.6^{+1.2}_{-0.9}\,^{\circ}$ respectively. They are consistent with the observed warp amplitude and support the interpretation that the planet perturbs the disk and drives the warp. Ref. \cite{Zwintz2019} measured the stellar inclination of $\beta$ Pic to be $87.8^{+1.6}_{-1.6}\,^{\circ}$ from the rotational splittings of non-radial stellar oscillation modes, indicating that both planets are consistent with alignment with the stellar spin axis.

\subsubsection*{9. $\epsilon$ Eri}

\begin{figure}[!t]
\centering
\includegraphics[width=\textwidth]{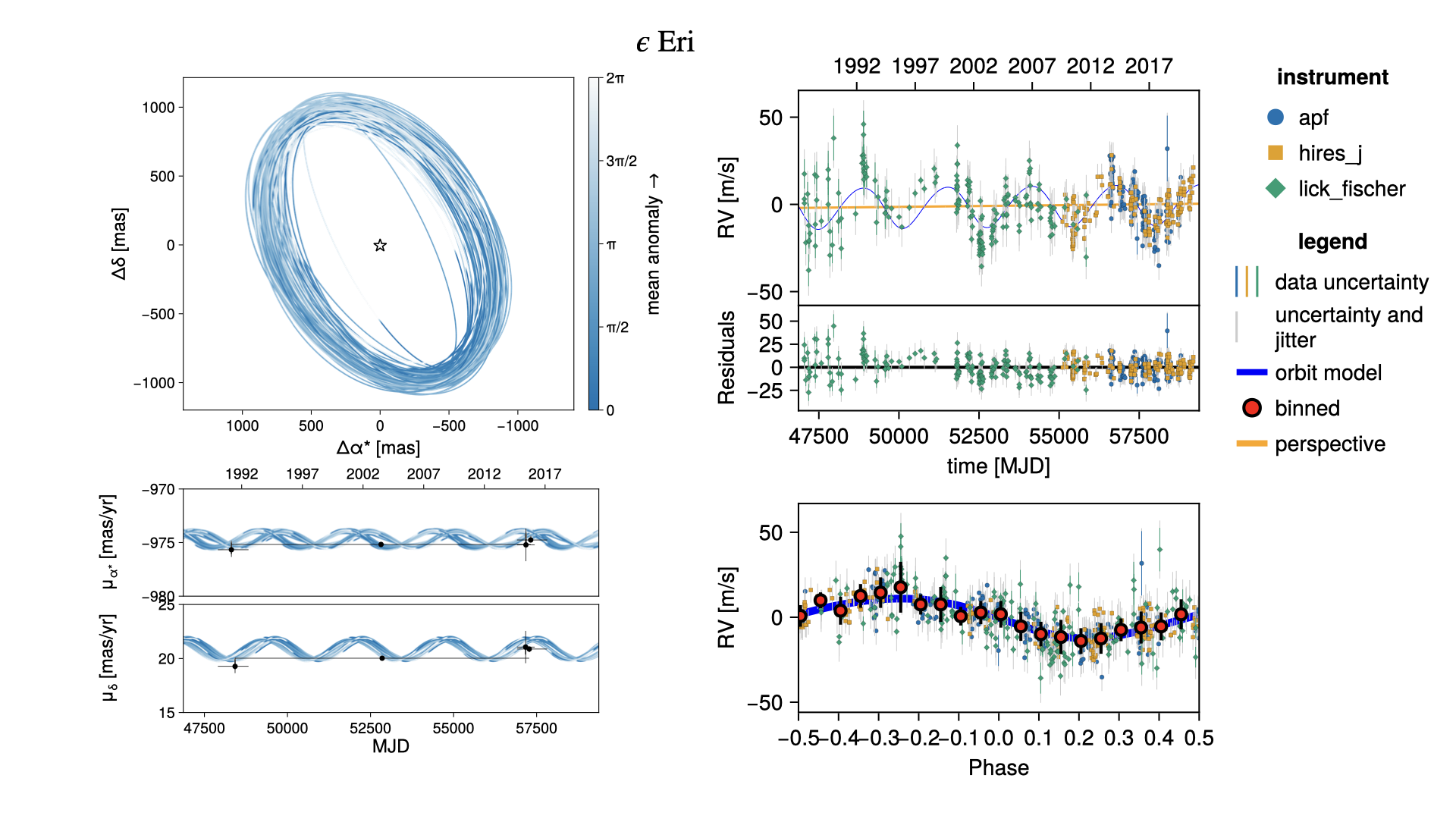}
\caption{\textbf{3-D Keplerian orbital model of $\epsilon$ Eri b.} Left: astrometric orbits of $\epsilon$ Eri b projected on the sky plane from our model (top) and observed and fitted absolute astrometry from Hipparcos and Gaia DR2 and DR3 in right ascension and declination (middle and bottom). The blue lines are randomly drawn from the posterior distribution. Right: observed and fitted RVs used in the fitting. }\label{fig_ft8}
\end{figure}

\begin{figure}
\centering
\includegraphics[width=0.8\textwidth]{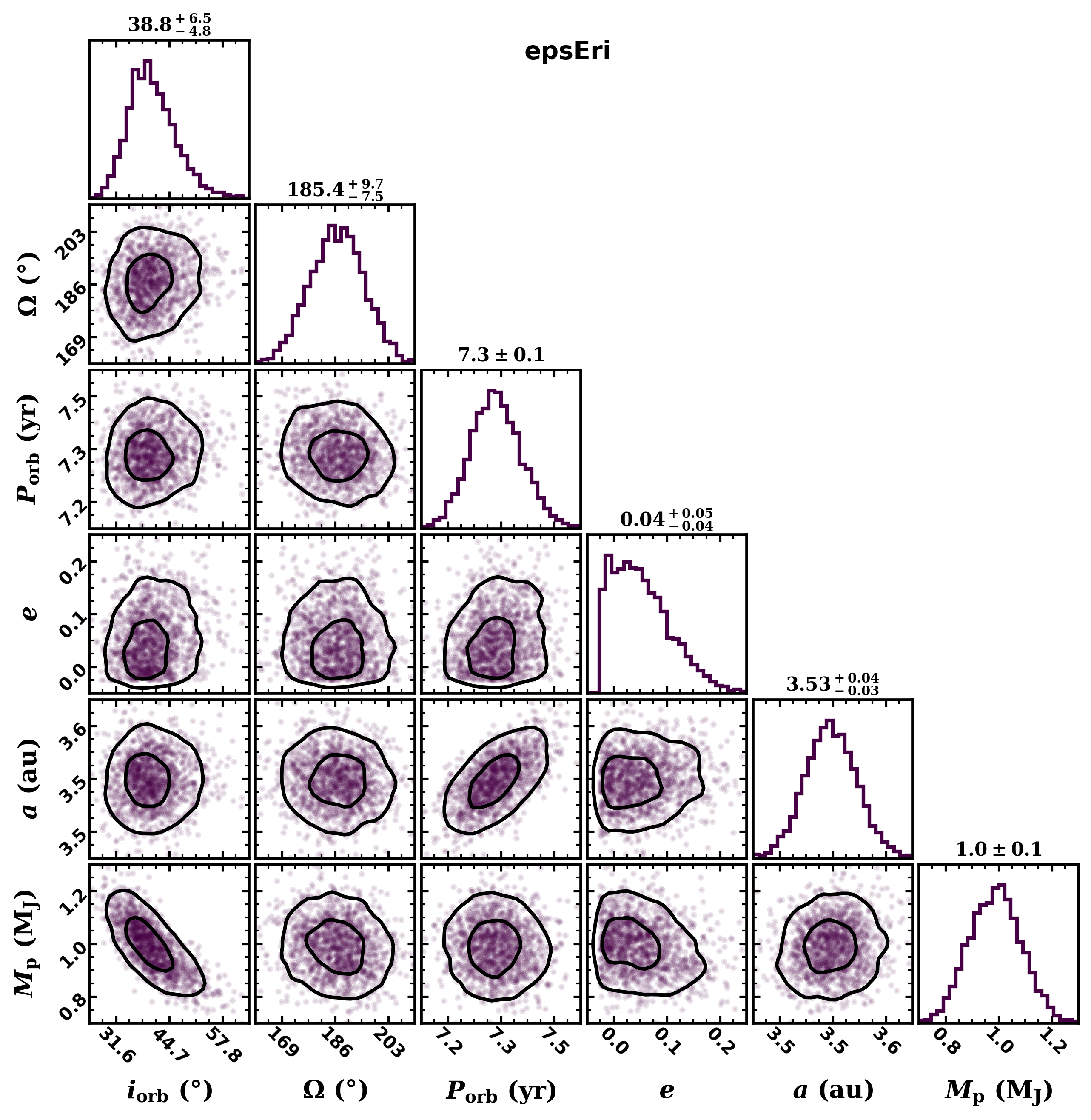}
\caption{\textbf{The joint posterior distributions from the orbital fit of $\epsilon$ Eri b.} The plotted parameters are orbital inclination $i_{\rm orb}$, longitude of ascending node $\Omega$, orbital period $P_{\rm {orb}}$, eccentricity, and semi-major axis $a$ and planet mass $M_p$.}\label{fig_cn8}
\end{figure}

$\epsilon$ Eri (HD 22049, HIP 16537) is a nearby K2V star ($M_*=0.82^{+0.02}_{-0.02}\ \rm{M_{\odot}}$, $m_v\sim3.7$) with a distance of $\sim3.9$ pc. The star has a age of $600^{+200}_{-200}$ Myr \cite{Llop-Sayson2021, TESSinput}. It hosts a giant planet at $\sim3.5$ AU with a minimus mass of $0.66\ \rm M_{J}$ \cite{Hatzes2000}. We used archival RVs with a baseline over 32 years (see   Table~\ref{tab:rv_tab1}).  Fig.~\ref{fig_ft8} presents the data and best-fit model for the $\epsilon$ Eri b. We obtained an orbital inclination of $38.9^{+6.5}_{-4.5}\,^{\circ}$ and longitude of ascending node of $185.5^{+9.9}_{-7.3}\,^{\circ}$. Our results agree with those from ref. \cite{Thompson2025}.  The joint posterior distributions for additional orbital parameters are presented in Fig.~\ref{fig_cn8}.

The debris disk around $\epsilon$ Eri was first resolved at submillimeter wavelengths with the James Clerk Maxwell Telescope \cite{Greaves1998}, and was later resolved in the far-infrared with \textit{Herschel}/PACS \cite{Greaves2014} and at submillimeter wavelengths with ALMA \cite{Booth2017}. We adopt the disk inclination and position angle from high-resolution ALMA observations of ref. \cite{Booth2017}, namely $33.7^{+0.5}_{-0.5}\,^{\circ}$ and $178.9^{+1.0}_{-1.0}\,^{\circ}$. The minimum disk-planet mutual inclination is $3.7^{+6.6}_{-3.7}\,^{\circ}$.

Multiple previous studies have found a stellar rotation period of approximately 12 d for $\epsilon$ Eri \cite{Baliunas1996,Cincunegui2007,Seli2022,Fuhrmeister2022}. From the TESS light curves, we independently identify a consistent stellar rotation period of $11.5 \pm 1$ d (see Fig.~\ref{fig_tl4}). We derive a stellar inclination of $41.4^{+6.2}_{-5.7}\,^{\circ}$, consistent with the result of ref. \cite{Hurt2023} within $1\sigma$. Based on this value, we derive $\Delta i_{p,*} = 2.5^{+6.5}_{-2.5}\,^{\circ}$.

\subsubsection*{10. HD 206893}

\begin{figure}[!t]
\centering
\includegraphics[width=\textwidth]{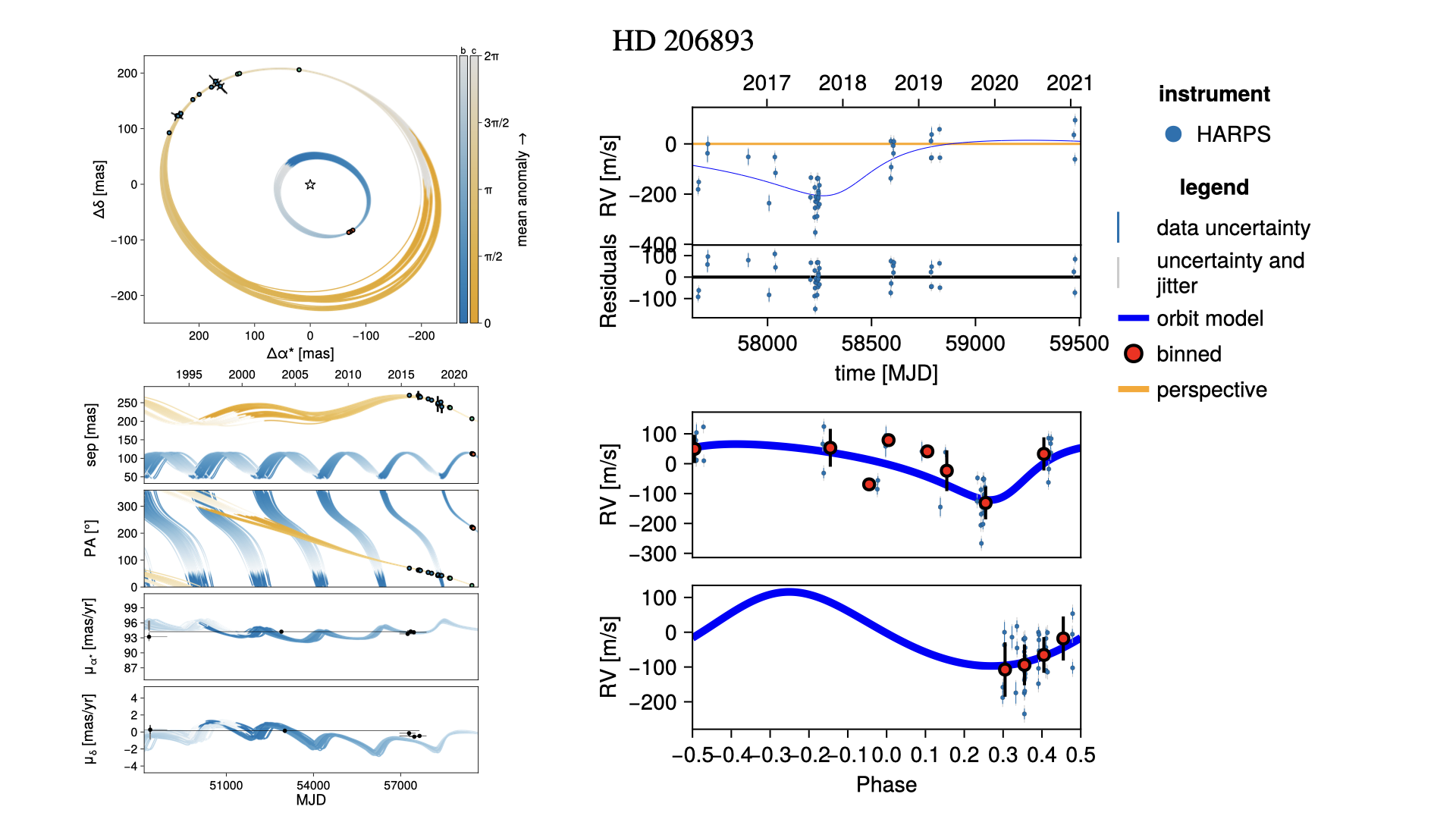}
\caption{\textbf{3-D Keplerian orbital model of HD 206893 b$\&$c.} Left: astrometric orbits of HD 206893 b$\&$c projected on the sky plane from our model with observed relative astrometries of the two planets (top) and observed and fitted absolute astrometry from Hipparcos and Gaia DR2 and DR3 in right ascension and declination (middle and bottom). The blue and yellow lines are randomly drawn from the posterior distribution. Right: observed and fitted RVs used in the fitting.}\label{fig_ft9}
\end{figure}

HD 206893 (HIP 107412) is a F5V star ($M_*=1.32^{+0.07}_{-0.05}\ \rm{M_{\odot}}$, $m_v\sim6.7$, $d\sim40.8$ pc) with an age of $155^{=15}_{-15}$ Myr \cite{Hinkley2023, TESSinput}. The star hosts a giant planet (HD 206893 c) at $\sim3.5$ AU \cite{Hinkley2023} and a brown dwarf companion (HD 206893 B) at $\sim 9.6$ AU \cite{Milli2017, Delorme2017,WardDuong2021}.  The giant planet HD 206893 c was discovered using RV observations, while the brown dwarf companion (HD 206893 B)  was detected from direct imaging observations. We used both archival RVs of the host stars and relative astrometry of the brown dwarf companion in the orbital fitting (see   Table~\ref{tab:rv_tab1} $\&$ \ref{tab_rel_astr}). We consider a two-companion model in the fitting. Fig.~\ref{fig_ft9} presents the data and best-fit model for the system. We obtained the orbital inclination and longitude of ascending node of the giant planet HD 206893 c as $149.9^{+2.2}_{-2.7}\,^{\circ}$ and $79.0^{+6.3}_{-2.4}\,^{\circ}$. The orbital inclination and longitude of ascending node of the brown dwarf HD 206893 B are $146.0^{+2.2}_{-1.7}\,^{\circ}$ and $73.1^{+1.7}_{-4.6}\,^{\circ}$. Our results shows the inner planets are nearly coplanar with the outer brown dwarf companion, with a mutual inclination of $i_{B,c}= 5.0^{+3.4}_{-2.9}\,^{\circ}$ between the two companions. The minimum mutual inclination between debris disk and HD 206893 c$\&$B are $5.8^{+3.4}_{-3.7}\,^{\circ}$ and $9.8^{+3.6}_{-4.2}\,^{\circ}$, respectively.

HD 206893 hosts a debris disk that was spatially resolved at far-infrared wavelengths with \textit{Herschel}/PACS \cite{Milli2017} and at submillimeter wavelengths with ALMA \cite{Marino2020}. We adopt the disk inclination of $40 \pm 3^{\circ}$ and position angle of $61 \pm 4^{\circ}$ from the ALMA observations from ref. \cite{Marino2020}. Owing to the disk-orientation degeneracy, which arises because the direction of motion of the disk material is not measured, the degenerate disk inclination solution is $140\pm3^{\circ}$. Considering the prograde configuration between planet and disk, we find the minimum mutual inclination between disk and 206893 c$\&$B are $5.8^{+3.4}_{-3.7}\,^{\circ}$ and $9.8^{+3.6}_{-4.2}\,^{\circ}$, respectively. We adopt the stellar rotation period of $0.996 \pm 0.003$ d from \cite{Delorme2017} and derive a stellar inclination of $151.6^{+1.7}_{-3.2}\,^{\circ}$. The stellar obliquities of HD 206893 c and the brown dwarf HD 206893 B are $5.2^{+2.8}_{-3.3}\,^{\circ}$ and $0.9^{+2.9}_{-0.9}\,^{\circ}$.

\subsubsection*{11. HD 95086}

HD 95086 (HIP 53524) is a young A2V star ($M_*=1.77^{+0.36}_{-0.21}\ \rm{M_{\odot}}$, $m_v\sim7.4$ ) with an age of $17^{+4}_{-4}$ Myr \cite{Rameau2013, TESSinput}. The star hosts a young giant planet at $\sim 55$ AU from direct imaging observations \cite{Rameau2013,Meshkat2013}. In our fitting, we used the archival relative astrometry spanning from Jan. 2012 to Mar. 2023, with a baseline over 11 years (see   Table~\ref{tab_rel_astr}). Fig.~\ref{fig_ft10} presents the data and best-fit model for HD 95086 b. We obtained the orbital inclination of HD 95086 b as $147.2^{+11.8}_{-10.9}\,^{\circ}$. We have also obtained two peaks in longitude of ascending node to be $76.8^{+55.1}_{-41.5}\,^{\circ}$ and $290.2^{+29.7}_{-55.0}\,^{\circ}$. 

The debris disk around HD 95086 was resolved at far-infrared wavelengths with \textit{Herschel}/PACS \cite{Moor2015} and at submillimeter wavelengths with ALMA \cite{su2017}. We adopted the disk inclination of $36^{+3}_{-2}\,^{\circ}$ and longitude of ascending node of $98^{+3}_{-3}\,^{\circ}$ deg from ALMA observations \cite{su2017}. Due to the disk-orientation degeneracy from two possible moving direction of the disk material, an degenerate disk inclination is $144^{+3}_{-2}\,^{\circ}$. For the pro-grade configuration, we find $\Delta i_{p,d} = 2.7^{+9.7}_{-2.5}\,^{\circ}$. Because we could not identify a reliable stellar rotation period measurement in the literature, nor did we detect one from the TESS light curve for HD 95086, the stellar inclination is not available.

\subsubsection*{12. HR 2562}

HR 2562 (HD 50571, HIP 32775) is a F5V star ($M_*=1.30^{+0.05}_{-0.05}\ \rm{M_{\odot}}$, $m_v\sim6.1$ ) with an age of $600^{+300}_{-300}$ Myr \cite{Konopacky2016, TESSinput}. It hosts a low-mass brown dwarf companion HR 2562 B at a separation of $\sim 20 $ AU from direct imaging observations \cite{Konopacky2016}.  We used the archival relative astrometry spanning from Jan. 2016 to Mar. 2024 in the orbital fitting (see   Table~\ref{tab_rel_astr}). Our results show that HR 2562 B orbits the host star in an edge-on orbit with an orbital inclination of $86.4^{+0.1}_{-0.2}\,^{\circ}$ and longitude of ascending node of $120.0^{+0.1}_{-0.1}\,^{\circ}$.

HR 2562 hosts a nearly edge-on debris disk that has been spatially resolved at $70$, $100$, and $160 \ \mu\mathrm{m}$ with \textit{Herschel}/PACS \cite{Moor2015}, and at submillimeter wavelengths with ALMA \cite{Zhang2023}. We adopt the disk geometry from the ALMA observations, which yield an inclination of $84.7^{+5.3}_{-5.3}\,^{\circ}$ and a position angle of $32.7^{+0.7}_{-0.8}\,^{\circ}$. The minimum mutual inclination between HR 2652 B and disk is $1.1^{+4.2}_{-1.1}\,^{\circ}$. 

From the TESS light curves, we identify a stellar rotation period of $2 \pm 0.5$ d (see Fig.~\ref{fig_tl5}), obtaining a stellar inclination of $86.8^{+3.0}_{-9.7}\,^{\circ}$. Therefore, the light-of-sight stellar obliquity of HR 2562 B is $1.9^{+7.1}_{-1.9}\,^{\circ}$.

\subsubsection*{13. AU Mic}

AU Mic (HD 197481, HIP 102409) is a nearby young M1V star ($M_*=0.64^{+0.04}_{-0.07}\ \rm{M_{\odot}}$, $m_v\sim8.8$) with an age of $19^{+10}_{-6}$ Myr \cite{Mallorquin2024, TESSinput}. It hosts three small transiting planets at orbital periods of 8.64 d, 12.74 d and 18.86 d, respectively \cite{Plavchan2020,Martioli2021,Wittrock2023}. We adopted their orbital inclinations constrained from transiting observations as $i_b=89.6^{+0.4}_{-0.4}\,^{\circ}$, $i_c=89.3^{+1.1}_{-1.1}\,^{\circ}$, $i_d=89.2^{+0.2}_{-0.2}\,^{\circ}$ from ref. \cite{Wittrock2023}. Their longitude of ascending node is not available from transiting observations. 

AU Mic hosts a nearly edge-on debris disk that has been spatially resolved in scattered light with coronagraphic imaging at the UH 2.2 m telescope \cite{Kalas2004}, with HST/ACS \cite{Krist2005}, and with SPHERE \cite{Engler2025}, as well as in thermal emission with ALMA \cite{MacGregor2013}. We adopt the disk inclination and position angle from the ALMA observations \cite{MacGregor2013}, which give $89.5^{+0.5}_{-0.5}\,^{\circ}$ and $128.4^{+0.1}_{-0.1}\,^{\circ}$, respectively. The minimum mutual inclination between disk and AU Mic b, c, d are $0.2^{+0.5}_{-0.2}\,^{\circ}$, $0.3^{+1.0}_{-0.3}\,^{\circ}$ and $0.1^{+0.5}_{-0.1}\,^{\circ}$. 

Multiple previous studies have reported a stellar rotation period of approximately $4.8$ d for AU Mic, consistent with our measurement of $4.86 \pm 0.5$ d using TESS light curves (see Fig.~\ref{fig_tl3}). From this, we derive a stellar inclination of $88.1^{+1.6}_{-7.7}\,^{\circ}$. the light-of-sight stellar obliquity of the three planets are $1.4^{+7.6}_{-1.4}\,^{\circ}$ ,  $1.5^{+7.3}_{-1.4}\,^{\circ}$ and $1.1^{+7.6}_{-1.1}\,^{\circ}$, respectively.

\subsubsection*{14. HD 114082}

HD 114082 (HIP 64184) is young F3V star ($M_*=1.47^{+0.07}_{-0.07}\ \rm{M_{\odot}}$) with an age of $15^{+6}_{-6}$ Myr \cite{Zakhozhay2022, TESSinput}. It hosts two transiting giant planets at $\sim0.5$ AU and $\sim 1$ AU \cite{Zakhozhay2022, Engler2025}. We adopted their orbital inclinations constrained from transiting observations as $i_b=89.8^{+0.3}_{-0.3}\,^{\circ}$, $i_c=89.7^{+0.2}_{-0.2}\,^{\circ}$ \cite{Engler2025}. Their longitude of ascending node is not available from transiting observations.

The debris disk around HD 114082 has been spatially resolved in scattered light with SPHERE \cite{Wahhaj2016, Engler2025}, revealing a nearly edge-on disk with an inclination of $83.2^{+1.1}_{-1.1}\,^{\circ}$ and a position angle of $105.7^{+1.4}_{-1.4}\,^{\circ}$. Relative to the planetary orbits, this implies a line-of-sight mutual inclination between planets and the debris disk of $\Delta i_{p,d}=6.6^{+1.0}_{-1.2}\,^{\circ}$. Because we could not identify a reliable stellar rotation period measurement in the literature, nor did we detect one from the TESS light curve for HD 114082, the stellar inclination is not available. 

\subsubsection*{15. HR 8799}

HR 8799 (HD 218396, HIP 114189) is a young A5V star with a mass of $M_*=1.50^{+0.05}_{-0.05}\,\rm{M_{\odot}}$ and an age of $60^{+100}_{-30}$ Myr \cite{Marois2008}. It hosts four giant planets at projected separations of approximately 68, 38, 24, and 16 AU, discovered through direct imaging observations \cite{Marois2008}.

Multiple studies have examined the dynamical interactions among the four planets, and most conclude that their orbits are nearly coplanar and likely locked in a 1:2:4:8 resonant chain \cite{FabryckyMurray2010,Konopacky2016,wang2018,GRAVITY2019,Godziewski2020,Zurlo2022}. At the same time, previous attempts to constrain the planetary orbits using relative astrometry have shown that the current observational time baseline is still too short to uniquely determine long-term stable orbital solutions for HR 8799 without using additional dynamical constraints \cite{wang2018,Godziewski2020,Zurlo2022}. A full orbital analysis of all four planets subject to dynamical stability constraints would require extensive computational resources and is beyond the scope of this work. We therefore adopt the orbital solution from ref. \cite{Zurlo2022}, which is based on 23 years of relative astrometry for the four planets. 

HR 8799 hosts a nearly face-on debris disk that has been resolved at far-infrared wavelengths with \textit{Herschel}/PACS \cite{Matthews2014} and at submillimeter wavelengths with ALMA \cite{Booth2016,Faramaz2021}. We adopt the disk geometry from the ALMA observations of ref.\cite{Faramaz2021}, which give an inclination of $30.8^{+2.8}_{-3.2}\,^{\circ}$ and a position angle of $52.4^{+5.4}_{-5.7}\,^{\circ}$. We therefore obtained a minimum mutual inclination between planet and disk as $0.9^{+2.7}_{-0.9}\,^{\circ}$. Adopting the stellar rotation period of 0.7 d from ref.\cite{Sepulveda2023}, we derive a stellar inclination of $26.8^{+2.0}_{-1.9}\,^{\circ}$, and the line-of-sight stellar obliquity to be $3.4^{+2.4}_{-3.4}\,^{\circ}$.

\subsubsection*{16. Solar System}

For the Solar System planets, we adopted the orbital inclinations and longitudes of ascending node from the JPL Planetary and Lunar Ephemerides DE440/DE441 \cite{park2021}, which were constructed by fitting numerically integrated orbital solutions to ground-based and space-based observations. All quoted orbital elements are referenced to the ecliptic plane. For the Kuiper Belt, we adopted the inclination and longitude of ascending node of the mean plane of the main belt from ref. \cite{Brown2004}, measured relative to the ecliptic plane, as $1.86^{+0.37}_{-0.37}\,^{\circ}$ and $86.7^{+0.20}_{-0.20}\,^{\circ}$, respectively. To get the stellar obliquities of the Solar System planets, we compare the Sun’s J2000 pole orientation from ref.\cite{Archinal2018} with each planet’s orbital plane. 

% If your supplement is very short you might need to uncomment the following line to avoid
% layout problems with the figures and tables.
\newpage

%%%%%%%%%%%%%%%% SUPPLEMENTARY FIGURES %%%%%%%%%%%%%%%


\clearpage

\clearpage

\clearpage

\clearpage

\clearpage

\clearpage

\clearpage

\clearpage

\clearpage
\begin{figure}
\centering
\includegraphics[width=\textwidth]{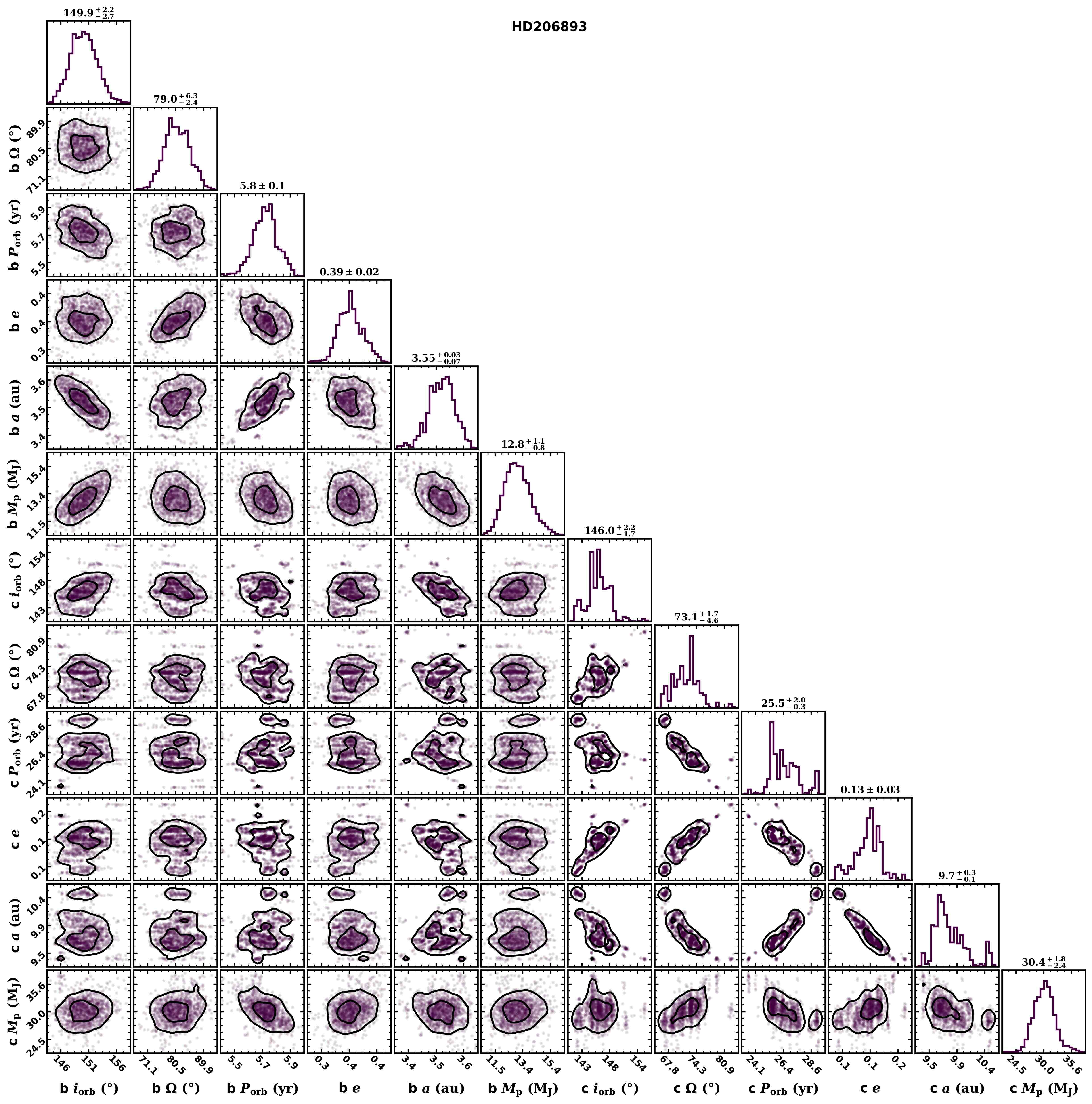}
\caption{\textbf{The joint posterior distributions from the orbital fit of HD 206893 b$\&$c.} The plotted parameters are orbital inclination $i_{\rm orb}$, longitude of ascending node $\Omega$, orbital period $P_{\rm {orb}}$, eccentricity, and semi-major axis $a$ and planet mass $M_p$.}\label{fig_cn9}
\end{figure}

\clearpage
\begin{figure}[!t]
\centering
\includegraphics[width=0.7\textwidth]{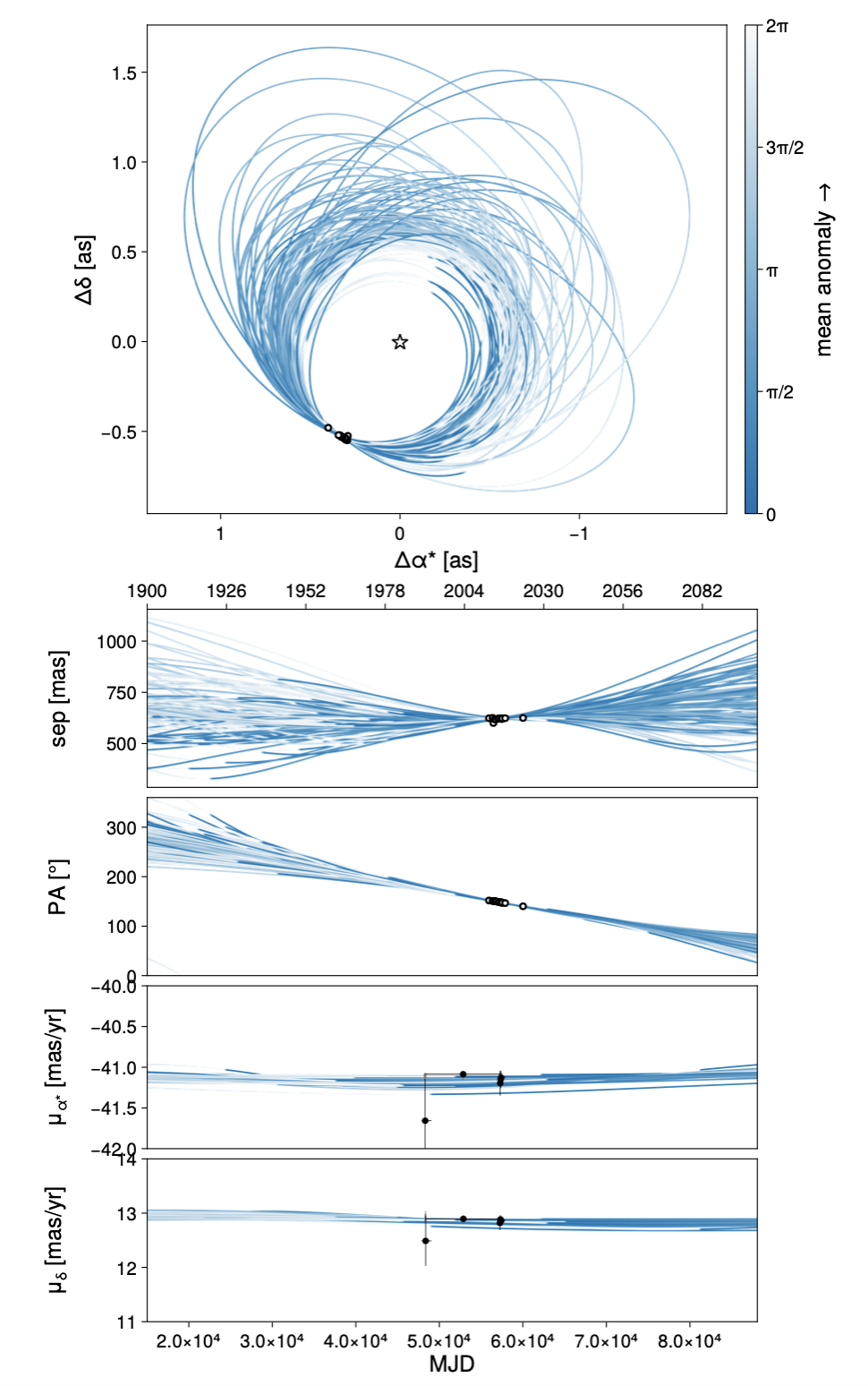}
\caption{\textbf{3-D Keplerian orbital model of HD 95086 b.} Left: astrometric orbits of HD 95086 b projected on the sky plane from our model (top) and observed and fitted absolute astrometry from Hipparcos and Gaia DR2 and DR3 in right ascension and declination (middle and bottom). The blue lines are randomly drawn from the posterior distribution. Right: observed and fitted RVs used in the fitting. }\label{fig_ft10}
\end{figure}
\begin{figure}
\centering
\includegraphics[width=\textwidth]{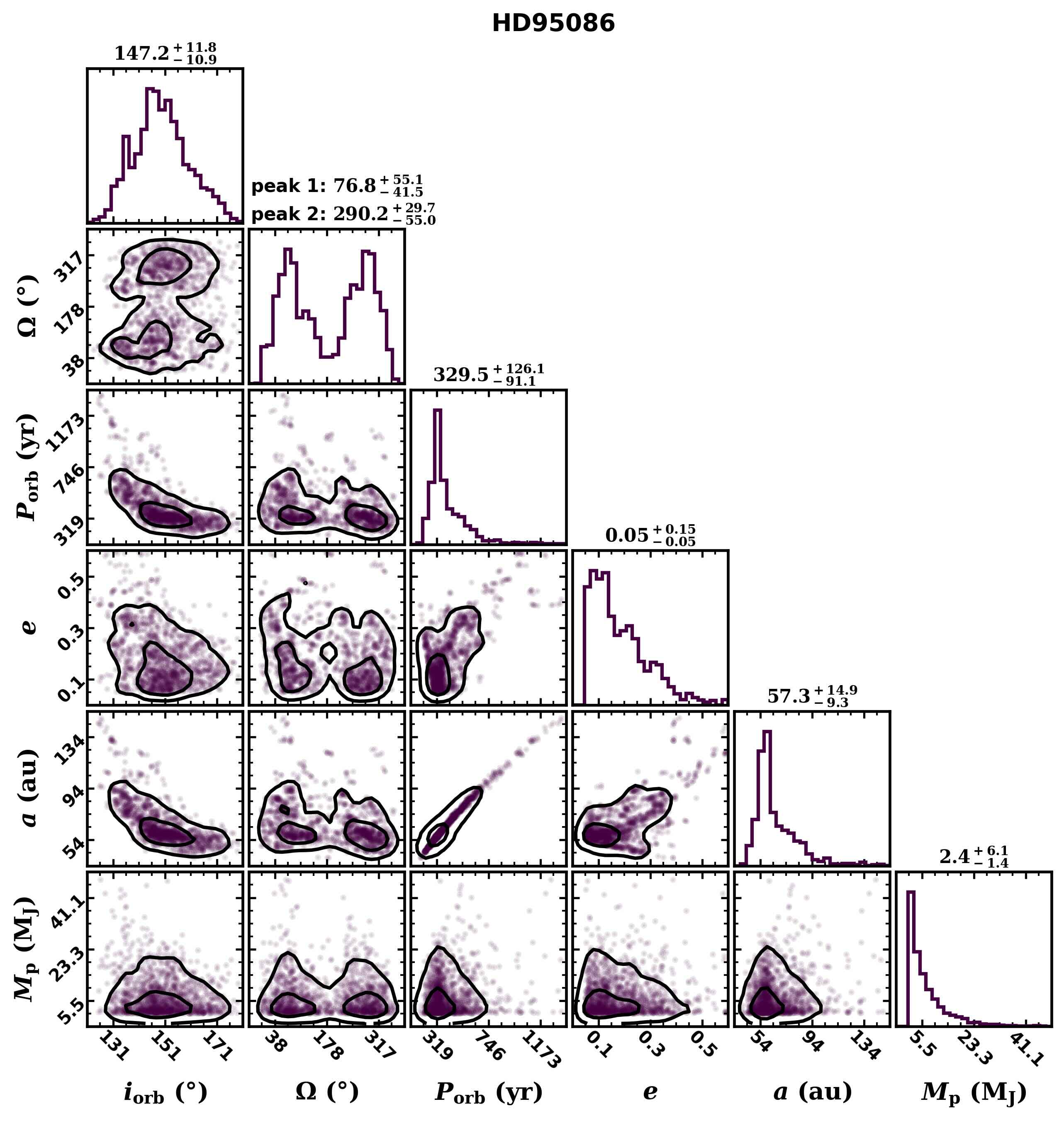}
\caption{\textbf{The joint posterior distributions from the orbital fit of HD 95086 b.} The plotted parameters are orbital inclination $i_{\rm orb}$, longitude of ascending node $\Omega$, orbital period $P_{\rm {orb}}$, eccentricity, and semi-major axis $a$ and planet mass $M_p$.}\label{fig_cn9}
\end{figure}

\clearpage
\begin{figure}[!t]
\centering
\includegraphics[width=0.66\textwidth]{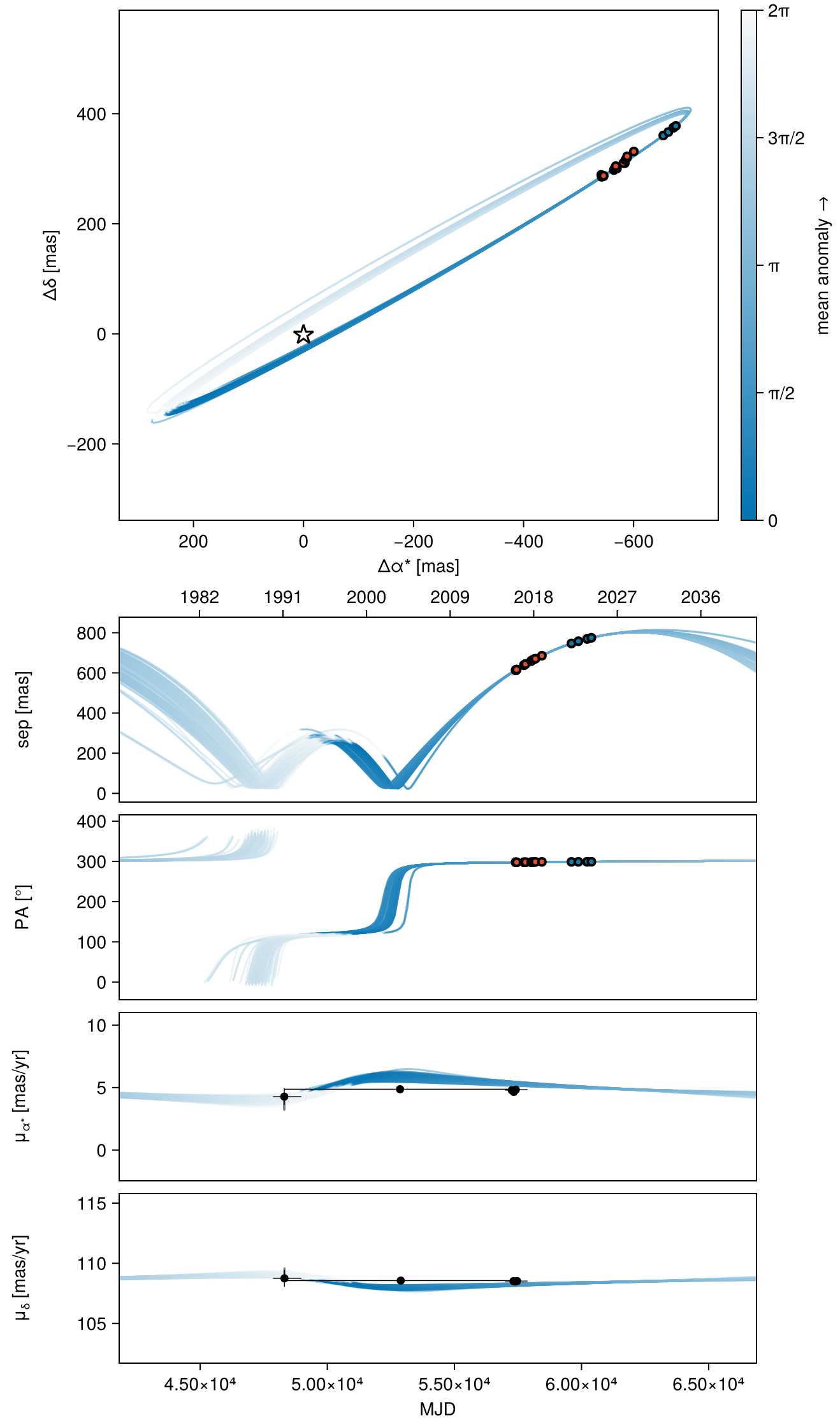}
\caption{\textbf{3-D Keplerian orbital model of HR 2562 B.} Left: astrometric orbits of HR 2562 B projected on the sky plane from our model with observed relative astrometries of the two planets (top) and observed and fitted absolute astrometry from Hipparcos and Gaia DR2 and DR3 in right ascension and declination (middle and bottom). The blue lines are randomly drawn from the posterior distribution.}\label{fig_ft18}
\end{figure}
\begin{figure}
\centering
\includegraphics[width=\textwidth]{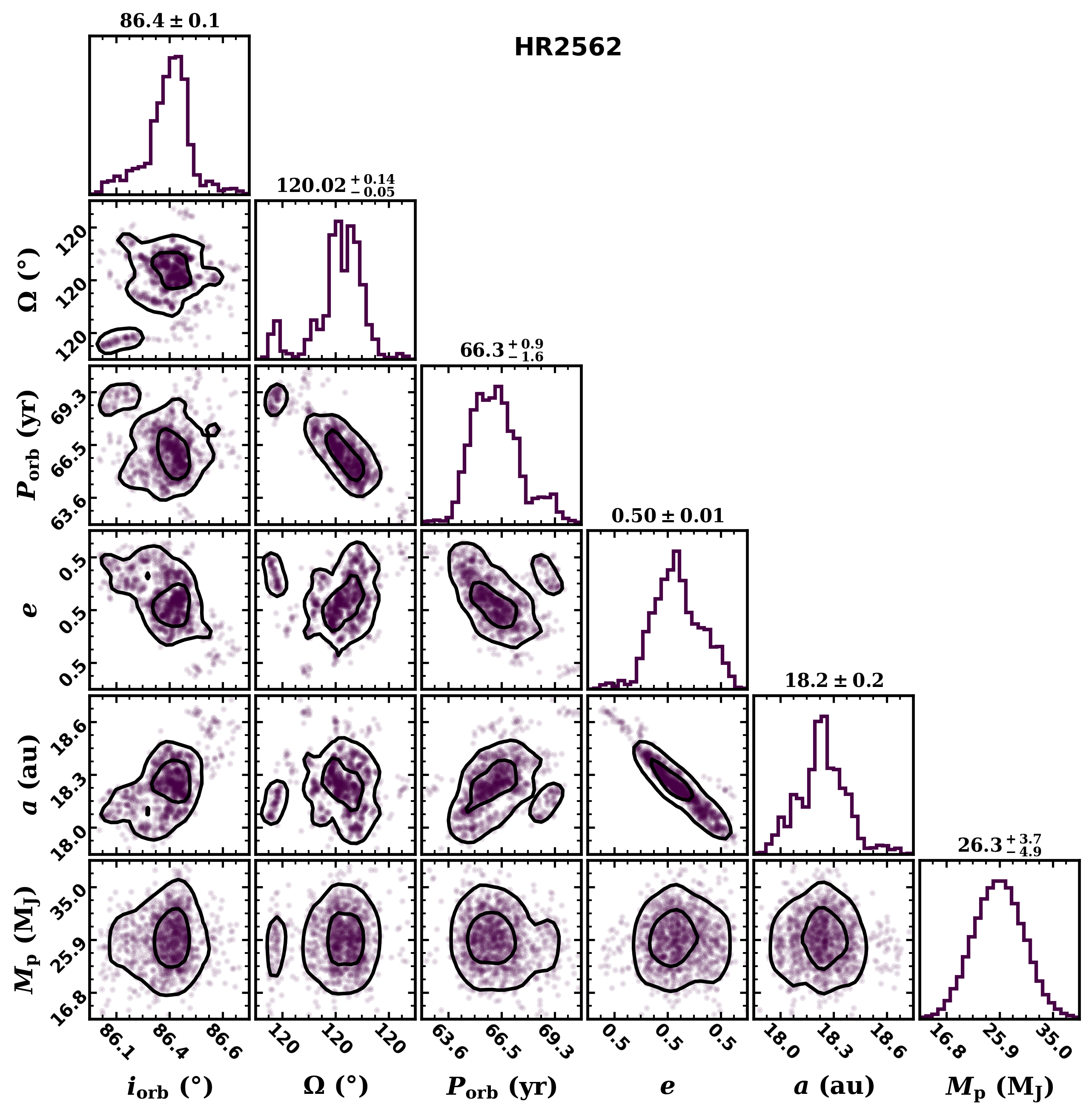}
\caption{\textbf{The joint posterior distributions from the orbital fit of HR 2562 B.} The plotted parameters are orbital inclination $i_{\rm orb}$, longitude of ascending node $\Omega$, orbital period $P_{\rm {orb}}$, eccentricity, and semi-major axis $a$ and planet mass $M_p$. }\label{fig_cn19}
\end{figure}
\clearpage
%\subsubsection*{Stellar rotation period from TESS light curves}

\begin{figure}
\centering
\includegraphics[width=\textwidth]{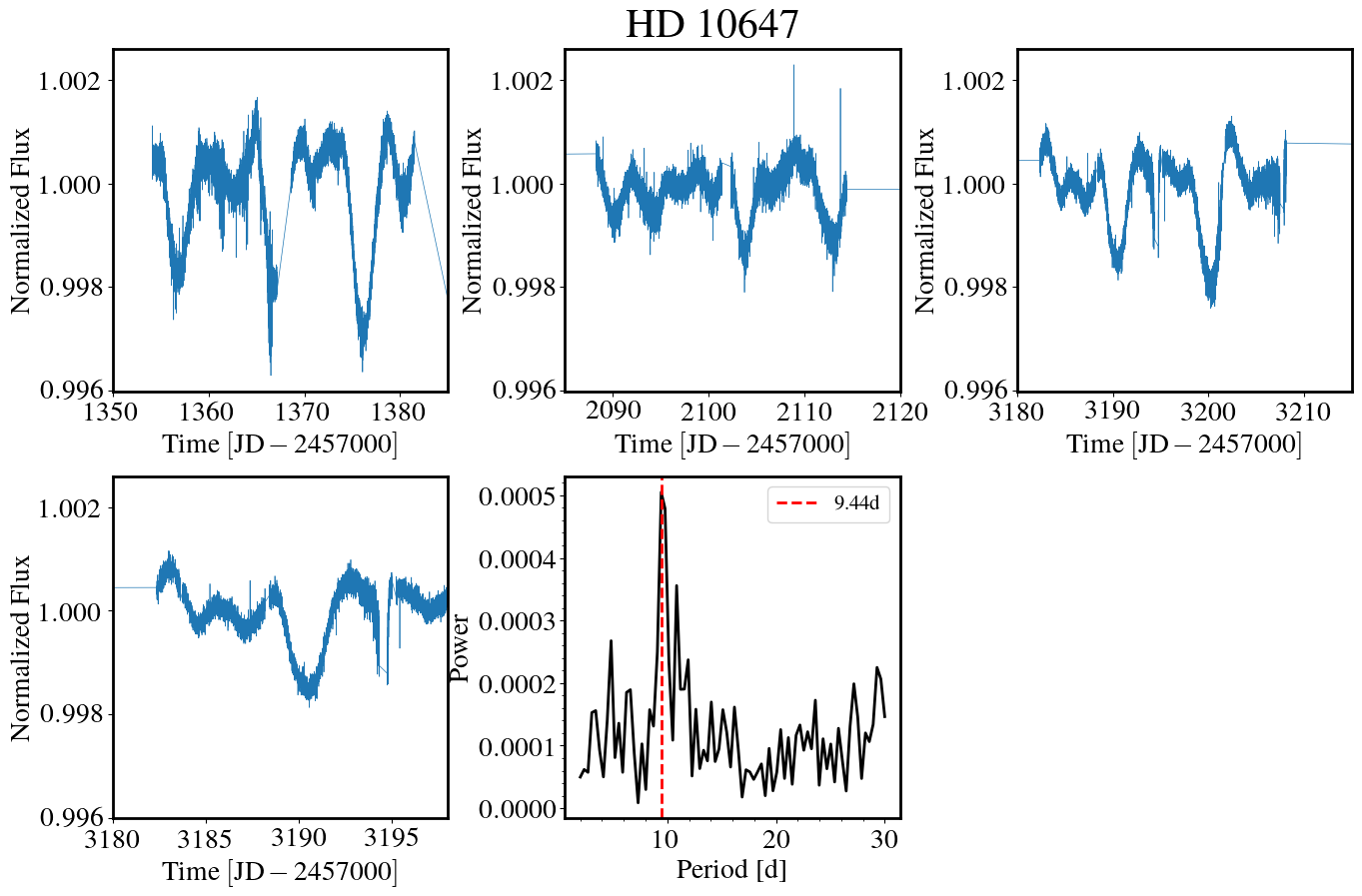}
\caption{\textbf{Measurements of stellar rotational period of HD 10647.} The first four panels show the TESS light curves, and the final panel shows the corresponding periodogram. The red dashed line marks the identified rotation period. }\label{fig_tl1}
\end{figure}
\begin{figure}
\centering
\includegraphics[width=\textwidth]{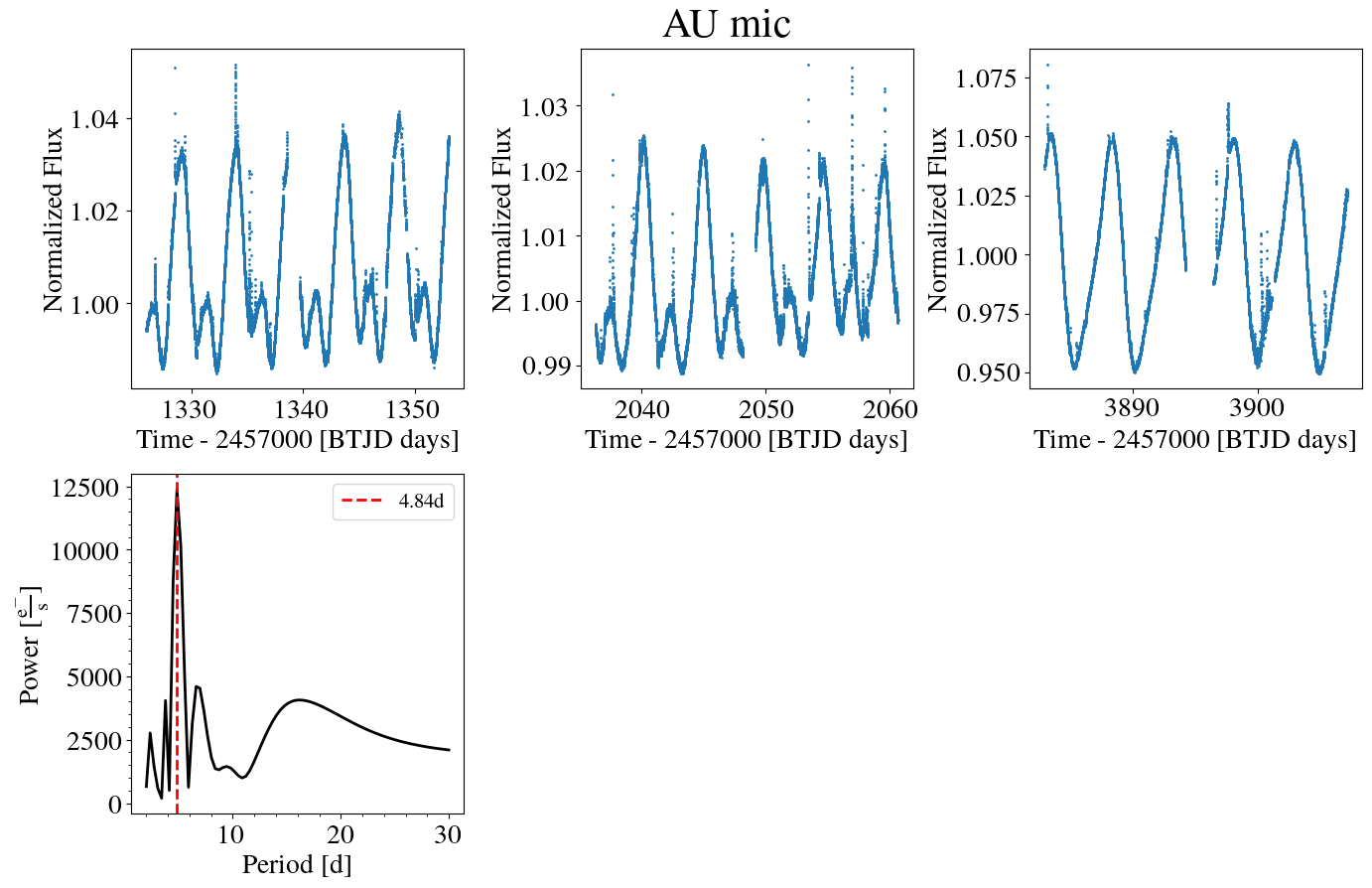}
\caption{\textbf{Measurements of stellar rotational period of AU Mic.} The three four panels show the TESS light curves, and the final panel shows the corresponding periodogram. The red dashed line marks the identified rotation period. }\label{fig_tl3}
\end{figure}
\clearpage
\begin{figure}[!b]
\centering
\includegraphics[width=\textwidth]{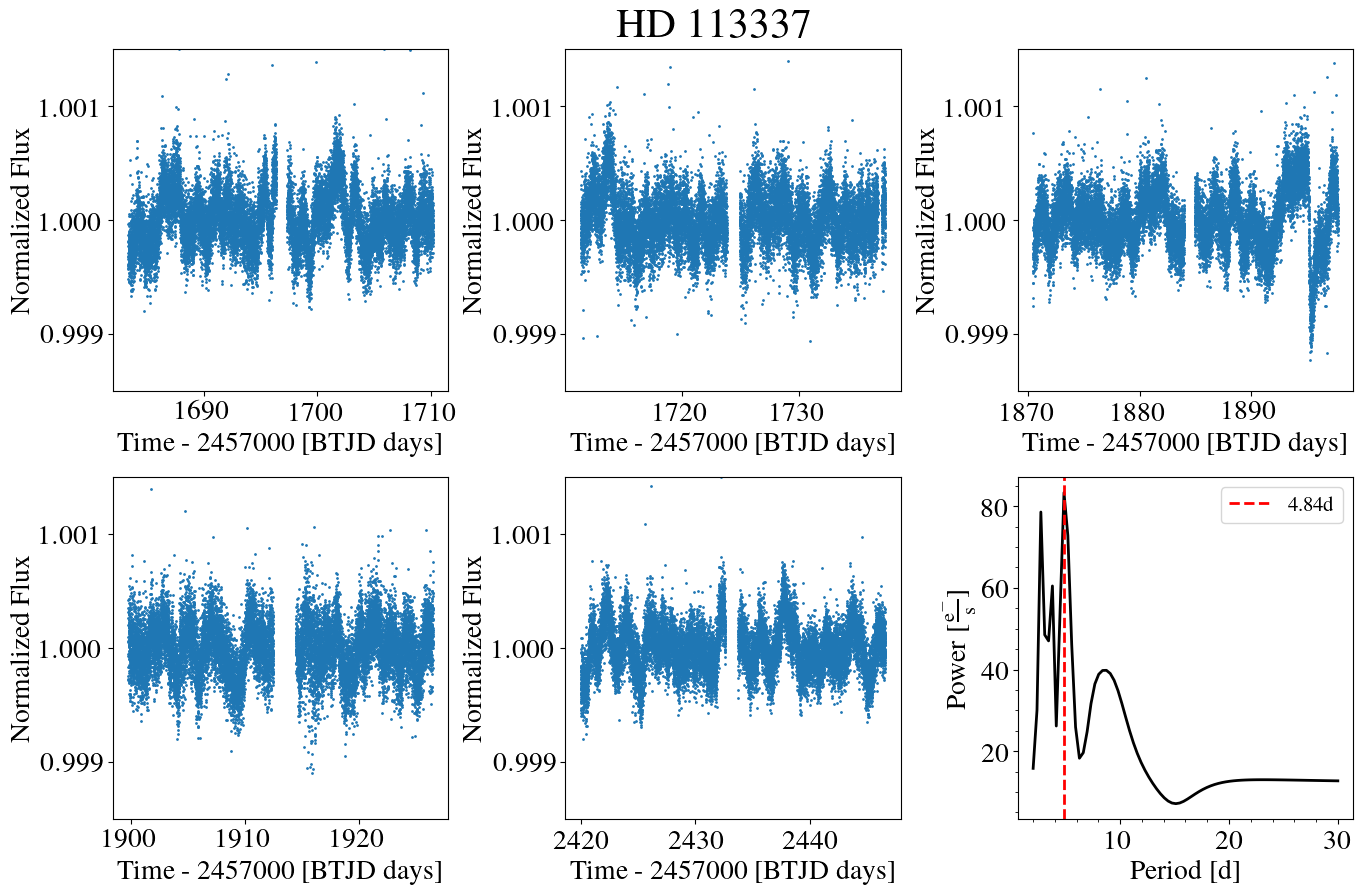}
\caption{\textbf{Measurements of stellar rotational period of HD 113337.} The first five panels show the TESS light curves, and the final panel shows the corresponding periodogram. The red dashed line marks the identified rotation period.}\label{fig_tl2}
\end{figure}

\begin{figure}
\centering
\includegraphics[width=\textwidth]{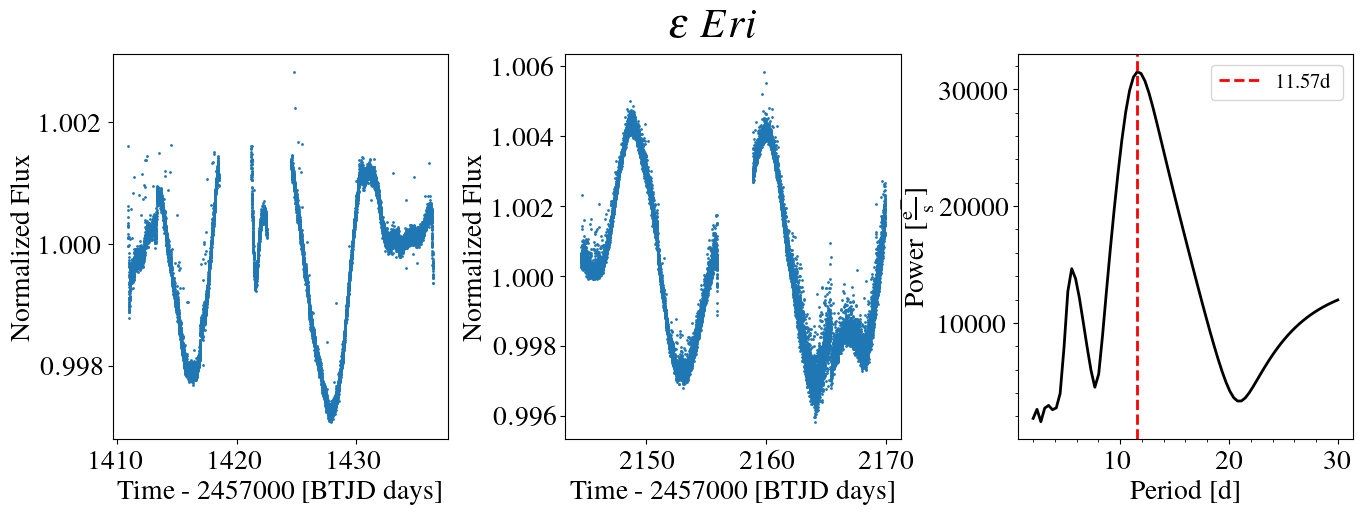}
\caption{\textbf{Measurements of stellar rotational period of $\epsilon$ Eri.} The first two panels show the TESS light curves, and the final panel shows the corresponding periodogram. The red dashed line marks the identified rotation period.}\label{fig_tl4}
\end{figure}
\begin{figure}
\centering
\includegraphics[width=\textwidth]{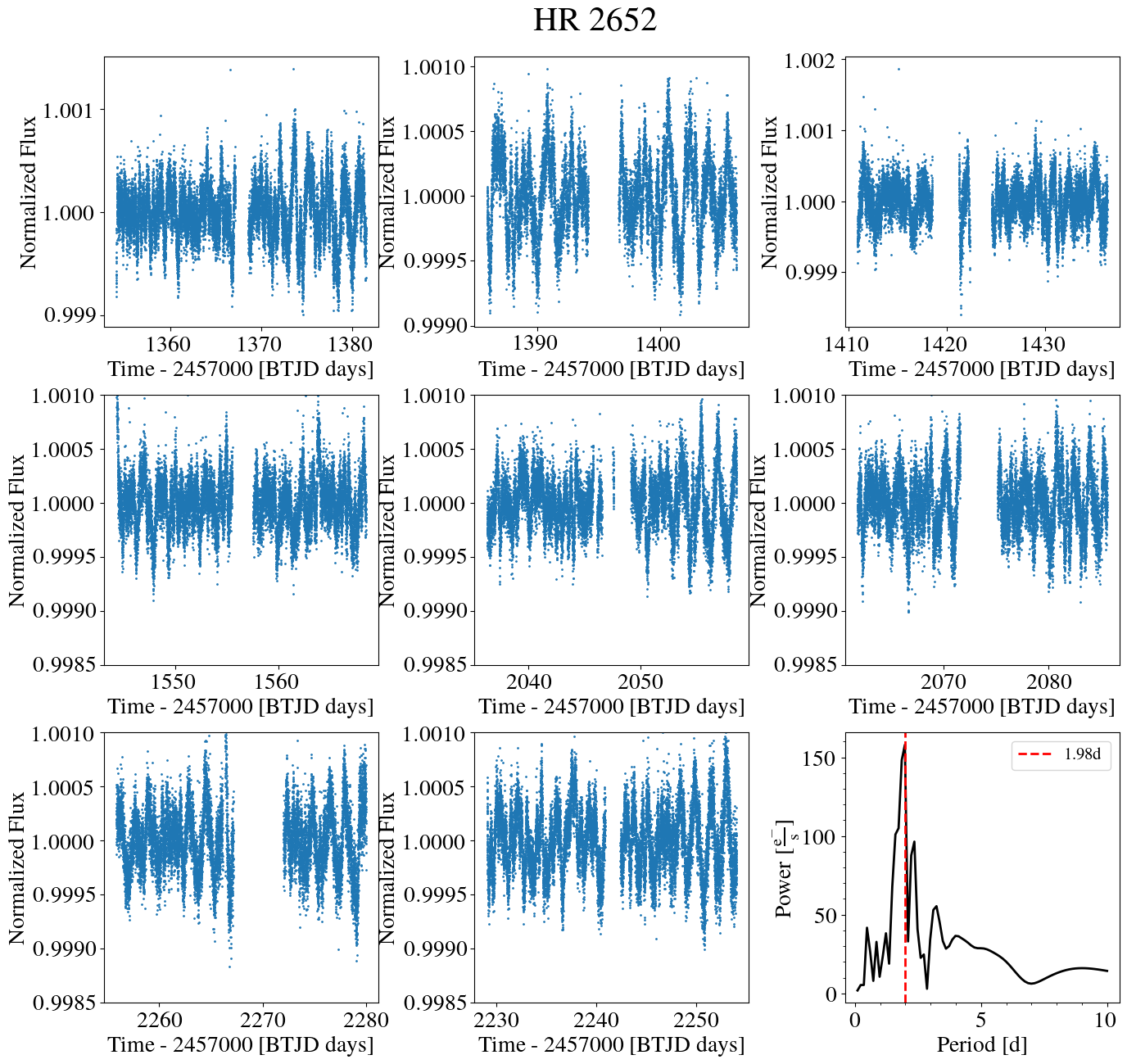}
\caption{\textbf{Measurements of stellar rotational period of HR 2562.} The first eight panels show the TESS light curves, and the final panel shows the corresponding periodogram. The red dashed line marks the identified rotation period.}\label{fig_tl5}
\end{figure}

%%%%%%%%%%%%%%%% SUPPLEMENTARY TABLES %%%%%%%%%%%%%%%
\clearpage
\begin{table}[h]
\caption{\textbf{Summary of stellar parameters in our sample}}\label{tab_stellar_parameter}%
\begin{tabular}{@{}lccccc@{}}
\hline
Name (other identifiers)   & Spectral type & $M_{*}$ & $\rm{T_{eff}}$ & age & Ref.\\
&  &($\rm M_{\odot}$)& (K) & (Myr) &\\
\hline
AU Mic (HD 197481, HIP 102409)  & M1V &    $0.64^{+0.04}_{-0.07}$ & $3540^{+120}_{-110}$& $19^{+10}_{-6}$& \cite{Mallorquin2024}\\
HD 38529 (HIP 2725) & G4IV&  $1.41^{+0.03}_{-0.03}$& $5573^{+31}_{-31}$ & $3090^{+170}_{-170}$& \cite{Jofre2015} \\
HD 114082 (HIP 64184)& F3V& $1.47^{+0.07}_{-0.07}$  & $6651^{+35}_{-35}$ & $15^{+6}_{-6}$ & \cite{Zakhozhay2022}\\
HD 82943 (HIP 47007)& G2V& $1.04^{+0.02}_{-0.02}$  & $6011^{+70}_{-70}$ & $2350^{+330}_{-330}$& \cite{Liu2014} \\
HD 38858 (HIP 27435)&G3V & $0.94^{+0.04}_{-0.04}$  & $5752^{+52}_{-52}$ & $5400^{+2680}_{-3080}$ & \cite{Takeda2007} \\
HD 113337 (HIP 63584) & F6V & $1.40^{+0.14}_{-0.14}$  & $6577^{+153}_{-150}$ & $150^{+100}_{-50}$& \cite{Borgniet2019}\\
HD 10647 ($q^{1}$ Eri, HIP 7978) & F9V & $1.11^{+0.02}_{-0.02}$  & $6218^{+20}_{-20}$  & $1400^{+900}_{-900}$& \cite{Marmier2013}  \\
$\kappa$ CrB (HD 142091, HIP 77655) & K0IV &  $1.45^{+0.09}_{-0.09}$ & $4889^{+48}_{-48}$ & $3286^{+554}_{-554}$& \cite{Sousa2015} \\
HD 216435 ($\tau^{1}$ Gruis, HIP 113044)& G0V & $1.30^{+0.10}_{-0.10}$ & $5993^{+35}_{-35}$ &$3500^{+1500}_{-500}$ &\cite{Ghezzi2010} \\
$\beta$ Pic (HD 39060, HIP 2732)& A5V & $1.76^{+0.18}_{-0.17}$ & $7099^{+250}_{-250}$ & $16.5^{+4.5}_{-4.5}$& \cite{Nielsen2014}\\
$\epsilon$ Eri (HD 22049, HIP 16537) & K2V & $0.82^{+0.02}_{-0.02}$ & $5065^{+131}_{-72}$  & $600^{+200}_{-200}$ & \cite{Llop-Sayson2021}\\
HD 206893 ((HIP 107412) & F5V& $1.32^{+0.07}_{-0.05}$  & $6500^{+100}_{-100}$ & $155^{+15}_{-15}$& \cite{Hinkley2023}\\
HR 8799 (HD 218396, HIP 11418) & A5V & $1.50^{+0.05}_{-0.05}$ & $7204^{+84}_{-123}$ & $60^{+100}_{-30}$ & \cite{Marois2008}\\
HR 2562 (HD 50571, HIP 3277) & F5V & $1.30^{+0.05}_{-0.05}$ & $6597^{+146}_{-134}$ & $600^{+300}_{-300}$ & \cite{Konopacky2016}\\
HD 95086 (HIP 5352)& A2V &  $1.77^{+0.36}_{-0.21}$& $7595^{+96}_{-156}$ & $17^{+4}_{-4}$ & \cite{Rameau2013} \\
\hline
\end{tabular}
\end{table}

\begin{table}
\caption{\textbf{Summary of the orientations of the planetary orbits, debris disks, and stellar spin axes.} The values reported here are MPA and $68\%$ highest probability region }\label{oritation_table}%
\begin{tabular}{@{}lllllllll@{}}
\hline
 Planet  &$\rm{i_{p}}$ & $\rm{\Omega_{p}}$ & $\rm{i_{disk}^{\dagger}}$  & $\rm{\Omega_{disk}^{\dagger\dagger}}$    & $i_*^{\dagger}$ & $\Delta i_{p,d}$& $\Delta i_{p,*}$   \\
Name & $(^{\circ})$  & $(^{\circ})$   &$(^{\circ})$ &$(^{\circ})$  & $(^{\circ})$& $(^{\circ})$ &   $(^{\circ})$ &  \\
\hline
AU Mic b  & $89.6^{+0.4}_{-0.4}$ & - & $89.5^{+0.5}_{-0.5}$   & $128.4^{+0.1}_{-0.1}$    & $88.1^{+1.6}_{-7.7}$ & $0.2^{+0.5}_{-0.2}$ & $1.4^{+7.6}_{-1.4}$ & \\
AU Mic c  &$89.3^{+1.1}_{-1.1}$ &-  &$89.5^{+0.5}_{-0.5}$   & $128.4^{+0.1}_{-0.1}$  &  $88.1^{+1.6}_{-7.7}$ & $0.3^{+1.0}_{-0.3}$ & $1.5^{+7.3}_{-1.4}$ &\\
AU Mic d  &$89.2^{+0.2}_{-0.2}$ & - & $89.5^{+0.5}_{-0.5}$   & $128.4^{+0.1}_{-0.1}$ &  $88.1^{+1.6}_{-7.7}$ & $0.1^{+0.5}_{-0.1}$ & $1.1^{+7.6}_{-1.1}$ & \\
HD 38529 c & $107.8^{+9.6}_{-5.6}$ & $63.2^{+12.8}_{-10.0}$ &  $109^{+10}_{-7}$ & $48^{+5}_{-5}$ &   $112.2^{+9.7}_{-14.2}$  & $2.98_{-2.6}^{+8.6}$ & $3.9^{+9.6}_{-3.9}$ & \\
HD 114082 b & $89.8^{+0.3}_{-0.3}$ & - & $83.2^{+1.1}_{-1.1}$ & $105.7^{+1.4}_{-1.4}$  & -& $6.6^{+1.0}_{-1.2}$ &-\\
HD 114082 c & $89.7^{+0.2}_{-0.2}$ & - & $83.2^{+1.1}_{-1.1}$ & $105.7^{+1.4}_{-1.4}$  &  -& $6.6^{+1.0}_{-1.2}$ &- \\
HD 82943 b & $19.1^{+6.5}_{-4.0}$ & $39.9^{+19.9}_{-25.3}$ & $27^{+4}_{-4}$ & $152^{+8}_{-8}$  & $19.9^{+3.7}_{-3.7}$ & $6.8^{+4.5}_{-5.4}$ & $1.9^{+4.9}_{-1.8}$\\
HD 82943 c & $23.3^{+7.2}_{-5.6}$ & $27.1^{+27.1}_{-21.7}$ & $27^{+4}_{-4}$ & $152^{+8}_{-8}$  & $19.9^{+3.7}_{-3.7}$ & $1.8^{+5.6}_{-1.8}$ &$2.3^{+6.4}_{-2.3}$   \\
HD 38858 b & $46.5^{+20.6}_{-23.2}$ & $41.6^{+296.6}_{-32.6}$ & $44^{+5}_{-5}$& $67^{+5}_{-5}$  &$33.00^{+10.96}_{-12.10}$ & $4.9^{+17.8}_{-4.9}$ &  $8.5^{+18.3}_{-8.4}$ \\
HD 113337 c & $24.4^{+2.0}_{-0.7}$ & $224.1^{+5.8}_{-3.5}$ & $13^{+10}_{-9}$ & $160^{+63}_{-38}$ & $25.5^{+4.8}_{-3.6}$ & $9.8^{+6.4}_{-9.8}$ &  $1.1^{+3.5}_{-1.1}$\\
HD 10647 b & $34.0^{+10.6}_{-8.3}$ & $281.2^{+19.2}_{-20.6}$ & $77.8^{+0.1}_{-0.1}$ & $57.3^{+0.1}_{-0.1}$  &$68.1^{+13.5}_{-8.8}$ & $43.6^{+8.2}_{-10.4}$ &  $37.0^{+11.6}_{-17.4}$ \\
$\kappa$ CrB b  & $16.2^{+1.7}_{-2.9}$ & $110.7^{+7.5}_{-6.7}$ & $62.3^{+3.4}_{-3.4}$ & $148^{+5}_{-5}$ & - & $45.6^{+3.5}_{-5.5}$  & - \\
HD 216435 b & $72.0^{+7.6}_{-13.5}$ & $259.5^{+18.5}_{-21.0}$ & $68.5^{+3.6}_{-5.5}$ & $66.1^{+6}_{-5}$  & - & $3.1^{+8.4}_{-3.1}$  & -\\
$\beta$ Pic b & $89.1^{+0.1}_{-0.1}$ & $32.1^{+0.1}_{-0.1}$ & $85.3_{-0.2}^{+0.2}$ & $30.4^{+0.1}_{-0.1}$  & $87.8^{+1.6}_{-1.6}$& $3.9^{+0.9}_{-1.2}$ &  $2.1^{+1.6}_{-1.8}$\\
$\beta$ Pic c & $89.0^{+0.5}_{-0.6}$ & $31.3^{+0.1}_{-0.1}$ & $85.3_{-0.2}^{+0.2}$ & $30.4^{+0.1}_{-0.1}$  & $87.8^{+1.6}_{-1.6}$& $3.6^{+1.2}_{-0.9}$ & $2.3^{+1.4}_{-2.1}$\\
$\epsilon$ Eri b  & $38.8^{+6.5}_{-4.8}$ & $185.4^{+9.7}_{-7.5}$ & $33.7^{+0.5}_{-0.5}$ & $178.9^{+1.0}_{-1.0}$ & $41.4^{+6.2}_{-5.7}$ & $3.7^{+6.6}_{-3.7}$ &  $2.5^{+6.5}_{-2.5}$\\
HD 206893 B & $149.9^{+2.2}_{-2.7}$ & $79.0^{+6.3}_{-2.4}$ & $140^{+3}_{-3}$ & $61^{+4}_{-4}$   & $151.6^{+1.7}_{-3.2}$ & $9.8^{+3.6}_{-4.2}$ & $0.9^{+2.9}_{-0.9}$ \\ 
HD 206893 c & $146.0^{+2.2}_{-1.7}$ & $73.1^{+1.7}_{-4.6}$ & $140^{+3}_{-3}$& $61^{+4}_{-4}$  & $151.6^{+1.7}_{-3.2}$  & $5.8^{+3.4}_{-3.7}$ & $5.2^{+2.8}_{-3.3}$\\
HR 8799 b & $26.9^{+0.2}_{-0.2}$ & $62.3^{+0.3}_{-0.3}$& $30.8^{+2.8}_{-3.2}$ & $52.4^{+5.7}_{-5.4}$  & $26.8^{+2.0}_{-1.9}$ & $0.9^{+2.7}_{-0.9}$ &   $3.4^{+2.4}_{-3.4}$\\
HR 8799  c & $26.9^{+0.2}_{-0.2}$ & $62.3^{+0.3}_{-0.3}$& $30.8^{+2.8}_{-3.2}$ & $52.4^{+5.7}_{-5.4}$ & $26.8^{+2.0}_{-1.9}$ & $0.9^{+2.7}_{-0.9}$ &  $3.4^{+2.4}_{-3.4}$\\
HR 8799  d &$26.9^{+0.2}_{-0.2}$  & $62.3^{+0.3}_{-0.3}$& $30.8^{+2.8}_{-3.2}$ & $52.4^{+5.7}_{-5.4}$  & $26.8^{+2.0}_{-1.9}$ &$0.9^{+2.7}_{-0.9}$ & $3.4^{+2.4}_{-3.4}$\\
HR 8799 e &$26.9^{+0.2}_{-0.2}$  & $62.3^{+0.3}_{-0.3}$& $30.8^{+2.8}_{-3.2}$ & $52.4^{+5.7}_{-5.4}$  & $26.8^{+2.0}_{-1.9}$ & $0.9^{+2.7}_{-0.9}$& $3.4^{+2.4}_{-3.4}$\\
HR 2562 b & $86.4^{+0.1}_{-0.2}$ & $120.0^{+0.1}_{-0.1}$ & $84.7^{+5.3}_{-5.3}$ & $122.7^{+0.7}_{-0.8}$&  $86.8^{+3.0}_{-9.7}$ &$1.1^{+4.2}_{-1.1}$ &  $1.9^{+7.1}_{-1.9}$  \\
HD  95086 b & $147.2^{+11.8}_{-10.9}$ & $76.8^{+55.1}_{-41.5}$ & $144^{+3}_{-2}$ & $98^{+3}_{-3}$ & - & $2.7^{+9.7}_{-2.5}$  & -\\
\hline
\end{tabular}
\vspace{2pt}
\begin{minipage}{\linewidth}
$*$ For HD 38858, $\kappa$ CrB, and HD 216435, we obtain bimodal posterior distributions in orbital inclination due to the prograde-retrograde degeneracy. Here, we report only the solutions with inclinations below $90^\circ$.\\
$^\dagger$ For disk inclination $i_{\rm disk}$ and stellar inclination $i_*$, there are degenerate solutions at $\pi-i_{\rm disk}$ and $\pi-i_*$. We report the values corresponding to the prograde configurations with respect to the planetary orbits.\\
$^\ddagger$ For disk position angle $\Omega_{\rm disk}$, there are degenerate solutions at $\Omega_{\rm disk}+\pi$.\\
$^\S$ References for disk parameters are: AU Mic\cite{MacGregor2013}, HD 38529\cite{xuan2020_disk}, HD 114082\cite{Engler2025}, HD 82943\cite{Kennedy2013}, HD 38858\cite{Kennedy2015}, HD 113337\cite{xuan2020_disk}, HD 10647\cite{Milli2026}, $\kappa$ CrB\cite{Lovell2022}, HD 216435\cite{Marshall2021}, $\beta$ Pic\cite{MMB2015}, $\epsilon$ Eri\cite{Booth2023}, HD 206893\cite{Marino2020} , HR 8799\cite{Faramaz2021} , HR 2562\cite{Zhang2023}, HD 95086\cite{su2017}. \\
\end{minipage}
%\footnotetext[2]{Example for a second table footnote. This is an example of table footnote.}
\end{table}
%\footnotetext{ $\dagger$ For  disk inclination $i_{\rm{disk}}$ and  stellar inclination $i_*$, there are degenerate solutions at $\pi-i_{\rm{disk}}$ and $\pi-i_*$. We report the values corresponding to the prograde configurations with respect to the planetary orbits.}
%\footnotetext{$\dagger\dagger$ For disk position angle $\Omega_{\rm{disk}}$, there are degenerate solutions at $\Omega_{\rm{disk}}+\pi$.  }
%\footnotetext{$\diamond$ Solar System planets parameters are from JPL/Horizons DE440 \cite{park2021}. The orbital inclination of planets and Kuiper belt are relative to ecliptic plane.  }
%\footnotetext{$\S$ For HD 38858, $\kappa$ CrB, and HD 216435, we obtain bimodal posterior distributions in orbital inclination due to the prograde–retrograde degeneracy. Here, we report only the solutions with inclinations below $90^\circ$.}
\begin{table}\caption{ \textbf{Stellar radii, projected rotational velocities, and rotation periods.}}\label{supp_table_1}
\begin{tabular}{lllllll}
\hline
Name & $\mathrm{R_*}$  & $\mathrm{R_*}$   & $\mathrm{v\sin{i_*}}$ &   $\mathrm{v\sin{i_*}}$   & $\mathrm{P_{rot}}$ & $\mathrm{P_{rot}}$ \\
 & ($\mathrm{R_{\odot}}$)  & ref.& ($ \mathrm{km\ s^{-1}}$) &  ref.  & (d) & ref.  \\
\hline
HD10647 & $1.10\pm0.07$ & \cite{TESSinput} & $5.6\pm0.5$ & \cite{Valenti2005} & $9.2\pm0.1$ & \cite{Hurt2023} \\
 &  &  & $5.8\pm1.9$ & \cite{Borisov2023} & $10\pm3$ & \cite{Marmier2013} \\
 &  &  & $4.9$ & \cite{Marmier2013} & $9.4\pm1$ & This work \\
 &  &  & $5.1$ & \cite{CostaSilva2020} &  &  \\
 &  &  & $5.3$ & \cite{DelgadoMena2015} &  &  \\
 &  &  & $6$ & \cite{Nordstrom2004} &  &  \\
Adopted & $1.10\pm0.07$ &  & $5.4\pm0.3$ &  & $9.4\pm1$ &  \\
\hline
HD38858 & $0.92\pm0.04$ & \cite{TESSinput} & $0.5\pm1$ & \cite{Hojjatpanah2020} & $24.0$ & \cite{Isaacson2010} \\
 &  &  & $0.3\pm0.5$ & \cite{Valenti2005} & $24.7$ & \cite{Fuhrmeister2022} \\
 &  &  & $1.4$ & \cite{CostaSilva2020} &  &  \\
 &  &  & $2.6$ & \cite{Mart2010} &  &  \\
 &  &  & $3$ & \cite{Nordstrom2004} &  &  \\
Adopted & $0.92\pm0.04$ &  & $0.94\pm0.4$ &  & $24.4\pm2.4$ &  \\
\hline
HD38529 & $2.83\pm0.14$ & \cite{TESSinput} & $3.5\pm0.5$ & \cite{Fischer2003} & $35.7$ & \cite{Fischer2003} \\
 &  &  & $3.2\pm0.5$ & \cite{Henry2013} & $37.0\pm0.4$ & \cite{Henry2013} \\
 &  &  & $3.9\pm0.5$ & \cite{Valenti2005} & $37.1\pm0.2$ & \cite{Benedict2010} \\
Adopted & $2.83\pm0.14$ &  & $3.5\pm0.3$ &  & $37\pm4$ &  \\
\hline
HD113337 & $1.52\pm0.07$ & \cite{TESSinput} & $6.3\pm1.0$ & \cite{Borgniet2019} & $3\pm1$ & \cite{Borgniet2019} \\
 &  &  & $8$ & \cite{Nordstrom2004} & $5\pm1$ & This work \\
Adopted & $1.52\pm0.07$ &  & $6.3\pm1.0$ &  & $5\pm1$ &  \\
\hline
AU Mic & $0.82\pm0.02$ & \cite{TESSinput} & $8.5\pm0.2$ & \cite{Donati2023} & $4.86\pm0.02$ & \cite{Messina2017} \\
 &  &  & $9.5\pm0.2$ & \cite{Hojjatpanah2019} & $4.86\pm0.5$ & This work \\
 &  &  & $9.2\pm0.8$ & \cite{Hirano2020} & $4.85$ & \cite{Iba2023} \\
 &  &  & $8.7\pm0.2$ & \cite{Plavchan2020} & $4.83$ & \cite{Lin2024} \\
 &  &  & $7.5\pm1.5$ & \cite{Marfil2021} & $4.86$ & \cite{Seli2022} \\
 &  &  & $7.8\pm0.3$ & \cite{Zicher2022} & $4.89$ & \cite{Fuhrmeister2023} \\
Adopted & $0.82\pm0.02$ &  & $8.7\pm0.2$ &  & $4.86\pm0.5$ &  \\
\hline
$\epsilon$ Eri & $0.76\pm0.05$ & \cite{TESSinput} & $2.4\pm0.5$ & \cite{Valenti2005} & $12$ & \cite{Baliunas1996} \\
 &  &  & $2.4\pm0.4$ & \cite{Marsden2014} & $11.68$ & \cite{Cincunegui2007} \\
 &  &  & $1.3\pm0.6$ & \cite{Soto2018} & $11.68$ & \cite{Seli2022} \\
 &  &  & $2.2\pm0.4$ & \cite{Glebocki2005} & $11.63$ & \cite{Fuhrmeister2022} \\
 &  &  & $2.2\pm0.3$ & \cite{Hojjatpanah2019} & $11.55\pm1$ & This work \\
 &  &  & $4.1$ & \cite{Mart2010} &  &  \\
 &  &  & $4.7$ & \cite{Abdurro2022} &  &  \\
 &  &  & $0.3$ & \cite{Borisov2023} &  &  \\
 &  &  & $1.5$ & \cite{CostaSilva2020} &  &  \\
 &  &  & $2.4$ & \cite{Chavero2019} &  &  \\
Adopted & $0.76\pm0.05$ &  & $2.2\pm0.2$ &  & $11.55\pm1$ &  \\
\hline
\end{tabular}
\end{table}
\begin{table}\addtocounter{table}{-1}\caption{ \textbf{Stellar radii, projected rotational velocities, and rotation periods (continued).}}
\begin{tabular}{lllllll}
\hline
Name & $\mathrm{R_*}$  & $\mathrm{R_*}$   & $\mathrm{v\sin{i_*}}$ &   $\mathrm{v\sin{i_*}}$   & $\mathrm{P_{rot}}$ & $\mathrm{P_{rot}}$ \\
 & ($\mathrm{R_{\odot}}$)  & ref.& ($ \mathrm{km\ s^{-1}}$) &  ref.  & (d) & ref.  \\
\hline
HD82943 & $1.18\pm0.05$ & \cite{TESSinput} & $1.3\pm0.5$ & \cite{Valenti2005} & $18$ & \cite{Mayor2004} \\
 &  &  & $0.7\pm0.4$ & \cite{Soto2018} & $17$ & \cite{Isaacson2010} \\
 &  &  & $1.7$ & \cite{Mayor2004} &  & \\
 &  &  & $3$ & \cite{Nordstrom2004} &  &  \\
 &  &  & $1.4$ & \cite{CostaSilva2020} &  &  \\
 &  &  & $2.8$ & \cite{DelgadoMena2015} &  &  \\
Adopted & $1.18\pm0.05$ &  & $1.1\pm0.3$ &  & $18\pm2$ &  \\
\hline
HR 8799 & $1.49\pm0.05$ & \cite{TESSinput} & $49.2\pm5.4$ & \cite{Glebocki2005} & $0.7$ & \cite{Sepulveda2023} \\
 &  &  & $50.5\pm1$ & \cite{Swastik2021} &  &  \\
 &  &  & $35$ & \cite{Gebran2016} &  &  \\
 &  &  & $49$ & \cite{Royer2007} &  &  \\
Adopted & $1.49\pm0.05$ &  & $48\pm3$ &  & $0.7\pm0.1$ &  \\
\hline
HR 2562 & $1.40\pm0.07$ & \cite{TESSinput} & $57.5\pm6.4$ & \cite{Glebocki2005} & $2\pm0.5$ & This work \\
 &  &  & $43$ & \cite{Zakhozhay2022} &  &  \\
 &  &  & $60$ & \cite{Nordstrom2004} &  &  \\
Adopted & $1.40\pm0.07$ &  & $55\pm5$ &  & $2\pm0.5$ &  \\
\hline
HD206893 & $1.31\pm0.06$ & \cite{TESSinput} & $32\pm2$ & \cite{Delorme2017} & $0.996\pm0.003$ & \cite{Delorme2017} \\
 &  &  & $29$ & \cite{Glebocki2005} &  &  \\
 &  &  & $32.7$ & \cite{Zakhozhay2022} &  &  \\
Adopted & $1.31\pm0.06$ &  & $32\pm2$ &  & $0.996\pm0.003$ &  \\
\hline
\end{tabular}
\end{table}
\begin{table}[h]
\caption{\textbf{Summary of RVs used in orbital fitting}}\label{tab:rv_tab1}%
\begin{tabular}{@{}lllll@{}}
\hline
Name & Instrument  & Number of RVs   & Observation Baseline& Reference\\
\hline
HD38529 & HIRES pre & 48 & 1996-12-01 to 2002-11-24 & \cite{Wright2009} \\
 & Lick Fischer & 136 & 1998-10-14 to 2011-10-12 & \cite{Fischer2001,Fischer2014}\\
 & HRS & 175 & 2004-12-02 to 2008-12-22 & \cite{Benedict2010}\\
 & HIRES post & 95 & 2006-01-15 to 2021-10-24 & \cite{Rosenthal2021} \\
& APF  & 314 & 2013-09-27 to 2020-02-08 & \cite{Rosenthal2021}\\
\hline
HD82943 & Lick Fischer & 7 & 2001-03-11 to 2009-02-01 & \cite{Fischer2014} \\
 & HIRES pre & 22 & 2001-04-07 to 2004-06-25 &  \cite{Lee2006}\\
 & HIRES post & 151 & 2005-01-27 to 2023-05-25 & \cite{Rosenthal2021}\\
\hline
HD38858 & HIRES pre & 17 & 1996-12-01 to 2003-09-15 & \cite{Bulter2006}\\
 & HARPS pre & 100 & 2003-10-29 to 2015-02-08 &  \cite{Kennedy2015}\\
 & HARPS post & 33 & 2015-10-07 to 2021-09-13 & \cite{Flores2018}\\
 & HIRES post & 71 & 2004-10-24 to 2021-08-19 & \cite{Rosenthal2021} \\
  & APF  & 127 & 2013-10-17 to 2020-02-08 & \cite{Rosenthal2021}\\
 & NEID & 69 & 2021-09-18 to 2024-01-20 & \cite{Gupta2025}\\
\hline
HD10647 & UCLES & 28 & 1998-01-17 to 2005-07-22 & \cite{Bulter2006}\\
 & COR98 & 97 & 1998-11-13 to 2007-01-21 &  \cite{Marmier2013}\\
  & COR07 & 16 & 2007-06-27 to 2011-11-03 & \cite{Marmier2013}\\
 & HARPS pre & 139 & 2003-11-06 to 2012-01-05 & \cite{Bisht2024}\\
 & HARPS post & 58 & 2018-10-17 to 2020-01-17 & \cite{Bisht2024} \\
\hline
HD113337 & SOPHIE & 207 & 2007-01-23 to 2022-07-15 & \cite{xuan2020_disk} \\
\hline
$\kappa$ CrB & Lick Fischer & 73 & 2004-04-26 to 2008-07-29 & \cite{Fischer2014}\\
 & HIRES post & 435 & 2010-03-01 to 2022-03-18 & \cite{Rosenthal2021} \\
 & APF & 332 & 2013-11-11 to 2021-12-17 & \cite{Rosenthal2021}\\
\hline
 HD216435 & UCLES & 58 & 1998-08-08 to 2005-11-28 &\cite{Bulter2006} \\
 & HARPS pre & 14 & 2004-10-11 to 2013-11-26 & \cite{Harada2025}\\
 & HARPS post & 40 & 2018-10-18 to 2021-09-15 &\cite{Harada2025} \\
\hline
$\epsilon$ Eri & Lick Fischer & 176 & 1987-09-08 to 2011-10-12 & \cite{Fischer2014}\\
 & HIRES post & 156 & 2009-10-06 to 2020-12-06 & \cite{Rosenthal2021}\\
 & APF & 177 & 2013-10-17 to 2019-11-13 & \cite{Rosenthal2021}\\
\hline
HD206893 & HARPS & 42 & 2016-10-05 to 2021-09-22 & \cite{Hinkley2023}\\
\hline
$\beta$ Pic & HARPS & 90 & 2010-12-12 to 2021-01-29 & \cite{Lagrange2019}\\
\hline
\end{tabular}
%\footnotetext{Source: This is an example of table footnote. This is an example of table footnote.}
%\footnotetext[1]{Example for a first table footnote. This is an example of table footnote.}
%\footnotetext[2]{Example for a second table footnote. This is an example of table footnote.}
\end{table}
\begin{table}[h]
\caption{\textbf{Summary of imaging relative astrometry used in orbital fitting}}\label{tab_rel_astr}%
\begin{tabular}{@{}lllll@{}}
\hline
Name & Number of  & Observation Baseline & Reference\\
 &Relative Astrometry& & \\
\hline
$\beta$ Pic b  &  40 & 2003-11-10 to 2018-11-19 & \cite{Lagrange2009,Chauvin2012, Nielsen2014,wang2016,Lagrange2019} \\
$\beta$ Pic c  &  3 & 2020-02-10 to 2020-03-08 & \cite{Nowak2020}\\
HD206893 B  & 13 & 2019-07-17 to 2025-06-08 & \cite{Milli2017,Delorme2017,WardDuong2021,Kral2026} \\
HD206893 c &  3 & 2021-09-03 to 2021-10-23 & \cite{Hinkley2023} \\
HR2562 B &  20 & 2016-01-25 to 2024-03-19 &   \cite{Konopacky2016,Maire2018,Roberts2025}\\
HD95086 b &  16 & 2012-01-12 to 2023-03-16 & \cite{Rameau2013,Desgrange2022,main2024}\\
\hline
\end{tabular}
\end{table}

%%%%%%%%%%% CAPTIONS FOR OTHER SUPPLEMENTARY FILES %%%%%%%%%%

%\subsection*{Auxiliary Supplementary Materials }
%\subsubsection*{Orbital fitting plots}

%%%%%%%%%%%%%%%% SUPPLEMENTARY REFERENCES %%%%%%%%%%%%%%%

% Do NOT include a reference list in the supplement.
% All references must be in a single list at the end of the main text.
% The copyeditors will ensure that the correct reference list appears with each version of the paper
% (print, HTML, PDF, mobile app, metadata for bibliographic databases etc.)

\end{document}